\documentclass[]{aiaa-tc}

\usepackage{threeparttable}
\usepackage{dcolumn}
  \newcolumntype{d}{D{.}{.}{-1}}
\usepackage{nomencl}
  \makeglossary
\usepackage{subfigure}
\usepackage{subfigmat}
\usepackage{fancyvrb}
  \fvset{fontsize=\footnotesize,xleftmargin=2em}
\usepackage{lettrine}
\usepackage[colorlinks]{hyperref}
\usepackage{graphicx}
\usepackage{dcolumn}
\usepackage{bm}
\usepackage{color, rotating, overpic}
\usepackage{amssymb,amsmath,amsbsy}

\definecolor{blue}{rgb}{0, 0.4470, 0.7410}
\definecolor{red}{rgb}{0.8500, 0.1250, 0.0480} 
\definecolor{green}{rgb}{0.4660, 0.6740, 0.1880}

\newcommand{\eq}{Eq.~}
\newcommand{\eqs}{Eqs.~}
\newcommand{\fig}{Fig.~}
\newcommand{\figs}{Figs.~}

\newcommand{\etal}{{\it{et al}.}}
\def\ip<#1,#2>{\left\langle #1,#2\right\rangle}

\usepackage{cite}

\newcommand{\dmdnext}[1]{{#1}^\#}
\newcommand{\mat}[1]{\bm{#1}}

\newcommand{\mSigma}{\mat{\Sigma}}

\usepackage{amsthm}
\newtheoremstyle{dotless}{}{}{\normalfont}{}{\bfseries}{}{ }{}
\theoremstyle{dotless}

\newtheorem*{inputs}{Inputs:}
\newtheorem*{outputs}{Outputs:}
\newtheorem*{strengths}{Strengths:}
\newtheorem*{weaknesses}{Weaknesses:}

\usepackage{bigfoot}
\DeclareNewFootnote[para]{B}[alph]
\DeclareNewFootnote{default}

\usepackage{multirow}
\usepackage{rotating}

\title{\bf Modal Analysis of Fluid Flows: An Overview}

\author{
  Kunihiko Taira\footnote{Associate Professor, Mechanical Engineering, Associate Fellow AIAA; 
  $^\dagger$Assistant Professor, Mechanical Engineering, Member AIAA;
  $^\ddag$Graduate Research Assistant, Mechanical and Aerospace Engineering, Student Member AIAA;
  $^\S$Professor, Mechanical and Aerospace Engineering, Associate Fellow AIAA;
  $^\P$Professor, Mechanical Engineering, Associate Fellow AIAA;
  $^\parallel$Professor, Aeronautics, Associate Fellow AIAA;
  $^{\ast\ast}$Postdoctoral Research Associate, Mechanical Engineering, Member AIAA;
  $^{\dag\dag}$Associate Professor, Aerospace and Mechanical Engineering, Associate Fellow AIAA;
  $^{\ddag\ddag}$Professor, Aerospace Engineering, Associate Fellow AIAA;
  $^{\S\S}$Associate Professor, Mechanical and Aerospace Engineering, Associate Fellow AIAA.
  }\\
  {\normalsize\itshape Florida State University, Tallahassee, FL 32310, USA}\\
  \and
  Steven L. Brunton$^\dagger$\\ 
  {\normalsize\itshape University of Washington, Seattle, WA, 98195, USA}\\
  \and  
  Scott T. M. Dawson,$^\ddag$ 
  \
  Clarence W. Rowley$^\S$\\ 
  {\normalsize\itshape Princeton University, Princeton, NJ 08544, USA}\\
  \and
  Tim Colonius$^\P$, 
  \
  Beverley J. McKeon$^{\parallel}$, 
  \ 
  Oliver T. Schmidt$^{\ast\ast}$ \\ 
  {\normalsize\itshape California Institute of Technology, Pasadena, CA, 91125, USA}\\
  \and
  Stanislav Gordeyev$^{\dag\dag}$\\ 
  {\normalsize\itshape University of Notre Dame, Notre Dame, IN 46556, USA}\\
  \and
  Vassilios Theofilis$^{\ddag\ddag}$ \\ 
  {\normalsize\itshape University of Liverpool, Brownlow Hill, L69 3GH, UK}\\
  \and  
  Lawrence S. Ukeiley$^{\S\S}$ \\ 
  {\normalsize\itshape University of Florida, Gainesville, FL, 32611, USA}
 }
 
\makeatletter
\renewcommand{\p@subsection}{\thesection~\Alph{subsection}\expandafter\@gobble}
\makeatother

\begin{document}

\maketitle


\section{Introduction}
\label{sec:intro}

Simple aerodynamic configurations under even modest conditions can exhibit complex flows with a wide range of temporal and spatial features.  It has become common practice in the analysis of these flows to look for and extract physically important features, or modes, as a first step in the analysis. This step typically starts with a modal decomposition of an experimental or numerical dataset of the flow field, or of an operator relevant to the system.  We describe herein some of the dominant techniques for accomplishing these modal decompositions and analyses that have seen a surge of activity in recent decades [\citen{Holmes96,Rowley:ARFM17, Kutz2013book,Kutz2016book, Theofilis:ARFM11, Schmid01, Taira:Nagare11a, Taira:Nagare11b}].  For a non-expert, keeping track of recent developments can be daunting, and the intent of this document is to provide an introduction to modal analysis that is accessible to the larger fluid dynamics community.  In particular, we present a brief overview of several of the well-established techniques and clearly lay the framework of these methods using familiar linear algebra.  The modal analysis techniques covered in this paper include the proper orthogonal decomposition (POD), balanced proper orthogonal decomposition (Balanced POD), dynamic mode decomposition (DMD), Koopman analysis, global linear stability analysis, and resolvent analysis.

In the study of fluid mechanics, there can be distinct physical features that are shared across a variety of flows and even over a wide range of parameters such as the Reynolds number and Mach number [\citen{Samimy03, Brown:JFM74}].  Examples of common flow features and phenomena include von K\'arm\'an shedding [\citen{Strouhal:APC78, Rayleigh:PM79,Benard:CAS08, Karman:GN11, Taneda:JPSJ56, Coutanceau:JFM77a, Canuto:JFM15}], Kelvin--Helmholtz instability [\citen{Helmholtz1868, Kelvin1871, Rosenhead1931}], and vortex pairing/merging [\citen{Winant:JFM74, Zaman:JFM80, Melander:JFM88}]. The fact that these features are often easily recognized through simple visual inspections of the flow even under the presence of perturbations or variations provides us with the expectation that the features can be extracted through some mathematical procedure [\citen{Bonnet:EF98}].  We can further anticipate that these dominant features provide a means to describe in a low-dimensional form that appears as complex high-dimensional flow.  Moreover, as computational techniques and experimental measurements are advancing their ability in providing large-scale high-fidelity data, the compression of a vast amount of flow field data to a low-dimensional form is ever more important in studying complex fluid flows and developing models for understanding and modeling their dynamical behavior. 

\begin{figure}[t]
   \centering
   \includegraphics[width=0.77\textwidth]{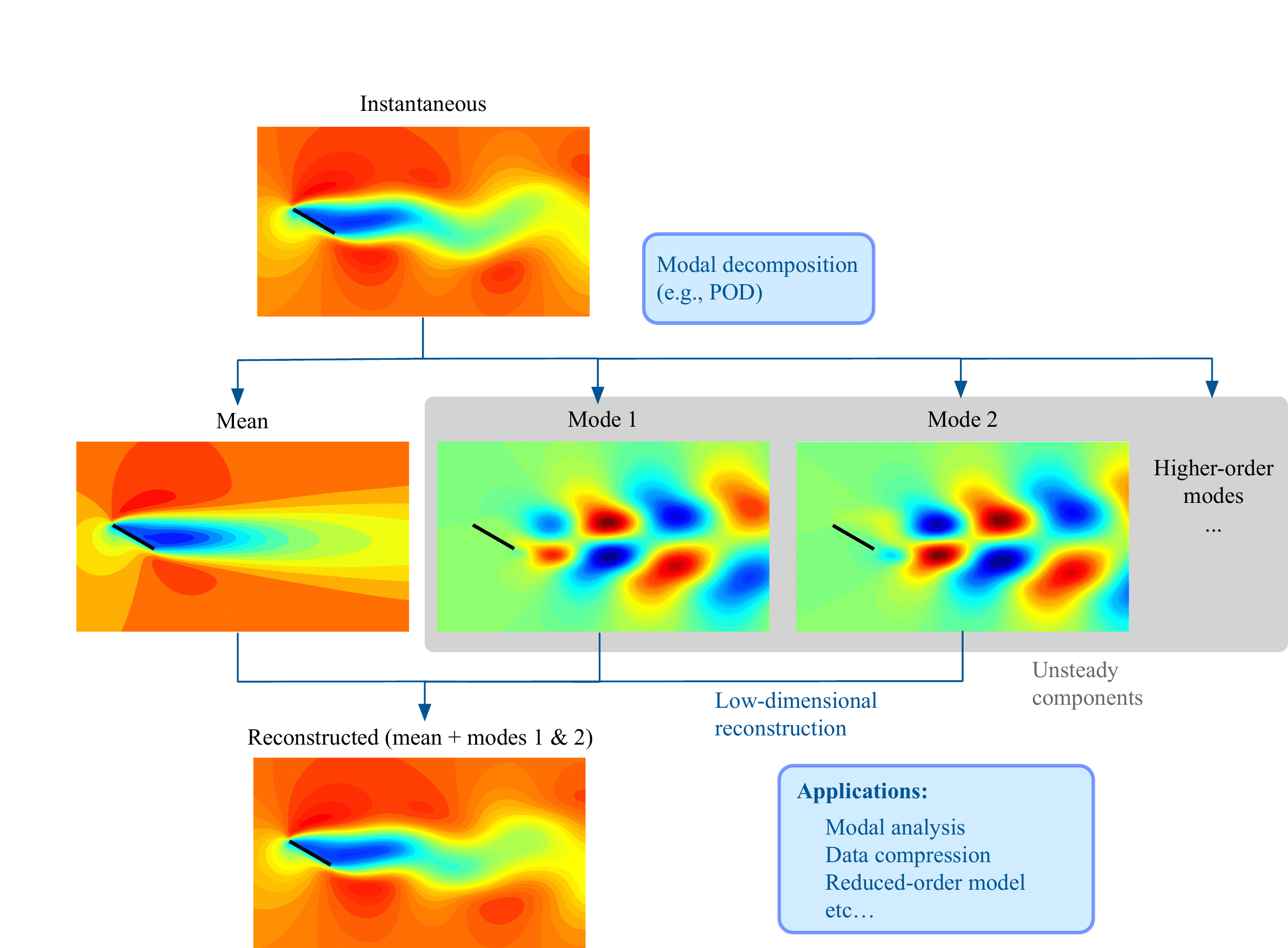}
   \caption{Modal decomposition of two-dimensional incompressible flow over a flat-plate wing [\citen{Taira:JFM09, Colonius:CMAME08}] ($Re= 100$ and $\alpha = 30^\circ$).  This example shows complex nonlinear separated flow being well-represented by only two POD modes and the mean flow field.  Visualized are the streamwise velocity profiles.}
   \label{fig:intro_POD}
\end{figure}

To briefly illustrate these ideas, let us provide a preview of modal decomposition.  In \fig \ref{fig:intro_POD}, we present a modal decomposition analysis of two-dimensional laminar separated flow over a flat-plate wing [\citen{Taira:JFM09, Colonius:CMAME08}].  By inspecting the flow field, we clearly observe the formation of a von K\'arm\'an vortex street in the wake as the dominant unsteady feature.  A modal decomposition method discussed later (proper orthogonal decomposition [\citen{Lumley1967, Holmes96, Sirovich:QAM87}]; see Section \ref{sec:pod}) can extract the important oscillatory modes of this flow.  Moreover, two of these most dominant modes and the mean represent (reconstruct) the flow field very effectively, as shown in the bottom figure.  Additional modes can be included to reconstruct the original flow more accurately, but their contributions are much smaller in comparison to the two unsteady modes shown in this example.  What is also encouraging is that the modes seen here share striking resemblance to the dominant modes for three-dimensional turbulent flow at a much higher Reynolds number of $23,000$ with a different airfoil and angle of attack (see \fig \ref{fig:naca0012}).  

We refer to {\it modal decomposition} as a mathematical technique to extract energetically and dynamically important features of fluid flows.  The spatial features of the flow are called (spatial) {\it modes} and they are accompanied by characteristic values, representing either the energy content levels or growth rates and frequencies.  These modes can be determined from the flow field data or from the governing equations. We will refer to modal decomposition techniques that take flow field data as input to the analysis as {\it data-based} techniques.  This paper will also present modal analysis methods that require a more theoretical framework or discrete operators from the Navier--Stokes equations, and we will refer to them as {\it operator-based} techniques.  

The origin of this document lies with an AIAA Discussion Group, {\it Modal Decomposition of Aerodynamic Flows}, formed under the auspices of the Fluid Dynamics Technical Committee (FDTC).  One of the initial charters for this group was to organize an invited session where experts in the areas of modal decomposition methods would provide an introductory crash course on the methods.  The intended audience for these talks was the non-specialist, e.g. a new graduate student or early-career researcher, who, in one afternoon, could acquire a compact, yet intensive introduction to the modal analysis methods.  This session (121-FC-5) appeared at the 2016 AIAA Aviation conference\footnote{Video recordings of this session have been made available by AIAA on Youtube.} (June 13-17, Washington, DC) and has also provided the foundation for the present overview article.

In this overview document, we present key modal decomposition and analysis techniques that can be used to study a range of fluid flows.  We start by reprising the basics of eigenvalue and singular value decompositions as well as pseudospectral analysis in Section \ref{sec:eig_svd}, which serve as the backbone for all decomposition and analysis techniques discussed here.  We then present data-based modal decomposition techniques: Proper Orthogonal decomposition (POD) in Section \ref{sec:pod}, Balanced POD in Section \ref{sec:bpod}, and Dynamic Mode Decomposition (DMD) in Section \ref{sec:dmd}.  These sections are then followed by discussions on operator-based modal analysis techniques.  The Koopman analysis is briefly discussed in Section \ref{sec:koopman} as a generalization of the DMD analysis to encapsulate nonlinear dynamics using a linear (but infinite-dimensional) operator-based framework.  Global linear stability analysis and resolvent analysis are presented in Sections \ref{sec:global} and \ref{sec:resolvent}, respectively.  Table \ref{table:summary} provides a brief summary of the techniques to facilitate comparison of the methods before engaging in details of each method.

\begin{table}[t]
\begin{center}
\caption{Summary of the modal decomposition/analysis techniques for fluid flows presented in the present paper.  L (linear), NL (nonlinear), C (computational), E (experimental), and NS (Navier--Stokes).}
\label{table:summary}
{\small
\begin{tabular}{c p{3.1cm} l p{2.52cm} p{7.2cm}}\hline
	& Techniques 			& Sections \!\!\!					& Inputs 								& General descriptions \\ \hline \hline
\multirow{3}{*}{\begin{sideways} data-based ~~~~~~ \end{sideways}}  
	& POD 				& \ref{sec:pod}						& data (L or NL flow; C \& E)			& Determines the optimal set of modes to represent data based on $L_2$ norm (energy). \\
	& Balanced POD 		& \ref{sec:bpod} 					& data (L forward \& L adjoint flow; C) 	& Gives balancing and adjoint modes based on input-output relation (balanced truncation). \\
	& DMD 			 \!\!\! & \ref{sec:dmd}  \!\! & data (L or NL flow; C \& E)		& Captures dynamic modes with associated growth rates and frequencies; linear approximation to nonlinear dynamics.  \\ \hline
\multirow{3}{*}{\begin{sideways} operator-based ~~~~~~~ \end{sideways}}
	& Koopman analysis	 \!\!\! & \ref{sec:koopman} \!\! 				& theoretical (also see DMD)			& Transforms nonlinear dynamics into linear representation but with an infinite-dimensional operator; Koopman modes are approximated by DMD modes. \\ 
	& Global linear stability analysis & \ref{sec:global}					& L NS operators \& base flow (C)	& Finds linear stability modes about a base flow (i.e., steady state); assumes small perturbations about base flow.\\
	& Resolvent analysis 	& \ref{sec:resolvent}					& L NS operators \& base flow (C)		& Provides forcing and response modes based on input-output analysis with respect to a base flow (including time-averaged mean flow); can be applied to turbulent flow. \\ \hline
\end{tabular}
}
\end{center}
\end{table}

For each of the methods presented, we provide subsections on overview, description, illustrative examples, and future outlook.   We offer in the Appendix an example of how the flow field data can be arranged into vector and matrix forms in preparation for performing the (data-based) modal decomposition techniques presented here.  At the end of the paper in Section \ref{sec:remarks}, we provide concluding remarks on modal decomposition and analysis methods.

\section*{Preliminaries}

\section{Eigenvalue and Singular Value Decompositions}
\label{sec:eig_svd}

The decomposition methods presented in this paper are founded on the eigenvalue and singular value decompositions of matrices or operators.  In this section, we briefly present some important fundamental properties of the eigenvalue and singular value decomposition techniques.  We also briefly discuss the concepts of pseudospectra and non-normality.   

Eigenvalue decomposition is performed on a square matrix whereas singular value decomposition can be applied on a rectangular matrix.  Analyses based on the eigenvalue decomposition are usually employed when the range and domain of the matrix or operator are the same [\citen{Trefethen97}].  That is, the operator of interest can take a vector and map it into the same space.  Hence, eigenvalue decomposition can help examine the iterative effects of the operator (e.g, $\boldsymbol{A}^k$ and $\exp(\boldsymbol{A} t) = \boldsymbol{I} + \boldsymbol{A} t + \frac{1}{2\!} \boldsymbol{A}^2 t^2 + \cdots$).  

The singular value decomposition on the other hand is performed on a rectangular matrix, which means that the domain and range spaces are not necessarily the same.  As a consequence, singular value decomposition is not associated with analyzing iterative operators.  That is, rectangular matrices cannot serve as propagators.  However, singular value decomposition can be applied on rectangular data matrices compiled from dynamical processes (see Section \ref{sec:EDvsSVD} and Appendix \ref{sec:App} for details). 

The theories and numerical algorithms for eigenvalue and singular value decompositions are not provided here but are discussed extensively in textbooks by Horn and Johnson [\citen{Horn85}], Golub and Loan [\citen{Golub96}], Trefethen and Embree [\citen{Trefethen97}], and Saad [\citen{Saad11}].  Numerical programs and libraries to perform eigenvalue and singular value decompositions are listed in Section \ref{sec:library}.

\subsection{Eigenvalue Decomposition}
\label{sec:eig}

Eigenvalues and eigenvectors of a matrix (linear operator) capture the directions in which vectors can grow or shrink.  For a given matrix $\boldsymbol{A} \in \mathbb{C}^{n\times n}$, a vector $\boldsymbol{v} \in \mathbb{C}^{n}$ and a scalar $\lambda \in \mathbb{C}$ are called an eigenvector and an eigenvalue, respectively, of $\boldsymbol{A}$ if they satisfy
\begin{equation}
   \boldsymbol{A} \boldsymbol{v} = \lambda \boldsymbol{v}.
   \label{eq:ED1}
\end{equation}
Note that the eigenvectors are unique only up to a complex scalar.  That is, if $\boldsymbol{v}$ is an eigenvector, $\alpha \boldsymbol{v}$ is also an eigenvector (where $\alpha \in \mathbb{C}$).  The eigenvectors obtained from computer programs are commonly normalized such that they have unit magnitude.  The set of all eigenvalues\footnote{If the square matrix $\boldsymbol{A}$ is real, its eigenvalues are either real or come in complex conjugate pairs.} of $\boldsymbol{A}$ is called a {\it spectrum} of $\boldsymbol{A}$.  

\begin{figure}[tb]
   \centering
   \includegraphics[width=0.65\textwidth]{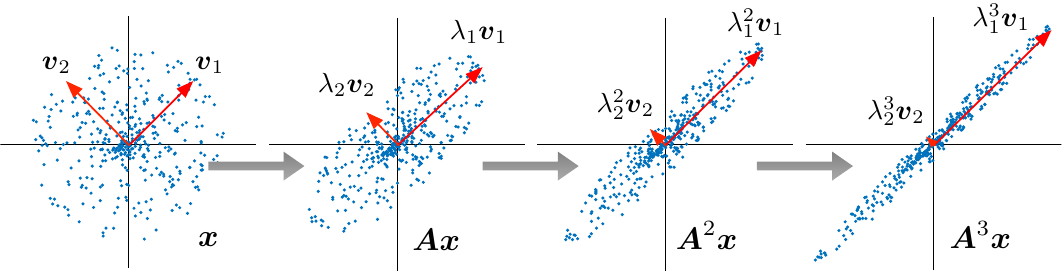}
   \caption{A collection of random points (vectors $\boldsymbol{x}$) stretched in the direction of the dominant eigenvector $\boldsymbol{v}_1$ with iterative operations $\boldsymbol{A}^k$ for matrix $\boldsymbol{A}$ that has eigenvalues of $\lambda_1 = 1.2$ and $\lambda_2 = 0.5$.}
   \label{fig:eig}
\end{figure}

While the above expression in \eq (\ref{eq:ED1}) appears simple, the concept of an eigenvector has great significance in describing the effect of premultiplying $\boldsymbol{A}$ on a vector.  The above expression states that if an operator $\boldsymbol{A}$ is applied to its eigenvector (eigendirection), that operation can be captured solely by the multiplication of a scalar $\lambda$, the eigenvalue associated with that direction.  The magnitude of the eigenvalue tells us whether the operator $\boldsymbol{A}$ will increase or decrease the size of the original vector in that particular direction.  If multiplication by $\boldsymbol{A}$ is performed in an iterative manner, the resulting vector from the compound operations can be predominantly described by the eigenvector having the eigenvalue with the largest magnitude as shown by the illustration in \fig \ref{fig:eig}.

If $\boldsymbol{A}$ has $n$ linearly independent eigenvectors $ \boldsymbol{v}_j$ with corresponding eigenvalues $\lambda_j$ ($j = 1,\dots,n$), then we have
\begin{equation}
   \boldsymbol{A} \boldsymbol{V} = \boldsymbol{V} \boldsymbol{\Lambda},
   \label{eq:ED2}
\end{equation}
where $\boldsymbol{V} = [\boldsymbol{v}_1~ \boldsymbol{v}_2~ \cdots~ \boldsymbol{v}_n] \in \mathbb{C}^{n\times n}$  and $\boldsymbol{\Lambda} = \text{diag}(\lambda_1, \lambda_2, \cdots, \lambda_n)  \in \mathbb{C}^{n\times n}$.  Post-multiplying $\boldsymbol{V}^{-1}$ to the above equation, we have
\begin{equation}
   \boldsymbol{A} = \boldsymbol{V} \boldsymbol{\Lambda} \boldsymbol{V}^{-1}.
   \label{eq:ED3}
\end{equation}
This is called the {\it eigenvalue decomposition}.  For the eigenvalue decomposition to hold, $\boldsymbol{A}$ needs to have a full set of $n$ linearly independent eigenvectors\footnote{In such case, $\boldsymbol{A}$ is called diagonalizable or non-defective.  If $\boldsymbol{A}$ is defective, we have $\boldsymbol{A} = \boldsymbol{V} \boldsymbol{J} \boldsymbol{V}^{-1}$ with $\boldsymbol{J}$ being the canonical Jordan form [\citen{Trefethen05, Golub96}].}.

\begin{figure}[tb]
   \centering
   \includegraphics[width=0.6\textwidth]{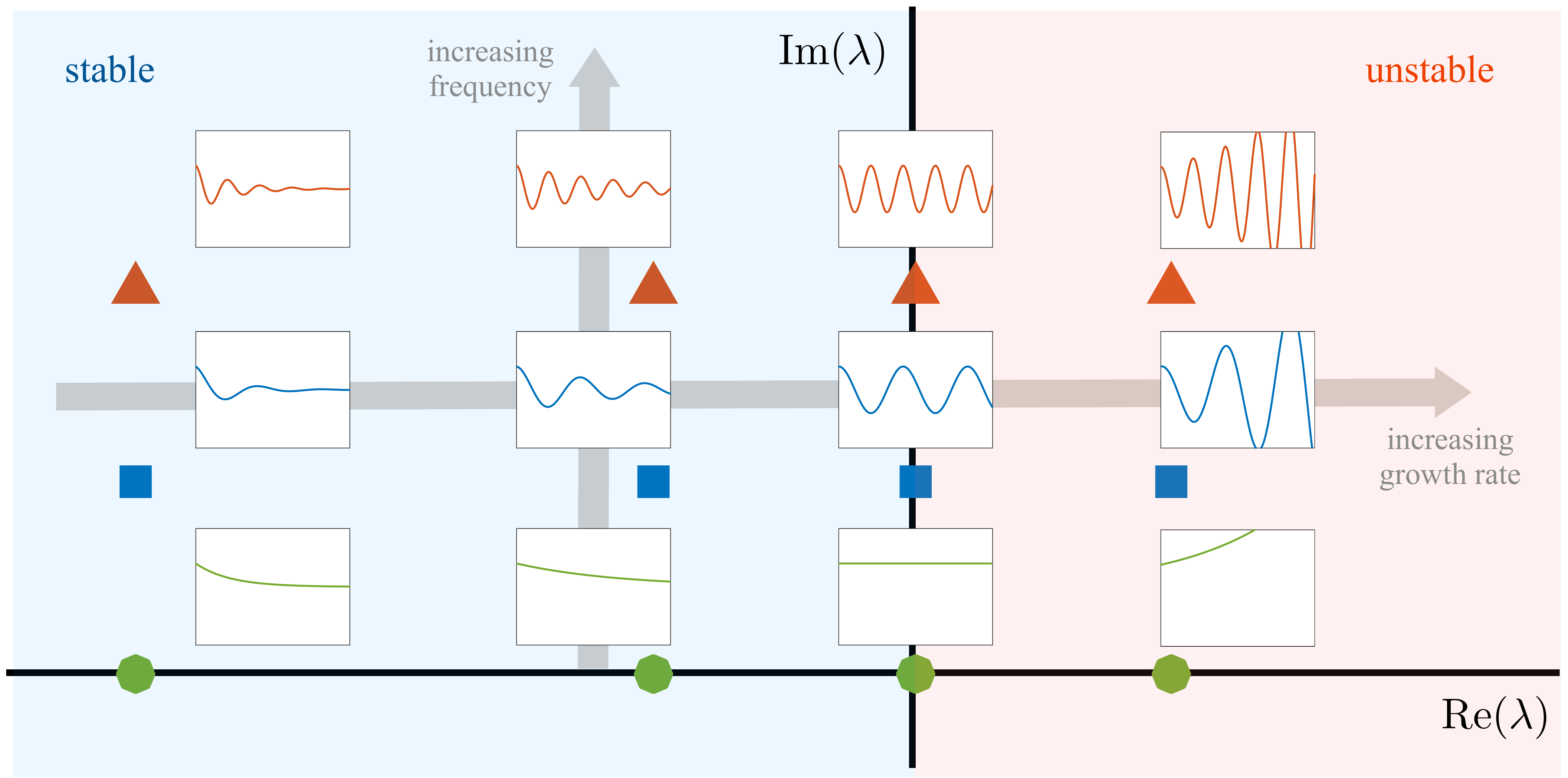}
   \caption{The dynamic response of a linear system characterized by the eigenvalues (stable: $Re(\lambda) < 0$ and unstable: $Re(\lambda) > 0$).  Location of example eigenvalues $\lambda$ are shown by the symbols with corresponding sample solutions $\exp(\lambda t)$ in inset plots.}
   \label{fig:eig_plane}
   \vspace{-0.2in}
\end{figure}

For linear dynamical systems, we often encounter systems for some state variable $\boldsymbol{x}(t) \in \mathbb{C}^n$ described by 
\begin{equation}
   \dot{\boldsymbol{x}}(t) = \boldsymbol{Ax}(t)
   \label{eq:xdoteqAx}
\end{equation}
with the solution of
\begin{equation}
   \boldsymbol{x}(t) = \exp(\boldsymbol{A}t) \boldsymbol{x}(0) = \boldsymbol{V} \exp( \boldsymbol{\Lambda}  t) \boldsymbol{V}^{-1} \boldsymbol{x}(0), 
   \label{eq:expAt}
\end{equation}
where $\boldsymbol{x}(0)$ denotes the initial condition.
Here, the eigenvalues characterize 
the long-term behavior of linear dynamical systems [\citen{Drazin81, Schmid01}] for $\boldsymbol{x}(t)$, as illustrated in \fig \ref{fig:eig_plane}.  The real and imaginary parts of $\lambda_j$ represent the growth (decay) rate and the frequency at which the state variable evolves in the direction of the eigenvector $\boldsymbol{v}_j$.  For a linear system to be stable, all eigenvalues need to be on the left-hand side of the complex plane, i.e., ${\rm Re}(\lambda_j) \le 0$ for all $j$.  

For intermediate dynamics, the pseudospectra [\citen{Trefethen:Science93, Trefethen05, Schmid:ARFM07}] can provide insights.  The concept of pseudospectra is associated with non-normality of operators and the sensitivity of the eigenvalues to perturbations.  We briefly discuss the pseudospectra in Section \ref{sec:pspec}.

For some problems, there can be a mass matrix $\boldsymbol{B} \in \mathbb{C}^{n\times n}$ that appears on the left-hand side of \eq (\ref{eq:xdoteqAx}):
\begin{equation}
   {\boldsymbol{B}} \dot{\boldsymbol{x}} = \boldsymbol{Ax}.
\end{equation}  
In such a case, we are led to a {\it generalized eigenvalue problem} of the form
\begin{equation}
   \boldsymbol{A} \boldsymbol{v} = \lambda \boldsymbol{B} \boldsymbol{v}.
   \label{eq:GED}
\end{equation}
If $\boldsymbol{B}$ is invertible, we can re-write the above equation as
\begin{equation}
   \boldsymbol{B}^{-1} \boldsymbol{A} \boldsymbol{v} = \lambda  \boldsymbol{v}
   \label{eq:GED2ED}
\end{equation}
and treat the generalized eigenvalue problem as a standard eigenvalue problem.  However, it may not be desirable to consider this reformulation if $\boldsymbol{B}$ is not invertible\footnote{The discretization of the incompressible Navier--Stokes equations can yield a singular $\boldsymbol{B}$ since the continuity equation does not have a time derivative term.  Also see Section \ref{sec:global}.} or if the inversion of $\boldsymbol{B}$ results in ill-conditioning (worsening of scaling) of the problem.  Note that generalized eigenvalue problems can also be solved with many numerical libraries, similar to the standard eigenvalue problem, \eq (\ref{eq:ED1}).  See Trefethen and Embree [\citen{Trefethen05}] and Golub \etal~[\citen{Golub96}] for additional details on the generalized eigenvalue problems.


\subsection{Singular Value Decomposition (SVD)}
\label{sec:SVD}

\begin{figure}[tb]
   \centering
   \includegraphics[width=0.48\textwidth]{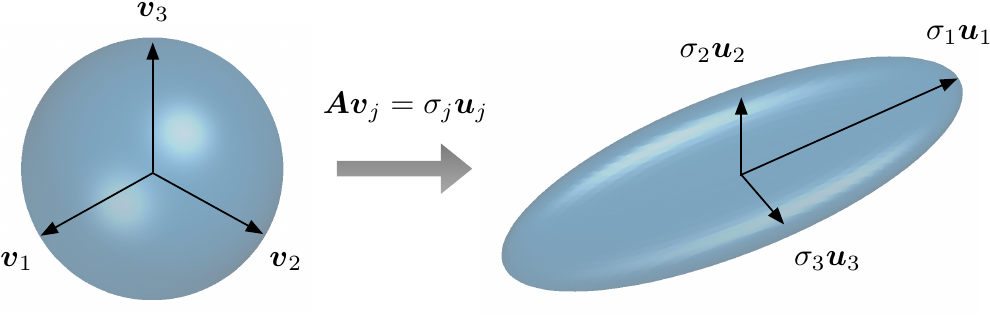}
   \caption{Graphical representation of singular value decomposition transforming a unit radius sphere, described by right singular vectors $\boldsymbol{v}_j$, to an ellipse (ellipsoid) with semiaxes characterized by the left singular vectors $\boldsymbol{u}_j$ and magnitude captured by the singular values $\sigma_j$.  In this graphical example, we take $\boldsymbol{A} \in \mathbb{R}^{3\times 3}$.}
   \label{fig:svd}
\end{figure}

The singular value decomposition is one of the most important matrix factorizations, generalizing the eigendecomposition to rectangular matrices.  
The SVD has many uses and interpretations, especially for dimensionality reduction, where it is possible to use the SVD to obtain {\it{optimal}} low-rank matrix approximations~[\citen{Eckart:Psycho36}].  
The singular value decomposition also reveals how a rectangular matrix or operator stretches and rotates a vector.  As an illustrative example, consider a set of vectors $\boldsymbol{v}_j \in \mathbb{R}^n$ of unit length that describe a sphere.  We can premultiply these unit vectors $\boldsymbol{v}_j$ with a rectangular matrix $\boldsymbol{A} \in \mathbb{R}^{m\times n}$ as shown in \fig \ref{fig:svd}.  The semiaxes of the resulting ellipse (ellipsoid) are represented by the unit vectors $\boldsymbol{u}_j$ and magnitudes $\sigma_j$.  Hence, we can view the singular values to capture the amount of stretching imposed by matrix $\boldsymbol{A}$ in the directions of the axes of the ellipse.

Generalizing this concept for complex $\boldsymbol{A} \in \mathbb{C}^{m\times n}$, $\boldsymbol{v}_j \in \mathbb{C}^n$, and $\boldsymbol{u}_j \in \mathbb{C}^m$, we have
\begin{equation}
   \boldsymbol{A} \boldsymbol{v}_j = \sigma_j \boldsymbol{u}_j.
   \label{eq:SVD1}
\end{equation}
In matrix form, the above relationship can be expressed as
\begin{equation}
   \boldsymbol{A} \boldsymbol{V} = \boldsymbol{U} \boldsymbol{\Sigma}, 
   \label{eq:SVD2}
\end{equation}
where $\boldsymbol{U}  = [\boldsymbol{u}_1~ \boldsymbol{u}_2~ \cdots~ \boldsymbol{u}_m] \in \mathbb{C}^{m \times m}$ and $\boldsymbol{V}  = [\boldsymbol{v}_1~ \boldsymbol{v}_2~ \cdots~ \boldsymbol{v}_n] \in \mathbb{C}^{n\times n}$ are unitary matrices\footnote{Unitary matrices $\boldsymbol{U}$ and $\boldsymbol{V}$ satisfy, $\boldsymbol{U}^* = \boldsymbol{U}^{-1}$, $\boldsymbol{V}^* = \boldsymbol{V}^{-1}$, with $^*$ denoting conjugate transpose.} and $\boldsymbol{\Sigma} \in \mathbb{R}^{m \times n}$ is a diagonal matrix with $\sigma_1 \ge \sigma_2 \ge \dots \ge \sigma_p \ge 0$ along its diagonal, where $p = \min(m,n)$.  Now, multiplying $\boldsymbol{V}^{-1} = \boldsymbol{V}^*$ from the right side of the above equation, we arrive at
\begin{equation}
   \boldsymbol{A} = \boldsymbol{U} \boldsymbol{\Sigma} \boldsymbol{V}^{*}, 
   \label{eq:SVD3}
\end{equation}
which is referred to as the {\it singular value decomposition} (SVD).  In the above equation, $^*$ denotes conjugate transpose.  The column vectors $\boldsymbol{u}_j$ and $\boldsymbol{v}_j$ of $\boldsymbol{U}$ and $\boldsymbol{V}$ are called the left and right singular vectors, respectively.  Both of the singular vectors can be determined up to a complex scalar of magnitude 1 (i.e, $e^{i\theta}$, where $\theta \in [0, 2\pi]$).   

Given a rectangular matrix $\boldsymbol{A}$, we can decompose the matrix with SVD in the following graphical manner
\begin{equation}
\includegraphics[width=0.42\textwidth]{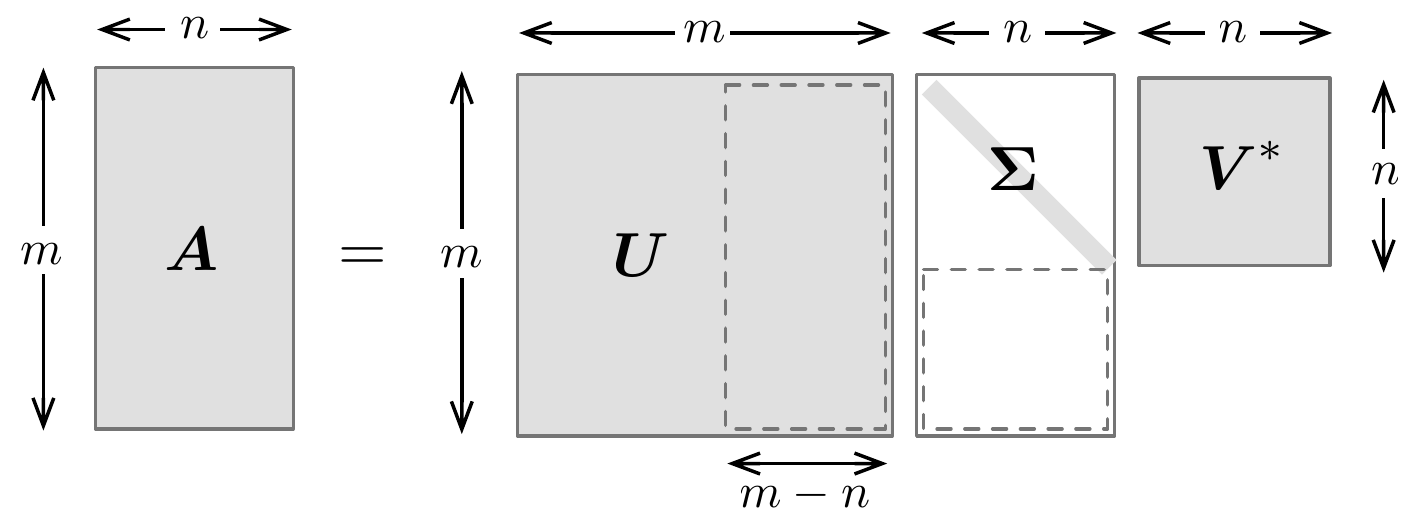}
\label{eq:econsvd}
\end{equation}
where we have taken $m > n$ in this example.  Sometimes the components in $\boldsymbol{U}$ enclosed by the broken lines are omitted from the decomposition, as they are multiplied by zeros in $\boldsymbol{\Sigma}$.  The decomposition that disregards the submatrices in the broken line boxes are called the {\it reduced SVD} ({\it economy-sized SVD}), as opposed to the full SVD.

In a manner similar to the eigenvalue decomposition, we can interpret SVD as a means to represent the effect of matrix operation merely through the multiplication by scalars (singular values) given the appropriate directions.  Because SVD is applied to a rectangular matrix, we need two sets of basis vectors to span the domain and range of the matrix.  Hence, we have the right singular vectors $\boldsymbol{V}$ that span the domain of $\boldsymbol{A}$ and the left singular vectors $\boldsymbol{U}$ that span the range of $\boldsymbol{A}$, as illustrated in \fig \ref{fig:svd}.  This is different from the eigenvalue decomposition of a square matrix, in which case the domain and the range is (generally) the same.  While the eigenvalue decomposition requires the square matrix to be diagonalizable, SVD on the other hand can be performed on {\it any} rectangular matrix.

\subsection{Relationship between Eigenvalue and Singular Value Decompositions}
\label{sec:EDvsSVD}

The eigenvalue and singular value decompositions are closely related.  In fact, the left and right singular vectors of $\boldsymbol{A} \in \mathbb{C}^{m \times n}$ are also the orthonormal eigenvectors of $\boldsymbol{A} \boldsymbol{A}^*$ and $\boldsymbol{A}^* \boldsymbol{A}$, respectively.  Furthermore, the nonzero singular values of $\boldsymbol{A}$ are the square roots of the nonzero eigenvalues of $\boldsymbol{A}\boldsymbol{A}^*$ and $\boldsymbol{A}^* \boldsymbol{A}$.  Therefore, instead of the SVD, the eigenvalue decomposition can be performed on $\boldsymbol{A}\boldsymbol{A}^*$ or $\boldsymbol{A}^* \boldsymbol{A}$ to solve for the singular vectors and singular values of $\boldsymbol{A}$.  For these reasons, the smaller of the square matrices of $\boldsymbol{A}\boldsymbol{A}^*$ and $\boldsymbol{A}^* \boldsymbol{A}$ is often chosen to perform the decomposition in a computationally inexpensive manner compared to the full SVD.  This property is taken advantage of in some of the decomposition methods discussed below since flow field data usually yields a rectangular data matrix that can be very high-dimensional in one direction (e.g., snapshot POD method [\citen{Sirovich:QAM87}] in Section \ref{sec:pod}).


\subsection{Numerical Libraries for Eigenvalue and Singular Value Decompositions}
\label{sec:library}

Eigenvalue and singular value decompositions can be performed with codes that are readily available.  We list a few standard numerical libraries to execute eigenvalue and singular value decompositions.

\paragraph{MATLAB}
In MATLAB\textsuperscript{\textregistered}, the command {\tt eig} finds the eigenvalues and eigenvectors for standard eigenvalue problems as well as generalized eigenvalue problems.   
The command {\tt svd} outputs the singular values and the left and right singular vectors.  It can also perform the economy-sized SVD.  For small to moderate size problems, MATLAB can offer a user-friendly environment to perform modal decompositions.  We provide in Table \ref{table:matlabcodes} some common examples of {\tt eig} and {\tt svd} in use for canonical decompositions.  For additional details, see the documentation available on \url{http://www.mathworks.com}.

\paragraph{LAPACK (Linear Algebra PACKage)}
LAPACK offers standard numerical library routines for a variety of basic linear algebra problems, including eigenvalue and singular value decompositions.  The routines are written in Fortran 90.  See \url{http://www.netlib.org/lapack} and the users' guide [\citen{LapackUsersGuide}].  

\paragraph{ScaLAPACK (Scalable LAPACK)}
ScaLAPACK is comprised of high-performance linear algebra routines for parallel distributed memory machines. ScaLAPACK solves dense and banded eigenvalue and singular value problems. See \url{http://www.netlib.org/scalapack/} and the users' guide [\citen{ScaLapackUsersGuide}].

\paragraph{ARPACK (ARnoldi PACKage)}
ARPACK is a numerical library, written in FORTRAN 77, specialized to handle large-scale eigenvalue problems as well as generalized eigenvalue problems.  It can also perform singular value decompositions.  The library is available both for serial and parallel computations.  See \url{http://www.caam.rice.edu/software/ARPACK} and the users' guide [\citen{Lehoucq98}].

\begin{table}[h]
\begin{center}
\caption{MATLAB {\tt eig} and {\tt svd} examples for eigenvalue and singular value decompositions.}
\label{table:matlabcodes}
\begin{tabular}{llll} \hline
Decomposition                   	& MATLAB code                       &                          & Ref.\\ 
\hline \hline
Eigenvalue decomposition        & {\tt [V,Lambda] = eig(A)}         & ({\tt A*V = V*Lambda})   & \eq (\ref{eq:ED2})\\ 
Generalized eigenvalue decomp.  & {\tt [V,Lambda] = eig(A,B)}       & ({\tt A*V = B*V*Lambda}) & \eq (\ref{eq:GED}) \\ 
\hline
Singular value decomposition    & {\tt [U,Sigma,V] = svd(A)}        & ({\tt A = U*Sigma*V'})   & \eq (\ref{eq:SVD3}) \\
Reduced (economy-sized) SVD     & {\tt [U,Sigma,V] = svd(A,'econ')} & ({\tt A = U*Sigma*V'})   & \eq (\ref{eq:econsvd}) \\
\hline
\end{tabular}
\end{center}
\end{table}


\subsection{Pseudospectra}
\label{sec:pspec}

Before we transition our discussion to the coverage of modal analysis techniques, let us consider the {\it pseudospectral analysis} [\citen{Trefethen:Science93, Trefethen05}], which reveals the sensitivity of the eigenvalue spectra with respect to perturbations to the operator.  This is also an important concept in studying transient and input--output dynamics, complementing the stability analysis based on eigenvalues.  Concepts from pseudospectral analysis appears later in resolvent analysis (Section \ref{sec:resolvent}).

For a linear system described by \eq (\ref{eq:xdoteqAx}) to exhibit stable dynamics, we require all eigenvalues of its operator $\boldsymbol{A}$ to satisfy ${Re}(\lambda_j(\boldsymbol{A})) < 0$, as illustrated in \fig \ref{fig:eig_plane}.  While this criterion guarantees the solution $\boldsymbol{x}(t)$ to be stable for large $t$, it does not provide insights into the transient behavior of $\boldsymbol{x}(t)$.  To illustrate this point, let us consider an example of $\boldsymbol{A} = \boldsymbol{V} \boldsymbol{\Lambda} \boldsymbol{V}^{-1}$ with stable eigenvalues of
\begin{equation}
   \lambda_1= -0.1, 
   \quad
   \lambda_2 = -0.2
   \label{ex:psec1}
\end{equation}
and eigenvectors of 
\begin{equation}
   \boldsymbol{v}_1 = \left[ \cos\left(\frac{\pi}{4}-\delta \right), \sin\left(\frac{\pi}{4}-\delta \right) \right]^T, \quad  
   \boldsymbol{v}_2 = \left[ \cos\left(\frac{\pi}{4}+\delta \right), \sin\left(\frac{\pi}{4}+\delta \right) \right]^T,
   \label{ex:psec2}
\end{equation}
where $\delta$ is a free parameter to choose.  Observe that as $\delta$ becomes small, the eigenvectors become nearly linearly dependent, which makes the matrix $\boldsymbol{A}$ ill-conditioned.  

\begin{figure}[tb]
   \centering
   \includegraphics[width=0.75\textwidth]{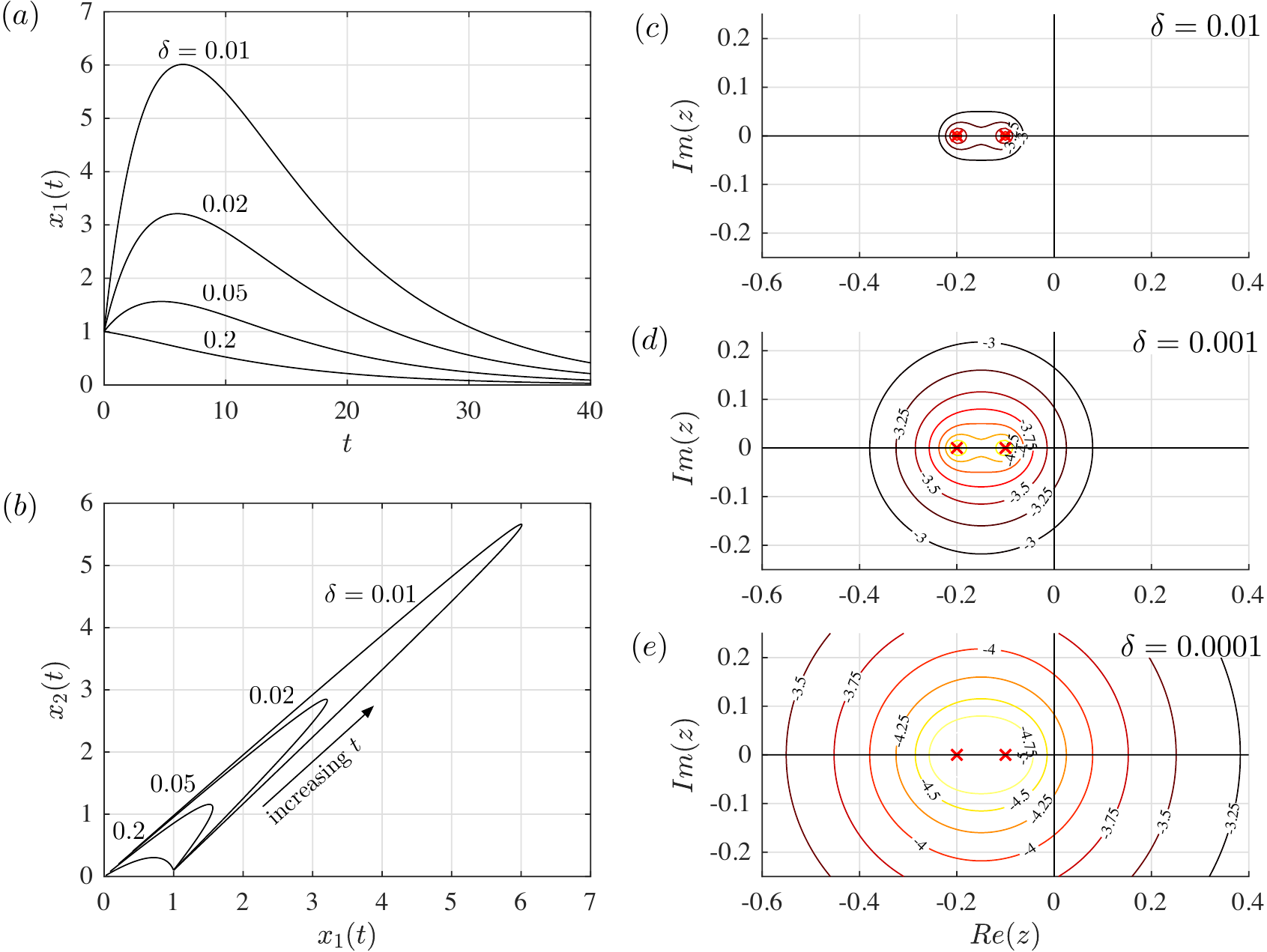}
   \caption{(a) Example of transient growth caused by increasing level of non-normality from decreasing $\delta$.  (b) The trajectories of $x_1(t)$ vs.~$x_2(t)$ exhibiting transient growth. (c)-(e) Pseudospectra expanding for different values of $\delta$.  The $\epsilon$-pseudospectra are shown with values of $\log_{10}(\epsilon)$ placed on the contours.  The stable eigenvalues are depicted with $\mathsf{x}$.}
   \label{fig:pspec}
\end{figure}

Providing an initial condition of $\boldsymbol{x}(t_0) = [1,~0.1]^T$,  we can solve Eq.~(\ref{eq:expAt}) for different values of $\delta$, as shown in \figs \ref{fig:pspec}(a) and (b).  Although all solutions decay to zero due to the stable eigenvalues, the transient growth of ${x}_1(t)$ and ${x}_2(t)$ become noticeable as $\delta \rightarrow 0$.  The large transient for small $\delta$ is caused by the eigenvectors becoming nearly parallel, which necessitates large coefficients to represent the solution (i.e., $\boldsymbol{x}(t) = \alpha_1(t) \boldsymbol{v}_1 + \alpha_2(t) \boldsymbol{v}_2$, where $|\alpha_1|$ and $|\alpha_2| \gg 1$ during the transient).  As such, the solution grows significantly during the transient before the decay from the negative eigenvalues starts to take over the solution behavior at large time.  Thus, we observe that the transient behavior of the solution is {\it{not}} controlled by the eigenvalues of {\it{A}}.  {\it Non-normal} operators (i.e., operators for which $\boldsymbol{AA}^* \neq \boldsymbol{A}^*\boldsymbol{A}$) have non-orthogonal eigenvectors and can exhibit this type of transient behavior.  Thus, it is important that care is taken when we examine transient dynamics caused by non-normal operators.  In fluid mechanics, the dynamics of shear-dominant flows often exhibit non-normality.

To further assess the influence of $\boldsymbol{A}$ on the transient dynamics, let us examine here how the eigenvalues are influenced by perturbations on $\boldsymbol{A}$.  That is, we consider 
\begin{equation}
  \Lambda_\epsilon(\boldsymbol{A}) 
  = \left\{ z\in \mathbb{C}: z \in \Lambda(\boldsymbol{A}+\Delta \boldsymbol{A})
  {\text{~where~}}
  \| \Delta \boldsymbol{A} \| \leq \epsilon \right\}.
  \label{eq:pspec1}
\end{equation}
This subset of perturbed eigenvalues is known as the {\it{$\epsilon$-pseudospectrum}} of $\boldsymbol{A}$.  It is also commonly known with the following equivalent definition:
\begin{equation}
  \Lambda_\epsilon(\boldsymbol{A}) 
  = \left\{ z\in \mathbb{C}: \|z\boldsymbol{I}-\boldsymbol{A}\|^{-1}\geq \epsilon^{-1} \right\}
  \label{eq:pspec2}  
\end{equation}
Note that as $\epsilon\rightarrow 0$, we recover the eigenvalues (0-pseudospectrum) and as $\epsilon\rightarrow\infty$, the subset $\Lambda_\infty(\boldsymbol{A})$ occupies the entire complex domain.  In order to numerically determine the pseudospectra, we can use the following definition based on the minimum singular value of $(z \boldsymbol{I} - \boldsymbol{A})$  
\begin{equation}
  \Lambda_\epsilon(\boldsymbol{A}) 
  = \left\{ z \in \mathbb{C}: \sigma_{\min} (z\boldsymbol{I}-\boldsymbol{A}) \leq \epsilon \right\}, 
  \label{eq:pspec3}
\end{equation}
which is equivalent to $\Lambda_\epsilon(\boldsymbol{A})$ described by \eqs (\ref{eq:pspec1}) and (\ref{eq:pspec2}).  If $\boldsymbol{A}$ is normal, the pseudospectrum $\Lambda_\epsilon(\boldsymbol{A})$ is the set of points away from $\Lambda_0(\boldsymbol{A})$ (eigenvalues) by only less than or equal to $\epsilon$ on the complex plane.  However, as $\boldsymbol{A}$ becomes non-normal, the distance between $\Lambda_0(\boldsymbol{A})$ and $\Lambda_\epsilon(\boldsymbol{A})$ may become much larger.  As discussed later, resolvent analysis in Section \ref{sec:resolvent} considers the pseudospectra along the imaginary axis [\citen{Schmid01}] (i.e., $z \rightarrow \omega i$, where $\omega \in \mathbb{R}$).

Let us return to the example given by \eqs (\ref{ex:psec1}) and (\ref{ex:psec2}) and compute the pseudospectra for decreasing $\delta$ of $0.01$, $0.001$, and $0.0001$, as shown in \figs \ref{fig:pspec}(c), (d), and (e), respectively.  Here, the contours of the $\epsilon$-pseudospectra are drawn for the same values of $\epsilon$.  With decreasing $\delta$, the matrix $\boldsymbol{A}$ becomes increasingly non-normal and susceptible to perturbations.  The influence of non-normality on the spectra is clearly visible with the expanding $\epsilon$-pseudospectra.  It should be noticed that some of the pseudospectra contours penetrate into the right-hand side of the complex plane suggesting that perturbations of such magnitude may thrust the system to become unstable even with stable eigenvalues.  This non-normal feature can play a role in destabilizing the dynamics with perturbations or nonlinearity.

The transient dynamics of $\dot{\boldsymbol{x}} = \boldsymbol{Ax}$ can be related to how the $\epsilon$-pseudospectrum of $\boldsymbol{A}$ expands from the eigenvalues as parameter $\epsilon$ is varied.  The pseudospectra of $\boldsymbol{A}$ can provide a lower bound on the amount of transient amplification by $\exp({\boldsymbol{A}t})$.  If $\Lambda_\epsilon(\boldsymbol{A})$ extends a distance $\eta$ into the right half-plane for a given $\epsilon$, it can be shown through Laplace transform that $\|\exp(\boldsymbol{A}t)\|$ must be as large as $\eta/\epsilon$ for some $t>0$.  If we let a constant $\kappa$ for $\boldsymbol{A}$ to be defined as the supremum of this ratio over all $\epsilon$, the lower bound for the solution can then be shown to take the form of [\citen{Trefethen:AN99}]
\begin{equation}
  \sup_{t\geq 0}\|\exp(\boldsymbol{A}t)\| \geq \kappa.
\end{equation}
This constant $\kappa$ is referred to as the Kreiss constant, which provides an estimate of how the solution, \eq (\ref{eq:expAt}), behaves during the transient.  This estimate is not obtained from the eigenanalysis but from the pseudospectral analysis.  The same concept applies to time-discretized linear dynamics [\citen{Higham:BIT93}].  Readers can find applications of pseudospectral analysis to fluid mechanics in Trefethen \etal~[\citen{Trefethen:Science93}], Trefethen and Embree [\citen{Trefethen05}], and Schmid and Henningson [\citen{Trefethen:Science93}].


\section*{Data-Based Modal Decomposition Methods}

In this section, we consider modal decomposition methods that use flow field data from numerical simulations or experiments.  Below, we discuss the proper orthogonal decomposition, balanced proper orthogonal decomposition, and dynamic mode decomposition.  These techniques require only the output data and do not necessitate the knowledge of the dynamics.

\section{Proper Orthogonal Decomposition (POD)}
\label{sec:pod}

The Proper Orthogonal Decomposition (POD) is a modal decomposition technique that extracts modes based on optimizing the mean square of the field variable being examined.  It was introduced to the fluid dynamics/turbulence community by Lumley [\citen{Lumley1967}] as a mathematical technique to extract coherent structures from turbulent flow fields.  The POD technique, also known as the Karhunen-Lo{\`e}ve (KL) procedure [\citen{Karhunen1946,Loeve1955}], provides an objective algorithm to decompose a set of data into a minimal number of basis functions or modes to capture as much energy as possible.  The method itself is known under a variety of names in different fields: POD, principal component analysis (PCA), Hotelling analysis, empirical component analysis, quasiharmonic modes, empirical eigenfunction decomposition and others. Closely related to this technique is factor analysis, which is used in psychology and economics. Roots of POD can be traced back to the middle of 19th century to the matrix diagonalization technique, which is ultimately related to SVD (Section II).  Excellent reviews on POD can be found in Refs.~[\citen{Berkooz1993}], [\citen{Holmes96}], and Chapter 3 of [\citen{cordier:hal-00417819}].

In applications of POD to a fluid flow, we start with a vector field $\boldsymbol{q}(\boldsymbol{\xi},t)$ (e.g., velocity) with its temporal mean $\overline{\boldsymbol{q}}(\boldsymbol{\xi})$ subtracted and assume that the unsteady component of the vector field can be decomposed in the following manner
\begin{equation}
   {\boldsymbol{q}}({\boldsymbol{\xi}},t) - \overline{\boldsymbol{q}}(\boldsymbol{\xi})
   =\sum_j a_j {\boldsymbol{\phi}}_j({\boldsymbol{\xi}},t),
   \label{eq:gen_fourier_old}
\end{equation}
where $\boldsymbol{\phi}_j (\boldsymbol{\xi},t)$ and $a_j$ represent the modes and expansion coefficients respectively. Here, $\boldsymbol{\xi}$ denotes the spatial vector\footnote{Because we reserve the symbol $\boldsymbol{x}$ to denote the data vector, we use $\boldsymbol{\xi}$ to represent the spatial coordinates in this paper.}.  This expression represents the flow field in terms of a generalized Fourier series for some set of basis functions $\boldsymbol{\phi}_j (\boldsymbol{\xi},t)$.  In the framework of POD, we seek the optimal set of basis functions for a given flow field data. In early applications of POD, this has typically led to modes that are functions of space and time/frequency [\citen{Herzog86, Aubry:JFM88, Moin:JFM82, Glauser87, Ukeiley:AIAAJ93}], as also discussed below.

Modern applications of modal decompositions have further sought to split space and time, hence only needing spatial modes.  In that context, the above equation can be written as 
\begin{equation}
   {\boldsymbol{q}}({\boldsymbol{\xi}},t) - \overline{\boldsymbol{q}}(\boldsymbol{\xi})
   =\sum_j a_j(t) {\boldsymbol{\phi}}_j({\boldsymbol{\xi}}),
   \label{eq:gen_fourier}
\end{equation}
where the expansion coefficients $a_j$ are now time dependent.  Note that Eq.~\eqref{eq:gen_fourier} explicitly employs a separation of variables, which may not be appropriate for all problems.  The application of two forms listed above should depend on the properties of the flow and the information one wishes to extract as discussed in Holmes \etal~[\citen{HLBR-11}].   In what follows, we will discuss the properties of POD assuming that the desire is to extract a spatially dependent set of modes.  

POD is one of the most widely used techniques in analyzing fluid flows.  There are a large number of variations of the POD technique with applications including fundamental analysis of fluids flows, reduced-order modeling, data compression/reconstruction, flow control, and aerodynamic design optimization.  Since POD serves as the basis and motivation for the development of other modal decomposition techniques, we provide a somewhat detailed overview of POD below.

\subsection{Description}

\subsubsection*{Algorithm}

\begin{inputs}
Snapshots of any scalar (e.g., pressure, temperature) or vector (e.g., velocity, vorticity) field, ${\boldsymbol{q}}({\boldsymbol{\xi}},t)$, over one, two, or three-dimensional discrete spatial points $\boldsymbol{\xi}$ at discrete times $t_i$.
\end{inputs}

\begin{outputs}
Set of orthogonal modes, ${\boldsymbol{\phi}}_j({\boldsymbol{\xi}})$, with their corresponding temporal coefficients, $a_j(t)$, and energy levels, $\lambda_j$, arranged in the order of their relative amount of energy. The fluctuations in the original field is expressed as a linear combination of the modes and their corresponding temporal coefficients, ${\boldsymbol{q}}({\boldsymbol{\xi}},t)- \overline{\boldsymbol{q}}(\boldsymbol{\xi})=\sum_j a_j(t){\boldsymbol{\phi}}_j({\boldsymbol{\xi}})$.
\end{outputs}

We discuss three main approaches to perform POD of the flow field data, namely the spatial (classical) POD method, snapshot POD method, and SVD.  Below, we briefly describe these three methods and discuss how they are related to each other.

\subsubsection*{Spatial (Classical) POD Method}

With POD, we determine the set of basis functions that optimally represents the given flow field data.  First, given the flow field ${\boldsymbol{q}}({\boldsymbol{\xi}},t)$, we prepare snapshots of the flow field stacked in terms of a collection of column vectors $\boldsymbol{x}(t)$.   That is, we consider a collection of finite-dimensional data vectors that represents the flow field
\begin{equation}
   \boldsymbol{x}(t) = \boldsymbol{q}(\boldsymbol{\xi},t)-\overline{\boldsymbol{q}}(\boldsymbol{\xi}) \in\mathbb{R}^n, \quad t = t_1, t_2, \dots, t_m.
\end{equation}
Here, $\boldsymbol{x}(t)$ is taken to be the fluctuating component of the data vector with its time-averaged value $\overline{\boldsymbol{q}}(\boldsymbol{\xi})$ removed.  While the data vector can be written as $\boldsymbol{x}(\boldsymbol{\xi},t)$, we simply write $\boldsymbol{x}(t)$ to emphasize that it is being considered as a snapshot at time $t$.  An example of forming the data vector $\boldsymbol{x}(t)$ for a given flow field is provided in Appendix \ref{sec:App}.  

The objective of the POD analysis is to find the optimal basis vectors that can best represent the given data.  In other words, we seek the vectors ${\boldsymbol{\phi}}_j({\boldsymbol{\xi}}) $ in \eq (\ref{eq:gen_fourier}) that can represent $\boldsymbol{q}(\boldsymbol{\xi})$ in an optimal manner and with the least number of modes.
The solution to this problem [\citen{Eckart:Psycho36}] can be determined by finding the eigenvectors $\phi_j$ and the eigenvalues $\lambda_j$ from 
\begin{equation}
  \boldsymbol{R} \boldsymbol{\phi}_j = \lambda_j \boldsymbol{\phi}_j, 
  \quad 
  \boldsymbol{\phi}_j \in \mathbb{R}^n, 
  \quad 
  \lambda_1 \ge \dots \ge \lambda_n \ge 0,
  \label{eqn:Reig}
\end{equation}
where $\boldsymbol{R}$ is the covariance matrix\footnote{Precisely speaking the covariance matrix is defined as $\boldsymbol{R}\equiv \boldsymbol{XX}^T/m$ or $\boldsymbol{XX}^T/(m-1)$.  For clarity of presentation, we drop the factor $1/m$ and note that it is lumped into the eigenvalue $\lambda_k$.} of vector $\boldsymbol{x}(t)$
\begin{equation}
   {\boldsymbol{R}} = \sum_{i=1}^m \boldsymbol{x}(t_i) \boldsymbol{x}^T(t_i) = \boldsymbol{XX}^T \in \mathbb{R}^{n\times n},
   \label{eq:cov_cont}
\end{equation}
where the matrix $\boldsymbol{X}$ represents the $m$ snapshot data being stacked into a matrix form of
\begin{equation}
   {\boldsymbol{X}} = \left[ {\boldsymbol{x}}(t_1)~{\boldsymbol{x}}(t_2)~\dots~{\boldsymbol{x}}(t_m) \right] 
   \in \mathbb{R}^{n\times m}.
   \label{eq:datamatrix}
\end{equation}
The size of the covariance matrix $n$ is based on the spatial degrees of freedom of the data.  For fluid flow data, $n$ is generally large and is equal to the number of grid points times the number of variables to be considered in the data, as illustrated in \eq (\ref{eq:stack}) of the Appendix.  See the Appendix for an example of preparing the data matrix from the velocity field data.  

The eigenvectors found from Eq.~(\ref{eqn:Reig}) are called the {\it POD modes}.  It should be noted that the POD modes are orthonormal.  That means that the inner product\footnote{For the sake of discussion, we consider the flow field data to be placed on a uniform grid such that scaling due to the size of the cell volume does not need to be taken into account.  In general, cell volume for each data point needs to be included in the formulation to represent this inner product (volume integral).  Consequently, the covariance matrix, \eqs (\ref{eqn:Reig}) and (\ref{eq:cov_cont}) should be written as $\boldsymbol{R} \equiv \boldsymbol{XX}^T \boldsymbol{W}$, where $\boldsymbol{W}$ holds the spatial weights.  The matrix $\boldsymbol{X}^T \boldsymbol{X}$ that later appears in \eq (\ref{eq:eig_snapshot}) for the method of snapshot would similarly be replaced by $\boldsymbol{X}^T \boldsymbol{WX}$.} between the modes satisfy
\begin{equation}
   \left< \boldsymbol{\phi}_j, \boldsymbol{\phi}_k \right> 
   \equiv \int_V \boldsymbol{\phi}_j \cdot \boldsymbol{\phi}_k {\rm d}V = \delta_{jk}, \quad j,k=1,\dots,n.
   \label{eq:IP}
\end{equation}
Consequently, the eigenvalues $\lambda_k$ convey how well each eigenvector $\boldsymbol{\phi}_k$ captures the original data in the $L_2$ sense (scaled by $m$).  When the velocity vector is used for $\boldsymbol{x}(t)$, the eigenvalues correspond to the kinetic energy captured by the respective POD modes.  If the eigenvalues are arranged from the largest to the smallest in decreasing order, the POD modes are arranged in the order of importance in terms of capturing the kinetic energy of the flow field.  

We can use the eigenvalues to determine the number of modes needed to represent the fluctuations in the flow field data.  
Generally, we retain only $r$ number of modes to express the flow such that 
\begin{equation}
  \sum_{j=1}^{r} \lambda_j / \sum_{j=1}^{n} \lambda_j \approx 1.
  \label{eq:energy_sum}
\end{equation}
With the determination of the important POD modes, we can represent the flow field only in terms of finite or truncated series,
\begin{equation}
   {\boldsymbol{q}}({\boldsymbol{\xi}},t) - \overline{\boldsymbol{q}}(\boldsymbol{\xi})
   \approx\sum_{j=1}^r a_j(t) {\boldsymbol{\phi}}_j(\boldsymbol{\xi})
\end{equation}
in an optimal manner, effectively reducing the high-dimensional ($n$) flow field to be represented only with $r$ modes.  The temporal coefficients are determined accordingly by 
\begin{equation}
   a_j(t) = \left< \boldsymbol{q}(\boldsymbol{\xi},t)-\overline{\boldsymbol{q}}(\boldsymbol{\xi}), \boldsymbol{\phi}_j(\boldsymbol{\xi}) \right> = \left< \boldsymbol{x}(t), \boldsymbol{\phi}_j\right>.
\end{equation}

\subsubsection*{Method of Snapshots}

When the spatial size of the data $n$ is very large, the size of the correlation matrix $\boldsymbol{R} = \boldsymbol{XX}^T$ becomes very large ($n \times n$), making the use of the classical spatial POD method for finding the eigenfunctions practically impossible. Sirovich [\citen{Sirovich:QAM87}] pointed out that the temporal correlation matrix will yield the same dominant spatial modes, while giving rise to a much smaller and computationally more tractable eigenvalue problem.  This alternative approach, called the {\it method of snapshots}, takes a collection of snapshots $\boldsymbol{x}(t_i)$ at discrete time levels $t_i$, $i = 1$, $2$, $\dots$, $m$, with $m \ll n$, and solves an eigenvalue problem of a smaller size ($m \times m$) to find the POD modes.  The number of snapshots $m$ should be chosen such that important fluctuations in the flow field are well resolved in time.

The method of snapshots relies on solving an eigenvalue problem of a much smaller size 
\begin{equation}
   {\boldsymbol{X}}^T {\boldsymbol{X}} \boldsymbol{\psi}_j = \lambda_j \boldsymbol{\psi}_j, 
   \quad \boldsymbol{\psi}_j \in \mathbb{R}^m, ~m\ll n,
   \label{eq:eig_snapshot}
\end{equation}
where ${\boldsymbol{X}}^T{\boldsymbol{X}}$ is of size ${m\times m}$ instead of the original eigenvalue problem of size $n \times n$, \eq (\ref{eqn:Reig}).  Although we are analyzing the smaller eigenvalue problem, the same nonzero eigenvalues are shared by ${\boldsymbol{X}}^T {\boldsymbol{X}}$ and ${\boldsymbol{X}} {\boldsymbol{X}}^T$ and the eigenvectors of these matrices can be related to each other (see Section \ref{sec:EDvsSVD}).  With the eigenvectors $ \boldsymbol{\psi}_j$ of the above smaller eigenvalue problem determined, we can recover the POD modes through
\begin{equation}
   \boldsymbol{\phi}_j = {\boldsymbol{X}} \boldsymbol{\psi}_j \frac{1}{\sqrt{\lambda_j}} \in \mathbb{R}^n,
   \quad
   j = 1, 2, \dots, m,
   \label{eq:eig_transform}
\end{equation}
which can equivalently be written in matrix form as
\begin{equation}
   {\boldsymbol{\Phi}} = {\boldsymbol{X}} {\boldsymbol{\Psi}} {\boldsymbol{\Lambda}}^{-1/2},
   \label{eq:eig_transform2}
\end{equation}
where ${\boldsymbol{\Phi}} = [\boldsymbol{\phi}_1~\boldsymbol{\phi}_2~ \dots ~\boldsymbol{\phi}_m]\in \mathbb{R}^{n\times m}$ and $\Psi = [\boldsymbol{\psi}_1~\boldsymbol{\psi}_2 \dots \boldsymbol{\psi}_m]\in \mathbb{R}^{m\times m}$.

Due to the significant reduction in the required computation and memory resource, the method of snapshots has been widely used to determine POD modes from high-dimensional fluid flow data.  In fact, this snapshot-based approach is presently the most widely used POD method in fluid mechanics.  

\subsubsection*{SVD and POD}

Let us consider the relation between POD and SVD, as discussed in Section \ref{sec:EDvsSVD}.  
Recall that SVD [\citen{Golub96, Trefethen97, Kutz2013book}] can be applied to a rectangular matrix to find the left and right singular vectors.  In matrix form, a data matrix $\boldsymbol{X}$ can be decomposed directly with SVD as
\begin{equation}
   {\boldsymbol{X}} = {\boldsymbol{\Phi}} {\boldsymbol{\Sigma}} {\boldsymbol{\Psi}}^T,
   \label{eq:X_svd}
\end{equation}
where ${\boldsymbol{\Phi}} \in \mathbb{R}^{n \times n}$, ${\boldsymbol{\Psi}} \in \mathbb{R}^{m \times m}$, and ${\boldsymbol{\Sigma}} \in \mathbb{R}^{n \times m}$, with $m < n$.  The matrices ${\boldsymbol{\Phi}}$ and ${\boldsymbol{\Psi}}$ contain the left and right singular vectors\footnote{The matrices ${\boldsymbol{\Phi}}$ and ${\boldsymbol{\Psi}}$ are orthonormal, i.e, ${\boldsymbol{\Phi}}^T {\boldsymbol{\Phi}} = {\boldsymbol{\Phi}} {\boldsymbol{\Phi}}^T =  \boldsymbol{I}$ and ${\boldsymbol{\Psi}}^T {\boldsymbol{\Psi}} = {\boldsymbol{\Psi}} {\boldsymbol{\Psi}}^T = \boldsymbol{I}$.} of ${\boldsymbol{X}}$ and matrix ${\boldsymbol{\Sigma}}$ holds the singular values ($\sigma_1$, $\sigma_2$, $\dots$, $\sigma_m$) along its diagonal.  These singular vectors ${\boldsymbol{\Phi}}$ and ${\boldsymbol{\Psi}}$ are identical to the eigenvectors of ${\boldsymbol{XX}}^T$ and ${\boldsymbol{X}}^T {\boldsymbol{X}}$, respectively.  Moreover, the singular values and the eigenvalues are related by $\sigma_j^2  = \lambda_j$.  This means that SVD can be directly performed on ${\boldsymbol{X}}$ to determine the POD modes ${\boldsymbol{\Phi}}$.  Note that SVD finds ${\boldsymbol{\Phi}}$ from \eq (\ref{eq:X_svd}) with a full size of $n \times n$ compared to $\boldsymbol{\Phi}$ from the snapshot approach with size of $n \times m$, holding only the leading $m$ modes.

The terms POD and SVD are often used interchangeably in the literature.  However, SVD is a decomposition technique for rectangular matrices and POD can be seen as a decomposition formalism for which SVD can be one of the approaches to determine its solution.  While the method of snapshot is preferred for handling large data sets, the SVD based technique to determine the POD modes is known to be robust against roundoff errors [\citen{Kutz2013book}].

\subsubsection*{Notes}

\paragraph{Optimality}
The POD modes are computed in the optimal manner in the $L_2$ sense [\citen{Holmes96}].  If the velocity or vorticity field is used to determine the POD modes, the modes are optimal to capture the kinetic energy or enstrophy, respectively, of the flow field.  Moreover, POD decomposition is optimal not only in terms of minimizing mean-square error between the signal and its truncated representation, but also minimizing the number of modes required to describe the signal for a given error [\citen{Algazi1969}].

The optimality (the fastest-convergent property) of POD reduces the amount of information required to represent statistically-dependent data to a minimum. This crucial feature explains the wide usage of POD in a process of analyzing data.  For this reason, POD is used extensively in the fields of detection, estimation, pattern recognition, and image processing. 

\paragraph{Reduced-order modeling}
The orthogonality of POD modes, $\left< \boldsymbol{\phi}_j, \boldsymbol{\phi}_k \right> = \delta_{jk}$, is an attractive property for constructing reduced-order models [\citen{Holmes96, Noack11}].  Galerkin projection can be utilized to reduce high-dimensional discretizations of partial differential equations into reduced-order ordinary-differential equation models for the temporal coefficients $a_j(t)$.  POD modes have been used to construct Galerkin projection based reduced-order models for incompressible [\citen{Aubry:JFM88, Ukeiley:JFM01, Noack:JFM03, Noack:JFM05}] and compressible [\citen{Rowley:PhysicaD04}] flows.

\paragraph{Traveling structures}
With real-valued POD modes, traveling structures cannot be represented as a single mode.  In general, traveling structures are represented by a pair of stationary POD modes, which are similar but appear shifted in the advection direction.  See, for instance, modes 1 and 2 in \figs \ref{fig:intro_POD} and \ref{fig:naca0012}.  One way to understand the emergence of POD mode pairs for traveling sturctures is to consider the following traveling sine wave example
\begin{equation}
  \sin(\xi-ct) = \cos (ct) \sin(\xi) - \sin(ct) \cos(\xi).
\end{equation}
Here, we have a pair of spatial modes, $\sin(\xi)$ and $\cos(\xi)$, which are shifted by a phase of $\pi/2$.  Note that we can combine the pair of modes into a single mode if a complex representation of POD modes is considered [\citen{Nair:arXiv17}].  There are variants of POD analysis, specialized for traveling structures [\citen{Glavaski:IEEE98, Noack:JFM16}].

\paragraph{Constraints}
With linear superposition of POD modes in representing the flow field, each and every POD mode also satisfies linear constraints, such as the incompressibility constraint and the no-slip boundary condition.  This statement assumes that the given data also satisfy these constraints.

\paragraph{Homogeneous directions}
For homogeneous, periodic, or stationary (translationally invariant) directions, POD modes reduce to Fourier modes [\citen{Lumley1970}].  

\paragraph{Spectral POD}
\label{sec:spod}

It is also possible to consider the use of POD in frequency domain.
Spectral POD provides time-harmonic modes at discrete frequencies from a set of realizations of the temporal Fourier transform of the flow field. 
This application of the POD provides an orthogonal basis of modes at discrete frequencies, that are optimally ranked in terms of energy, since the 
POD reduces to harmonic analysis over directions that are stationary or periodic [\citen{Lumley1970, George88}].  Hence, the method is effective at extracting coherent structures from statistically 
stationary flows and has been successfully applied in the early applications of the POD, as well as other flows including turbulent 
jets [\citen{CitrinitiGeorge2000, PicardDelville2000, Suzuki2006, Gudmundsson2011, SchmidtEtAl16}]. This frequency based approach to POD overcomes 
some of the weaknesses, and the discussions in the strengths and weaknesses do not directly apply here [\citen{TowneEtAl2017inPreparation}].

Spectral POD can be estimated from a time-series of snapshots in the form of Eq.~(\ref{eq:datamatrix}) using Welch's method [\citen{Welch1967}].   
First, the data is segmented into a number of $n_b$ (of potentially overlapping) blocks or realizations, consisting of $m_{\text{FFT}}$ snapshots each. 
Under the ergodicity hypothesis, each block can be regarded as statistically independent realization of the flow. We proceed by calculating the temporal 
Fourier transform 
\begin{equation} \label{eq:blockfft}
  \hat{\boldsymbol{X}}^{(l)} = \left[ \hat{\boldsymbol{x}}(\omega_1)^{(l)}~\hat{\boldsymbol{x}}(\omega_2)^{(l)}~\dots~\hat{\boldsymbol{x}}(\omega_{m_\text{FFT}})^{(l)} \right] 
  \in \mathbb{R}^{n\times m_{\text{FFT}}}, \end{equation} of each block, where superscript $l$ denotes the $l$-th block. All realizations of the Fourier transform at a specific frequency $\omega_k$ are now collected into a new data matrix 
\begin{equation}
   \hat{\boldsymbol{X}}_{\omega_k}
   = \left[ \hat{\boldsymbol{x}}(\omega_k)^{(1)}~\hat{\boldsymbol{x}}(\omega_k)^{(2)}
   ~\dots~\hat{\boldsymbol{x}}(\omega_{k})^{(n_b)} \right].
\end{equation}
The product $\hat{\boldsymbol{X}}_{\omega_k}\hat{\boldsymbol{X}}_{\omega_k}^T$ forms the cross-spectral density matrix, and its eigenvalue decomposition \begin{equation}
 \hat{\boldsymbol{X}}_{\omega_k}\hat{\boldsymbol{X}}_{\omega_k}^T \boldsymbol{\phi}_{\omega_{k,j}} = \lambda_{\omega_{k,j}} \boldsymbol{\phi}_{\omega_{k,j}},
 \quad
 \boldsymbol{\phi}_{\omega_{k,j}} \in \mathbb{R}^n,
 \quad
 \lambda_{\omega_k,1} \ge \dots \ge \lambda_{\omega_k,n_b} \ge 0,
 \label{eqn:Reig}
\end{equation}
yields the Spectral POD modes $\boldsymbol{\phi}_{\omega_{k,j}}$ and corresponding modal energies $\lambda_{\omega_{k,j}}$, respectively. As an 
example, the most energetic Spectral POD mode of a turbulent jet at one frequency is shown in \fig \ref{fig:jet_resolvent}(d). Spectral POD as described 
here dates back to the early work of Lumley [\citen{Lumley1970}] and is not related to the method under the same name proposed by 
Sieber \etal~[\citen{Sieber:JFM16}] whose approach blends between POD and DFT modes by filtering the temporal correlation matrix.

\subsubsection*{Strengths and Weaknesses}

\begin{strengths}
\end{strengths}
\begin{itemize} 
   \item POD gives an orthogonal set of basis vectors with the minimal dimension.  This property is useful in constructing a reduced-order model of the flow field.
   \item POD modes are simple to compute using either the (classical) spatial or snapshot methods.  The method of snapshots is especially attractive for high-dimensional spatial data sets.
   \item Incoherent noise in the data generally appears as high-order POD modes, provided that the noise level is lower than the signal level.  POD analysis can be used to practically remove the incoherent noise from the dataset by simply removing high-order modes from the expansion.
   \item POD (PCA) analysis is very widely used in a broad spectrum of studies.  It is used for pattern recognition, image processing, and data compression. 
\end{itemize}

\begin{weaknesses}
\end{weaknesses}
\begin{itemize} 
   \item As POD is based on second-order correlation, higher-order correlations are ignored.
   \item The temporal coefficients of spatial POD modes generally contain a mix of frequencies.  Spectral POD discussed above addresses this issue.
   \item POD arranges modes in the order of energy contents, and not in the order of the dynamical importance.  This point is addressed by Balanced POD and DMD analyses. 
   \item It is not always clear how many POD modes should be kept, and there are many different truncation criteria.  
\end{itemize}

\subsection{Illustrative Examples}

\subsubsection*{Turbulent Separated Flow over an Airfoil} 

\begin{figure}[tb]
\begin{center}
  \includegraphics[width=0.7\textwidth]{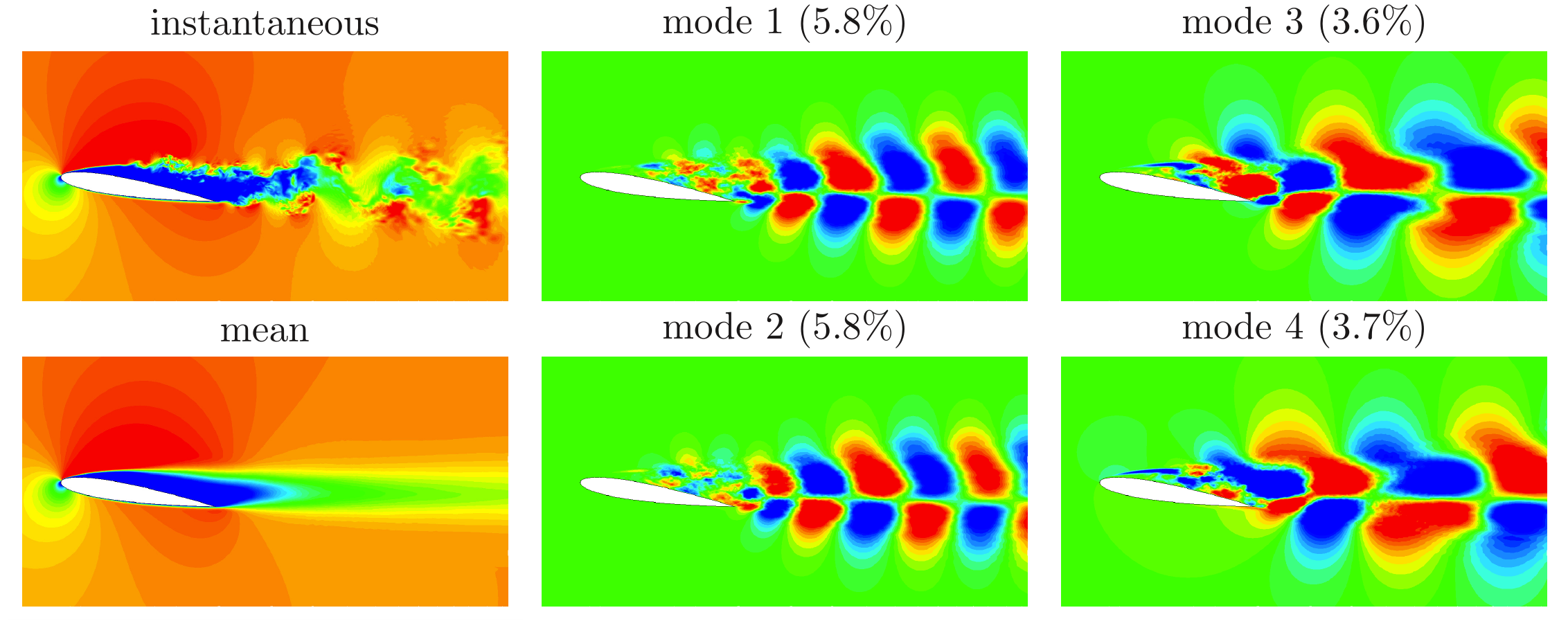} 
  \caption{POD analysis of turbulent flow over a NACA0012 airfoil at $Re = 23,000$ and $\alpha = 9^\circ$.  Shown are the instantaneous and time-averaged streamwise velocity fields and the associated four most dominant POD modes [\citen{Munday:AIAAJXX,Kajishima17}]. Reprinted with permission from Springer.}
  \label{fig:naca0012}
  \end{center}
\end{figure}

We present an example of applying the POD analysis on the velocity field obtained from three-dimensional large-eddy simulation (LES) of turbulent separated flow over a NACA 0012 airfoil [\citen{Munday:AIAA14}].  The flow is incompressible with spanwise periodicity at $Re = 23,000$ and $\alpha = 9^\circ$.  Visualized in \fig \ref{fig:naca0012} (left) are the instantaneous and time-averaged streamwise velocity on a spanwise slice.  We can observe that there are large-scale vortical structures in the wake from von K\'arm\'an shedding\index{von K\'arm\'an shedding}, yielding spatial and temporal fluctuations about the mean flow.  Also present are the finer-scale turbulent structures in the flow.  Performing POD on the flow field data, we can find the dominant modes [\citen{Kajishima17}].  Here, the first four dominant POD modes with the percentage of kinetic energy held by the modes are shown in \fig \ref{fig:naca0012} (middle and right).  The shown four modes together capture approximately 19\% of the unsteady fluctuations over the examined domain.  Modes 1 and 2 (first pair) represent the most dominant fluctuations in the flow field possessing equal level of kinetic energy, amounting to oscillatory (periodic) modes.  Modes 3 and 4 (second pair) represent the subharmonic spatial structures of modes 1 and 2 in this example. 
Compared to the laminar case shown in \fig \ref{fig:intro_POD}, the number of modes required to reconstruct this turbulent flow is increased, due to the emergence of multiple spatial scales and higher dimensionality of the turbulent flow.
The dominant features of the shown POD modes share similarities with the laminar flow example shown in \fig \ref{fig:intro_POD}, despite the large difference in the Reynolds numbers.

\subsubsection*{Compressible Open-Cavity Flows}

\begin{figure}[tb]
\centering 
  \includegraphics[width=0.56\textwidth]{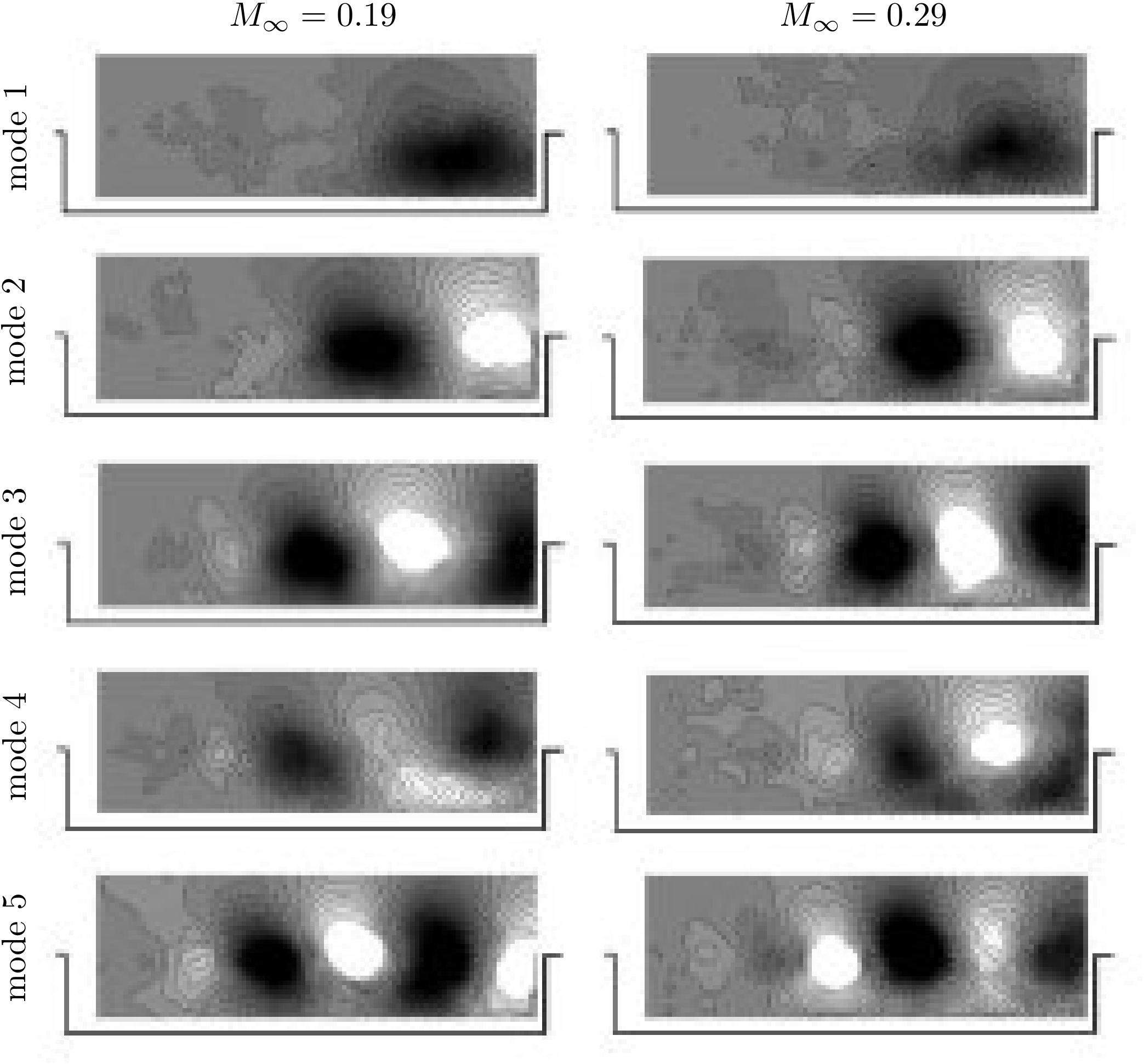}
  \caption{POD modes of the vertical fluctuating velocity for flows over a rectangular cavity at Mach numbers $M_\infty = 0.19$ and $0.29$ [\citen{murray2009}].  Reprinted with permission from AIP Publishing.}
  \label{fig:cavity}
\end{figure}

As the second sample application of POD, let us look at an analysis of the velocity field from open-cavity flow experiments [\citen{murray2009}].  In this study, the snapshot POD was applied to two-component PIV data acquired with several different free stream Mach numbers from 0.2 through 0.73.  Here, the dominant mode (mode 1) contains between 15 and 20 percent of the energy with 50 percent represented by the first 7 modes.  This study reveals the similarity of the modes amongst 4 of the free stream Mach numbers investigated even though there were differences in the mean flow patterns.  The first 5 POD modes associated with the vertical velocity component are presented in \fig \ref{fig:cavity}.  In this figure, we can observe a representation of the vortical structures in the cavity shear layer with similar wavelengths regardless of the free stream Mach number.  Further quantitative analysis of the similarity of the modes was verified by checking the orthogonality between the modes for the various applications.  The similarity amongst the modes implies that the underlying turbulence had the same structure, at least over the range of free stream Mach numbers investigated, regardless of the mean flow differences.     

\subsection{Outlook}

POD has been the bedrock of modal decomposition techniques to extract coherent structures for unsteady fluid flows.  To address some of the shortcomings of standard POD analysis, many variations have emerged; namely the Balanced POD [\citen{Rowley-ijbc05}] (see Section \ref{sec:bpod}), Split POD [\citen{Camphouse2008}], Sequential POD [\citen{Jorgensen:TCFD03,delSastre:06}], Temporal POD [\citen{Gordeyev2013}], and Joint POD [\citen{Gordeyev2014}], amongst others.   A number of overarching studies have emerged to bridge the gap between POD and other decomposition methods, revisiting some of the early POD discussions by Lumley [\citen{Lumley1967}] and George [\citen{George88}].  Recently, theoretical connections have been made between Spectral POD and several other methods, including resolvent analysis (Section \ref{sec:resolvent}) [\citen{TowneEtAl2015,SemeraroEtAl2016,TowneEtAl2017inPreparation}] and other data-based methods including the spatial POD described earlier and dynamic mode decomposition (Section \ref{sec:dmd}) [\citen{TowneEtAl2017inPreparation}].  

One of the most attractive properties of POD modes is orthogonality.  This feature allows us to develop models that are low in order and sparse.  Taking advantage of such property, there have been efforts to construct reduced-order models based on Galerkin projection to capture the essential flow physics [\citen{Aubry:JFM88, Holmes96, Ukeiley:JFM01, Noack:JFM03, Noack:JFM05}] and to implement model-based closed-loop flow control [\citen{Samimy:JFM07, Noack11}].  More recently, there are emerging approaches that take advantage of the POD modes to model and control fluid flows leveraging cluster-based analysis [\citen{Kaiser:JFM14}] and networked-oscillator representation [\citen{Nair:arXiv17}] of complex unsteady fluid flows.


\section{Balanced Proper Orthogonal Decomposition (Balanced POD)}
\label{sec:bpod}

Balanced Proper Orthogonal Decomposition (Balanced POD) is a modal decomposition technique that can extract two sets of modes for specified {\it inputs} and
{\it outputs}.  Here, the inputs are typically external disturbances or actuation used
for flow control.  The outputs are typically the available sensor measurements
or the quantities we want to capture with a model (for instance,
they could be amplitudes of POD modes).

This method is an approximation of a technique called {\it balanced
  truncation} [\citen{Moore-81}], a standard method used in control
theory, that balances the properties of {\it controllability} and {\it
  observability}. The most controllable states correspond to those that are most
easily excited by the inputs, and the most observable states correspond to those
that excite large future outputs.  In a reduced-order model, we wish to
retain both the most controllable modes and the most observable modes, but the
difficulty is that for some systems (particularly systems that are non-normal
which arise in many shear flows), states that have very small controllability might
have very large observability, and vice versa.  {\it Balancing} involves
determining a coordinate system in which the most controllable directions in
state space are also the most observable directions.  We then truncate the
states that are the least controllable/observable.

Balanced POD is closely related to POD: both procedures produce a set of modes
that describe the coherent structures in a given fluid flow, and the
computations required are also similar (both involve SVD).  
However, there are some important differences.  POD provides a
single set of modes that are orthogonal and are ranked by energy content. 
In contrast, Balanced POD provides two sets of modes, {\it balancing modes} and
{\it adjoint modes}, which form a bi-orthogonal set, and are ranked by
controllability/observability (which we can think of as the {\it importance to the
input-output dynamics}).  With both POD and Balanced POD, a quantity $\boldsymbol{q}(\boldsymbol{\xi},t)$ is
expanded as $\boldsymbol{q}(\boldsymbol{\xi},t) = \sum_{j=1}^n a_j(t) \boldsymbol{\phi}_j(\boldsymbol{\xi})$, where $\boldsymbol{\phi}_j$ are the POD modes or direct balancing modes, and $a_j(t)$ are scalar temporal coefficients.  
For POD, the modes $\boldsymbol{\phi}_j$ are orthonormal (which means
$\ip<\boldsymbol{\phi}_j, \boldsymbol{\phi}_k>=\delta_{jk}$), so the coefficients $a_j$ are computed
by $a_j (t) = \ip<\boldsymbol{q},\boldsymbol{\phi}_j>$.  With Balanced POD, bi-orthogonality of the
balancing modes $\boldsymbol{\phi}_j$ and adjoint modes $\boldsymbol{\psi}_k$ means that these satisfy
$\ip<\boldsymbol{\phi}_j,\boldsymbol{\psi}_k>=\delta_{jk}$, and the coefficients $a_j$ are then
determined by $a_j (t) = \ip<\boldsymbol{q},\boldsymbol{\psi}_j>$.

The dataset used for Balanced POD
is also quite specific: it consists of the linear responses of the system to impulsive
inputs (one time series for each input), as well as impulse responses of an
adjoint system (one adjoint response for each output).  It is these adjoint
simulations that enable Balanced POD to determine the observability (or
sensitivity) of different states, which makes the procedure so effective for
non-normal systems.  However, because adjoint information is required, it is
usually not possible to apply Balanced POD to experimental data.  It
has been shown that a system identification method called the Eigensystem
Realization Algorithm (ERA) [\citen{JuPa-85}] produces reduced-order models that
are equivalent to Balanced POD-based models, without the need for adjoint responses,
and can therefore be used on experimental data [\citen{Ma:TCFD09}].  
For the full details of Balanced POD, see Rowley [\citen{Rowley-ijbc05}] or the second
edition of Holmes {\etal} [\citen{Holmes96}] (Chapter~5).
 A description of a related method is also given in Willcox and Peraire [\citen{willcox:2002}].

\subsection{Description}

\subsubsection*{Algorithm}

As the Balanced POD analysis is founded on linear state-space systems, the inputs to Balanced POD should be obtained from linear dynamics.

\begin{inputs}
Two sets of snapshots from a linearized forward simulation and a companion adjoint simulation.
\end{inputs}

\begin{outputs}
Sets of balancing modes and adjoint modes ranked in the order of the Hankel singular values.  These modes comprise a coordinate transform that balances the controllability and observability of the system.
\end{outputs}

Balanced POD is based on the concept of balanced truncation that provides a balancing measure between controllability and observability in the transformed coordinate.  This approach seeks for the balancing transform $\boldsymbol{\Phi}$ and its inverse transform $\boldsymbol{\Psi}$ that can diagonalize and equate the (empirical) controllability and observability Gramians, $\boldsymbol{W}_c$ and $\boldsymbol{W}_o$, respectively, such that $\boldsymbol{\Psi}^* \boldsymbol{W}_c \boldsymbol{\Psi} = \boldsymbol{\Phi} \boldsymbol{W}_o \boldsymbol{\Phi}^* = \boldsymbol{\Sigma}$ is a diagonal matrix.  While the controllability Gramian can be determined from the forward simulation, the observability Gramian requires results from the adjoint simulation.
  
Now, let us introduce the forward and adjoint linear systems and present the snapshot-based Balanced POD technique [\citen{Rowley-ijbc05}] that is analogous to the snapshot-based POD method.  Following standard state-space notation from control theory, the forward and adjoint simulations solve
\begin{equation}
   \left\{
   \begin{split}
   \dot{\boldsymbol{x}} &= \boldsymbol{A} \boldsymbol{x} + \boldsymbol{B} \boldsymbol{u}\\
   \boldsymbol{y} 		&= \boldsymbol{C} \boldsymbol{x}
   \end{split}
   \right.
   \label{eq:forward}
\end{equation}
and
\begin{equation}
   \left\{
   \begin{split}
   \dot{\boldsymbol{z}} & = \boldsymbol{A}^T \boldsymbol{z} + \boldsymbol{C}^T \boldsymbol{v}\\
   \boldsymbol{w} 		&= \boldsymbol{B}^T \boldsymbol{z}.
   \end{split}
   \right.
   \label{eq:adjoint}
\end{equation}
For the forward dynamics, \eq (\ref{eq:forward}), $\boldsymbol{x}(t) \in \mathbb{R}^{n}$, $\boldsymbol{u}(t) \in \mathbb{R}^{p}$, $\boldsymbol{y}(t) \in \mathbb{R}^{q}$ are the state, input, and output vectors, and $\boldsymbol{A} \in \mathbb{R}^{n \times n}$, $\boldsymbol{B} \in \mathbb{R}^{n \times p}$, and $\boldsymbol{C} \in \mathbb{R}^{q \times n}$ are the state, input, and output matrices, respectively.  For the adjoint system, \eq (\ref{eq:adjoint}),
$\boldsymbol{z}(t) \in \mathbb{R}^{n}$, $\boldsymbol{v}(t) \in \mathbb{R}^{q}$, $\boldsymbol{w}(t) \in \mathbb{R}^{p}$ are the adjoint state, input, and output vectors.  

Based on the solution of linear and adjoint simulations, we can construct the data matrices $\boldsymbol{X}$ and $\boldsymbol{Z}$ (also see Appendix \ref{sec:App}).  For simplicity, let us consider below a single-input single-output (SISO) system with $p = q = 1$, for which we can construct the data matrices as
\begin{equation}
   \boldsymbol{X} = \left[ \boldsymbol{x}(t_1)~
   \boldsymbol{x}(t_2)~
   \dots~
   \boldsymbol{x}(t_m)
   \right] \in \mathbb{R}^{n \times m}
\end{equation}
and 
\begin{equation}
   \boldsymbol{Z} = \left[ \boldsymbol{z}(t_1)~
   \boldsymbol{z}(t_2)~
   \dots~
   \boldsymbol{z}(t_m)
   \right] \in \mathbb{R}^{n \times m}.
\end{equation}
For a multi-input multi-output (MIMO) system, the solutions from the linear and adjoin simulations can be stacked in an analogous manner to construct $\boldsymbol{X}$ and $\boldsymbol{Z}$, as discussed in Rowley [\citen{Rowley-ijbc05}].  Consequently, the empirical controllability and observability Gramians are given by 
$\boldsymbol{W}_c \approx \boldsymbol{X X}^T$ and  
$\boldsymbol{W}_o \approx \boldsymbol{ZZ}^T$,
respectively.  The balancing transform $\boldsymbol{\Phi}$ and its inverse $\boldsymbol{\Psi}$ can simply be found as
\begin{equation}
   \boldsymbol{\Phi} = \boldsymbol{XV \Sigma}^{-1/2}
   \quad \text{and} \quad
   \boldsymbol{\Psi} = \boldsymbol{ZU \Sigma}^{-1/2},
\end{equation}
where $\boldsymbol{U}$, $\boldsymbol{\Sigma}$, are $\boldsymbol{V}$ are determined from the SVD of the matrix product $\boldsymbol{Z}^T \boldsymbol{X}$:
\begin{equation}
   \boldsymbol{Z}^T \boldsymbol{X} = \boldsymbol{U} \boldsymbol{\Sigma} \boldsymbol{V}^T.
\end{equation}
The columns $\boldsymbol{\phi}_j$ and $\boldsymbol{\psi}_j$ of $\boldsymbol{\Phi}$ and $\boldsymbol{\Psi}$ correspond to the balancing and adjoint modes ranked in the order of $\boldsymbol{\Sigma} = \text{diag}(\sigma_1, \sigma_2, \dots, \sigma_m)$, which are referred to as the Hankel singular values.  

\subsubsection*{Notes}

\paragraph{Bi-orthogonality}
The property of bi-orthogonality ($\left< {\boldsymbol{\phi}}_j, {\boldsymbol{\psi}}_k\right> = \delta_{jk}$) can be used as a projection to derive a reduced-order model known as the Petrov--Galerkin model [\citen{Ilak:PF08}].  

\paragraph{Eigenvalue Realization Algorithm (ERA)}
If we are only interested in deriving a reduced-order model based on balanced truncation without the need to access the balancing and adjoint modes, we can use ERA and remove the requirement to perform the adjoint simulation.  For additional details on the derivation of the models and ERA, the readers can refer to the work of Juang and Papas [\citen{JuPa-85}] and Ma \etal~[\citen{Ma:TCFD09}].

\subsubsection*{Strengths and Weaknesses}

Strengths:
\begin{itemize}
    \item Balanced POD is particularly attractive for capturing the dynamics of non-normal systems with large transient growth [\citen{Ilak:PF08}].  Since POD ranks the modes based on energy content, non-normal characteristics of the flow may not be captured.  In contract, with Balanced POD models capture small-energy perturbations that are highly observable (typically through the adjoint modes).  For this reason, Balanced POD models typically perform better than POD models for non-normal systems.
    \item Balanced POD provides an input-output model suitable for feedback control [\citen{Ilak:PF08}].
\end{itemize}

\noindent Weaknesses:
\begin{itemize}
   \item Snapshots from adjoint simulations are needed, which makes Balanced POD analysis difficult or impossible to perform with experimental measurements.  However, ERA can be used instead if only the Balanced POD-based model is sought for without access to the balancing and adjoint modes [\citen{JuPa-85,Ma:TCFD09}].
   \item Both forward and adjoint simulations should be based on linear dynamics (though various extensions
to nonlinear systems have been introduced [\citen{LallMarsden-02,ilak2010gl}]).
\end{itemize}

\subsection{Illustrative Example}

\subsubsection*{Control of Wake Behind a Flat-Plate Wing}

\begin{figure}
   \centering
   \includegraphics[width=0.97\textwidth]{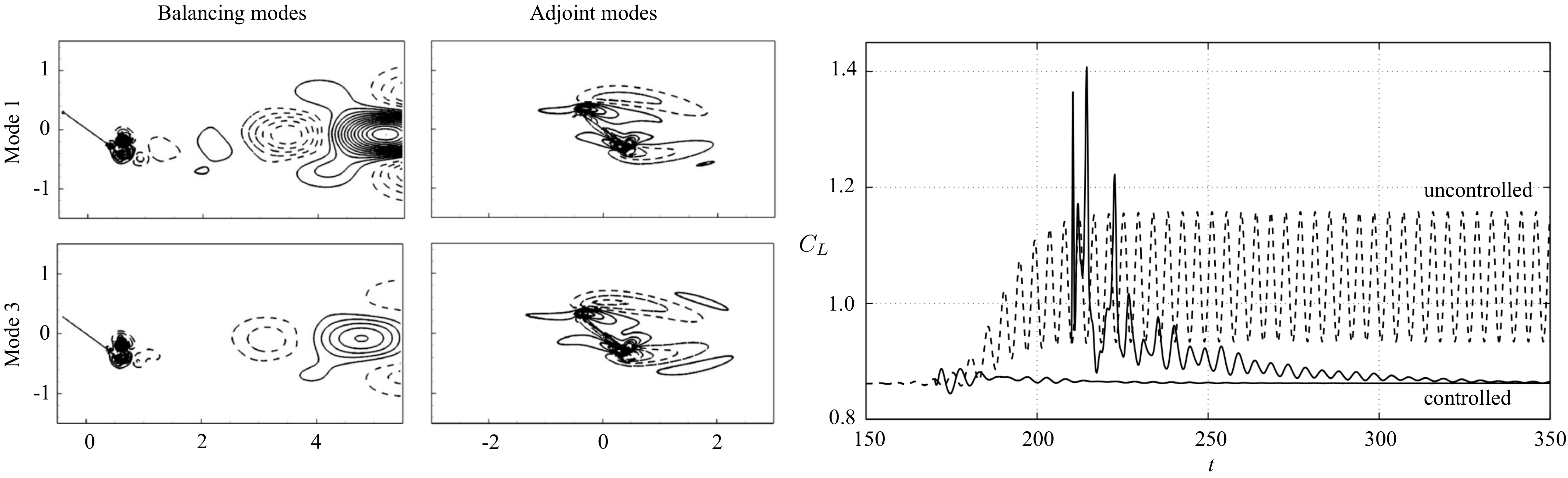}
   \caption{Use of Balanced POD analysis for feedback stabilization of the unstable wake behind a flat plate ($Re = 100$, $\alpha = 35^\circ$) [\citen{Ahuja:JFM10}].  First and third balancing and adjoint modes are shown on the left.  Baseline lift history is shown in dashed line and controlled cases with different initiation times of feedback control using Balanced POD based reduced-order model are plotted in solid lines on the right.  Reprinted with permission from Cambridge University Press.}
   \label{fig:stabilization}
\end{figure}

Balanced POD has been applied to analyze and control the unsteady wake behind a flat-plate wing at $Re= 100$.  In the work of Ahuja and Rowley [\citen{Ahuja:JFM10}], they performed linearized forward and adjoint simulations of flow over a flat-plate wing to determine balancing and adjoint modes as shown in \fig \ref{fig:stabilization} (left). The balancing modes resemble those of the traditional POD modes but the adjoint modes highlight regions of the flow that can trigger large perturbations downstream.  These modes are used to develop models and closed-loop controllers to stabilize the naturally unstable fluid flows, as depicted in \fig \ref{fig:stabilization} (right).   While their work necessitated snapshots from the forward and adjoint simulations, the requirement for adjoint simulations was later removed by the use of ERA in the work by Ma \etal~[\citen{Ma:TCFD09}].  Although the balancing and adjoint modes are not revealed from ERA, the resulting reduced-order model was shown to be identical to the Balanced POD based model.  This was numerically demonstrated using the same flat-plate wing problem along with a successful implementation of observer-based feedback control.

\subsection{Outlook}

While Balanced POD has been successfully applied to model and control a number of fluids systems [\citen{Ahuja:JFM10,Ilak:PF08,Barbagallo:2009,bagheri:2009,semeraro2011feedback}], further research could validate and extend its use to a wider range of fluids applications. Work towards this goal currently includes application to the control of nonlinear systems [\citen{semeraro2013TSwaves}], direct application to unstable systems [\citen{Flinois2015unstableBPOD}], and use with harmonically forced data [\citen{derghan2011frequential}].
Further work could also seek algorithmic variants that improve efficiency. 
Examples of work in this direction include analytic treatment of impulse response tails [\citen{tu2012improved}] and the use of randomized methods [\citen{yu2015randomized,halko2011random}]. 
While at present the need for adjoint simulations makes Balanced POD unsuitable for experimental data, future work could possibly remove this restriction, by making use of the connections that Balanced POD shares with methods such as ERA and DMD.


\section{Dynamic Mode Decomposition (DMD)}
\label{sec:dmd} 

Dynamic mode decomposition (DMD) provides a means to decompose time-resolved data into modes, with each mode having a single characteristic frequency of oscillation and growth/decay rate.   
DMD is based on the eigendecomposition of a {\it best-fit} linear operator that approximates the dynamics present in the data.  This technique was first introduced to the fluids community in an APS talk [\citen{schmid2008}], subsequently followed up with an archival paper by Schmid
[\citen{schmid2010DMD}]. The connections with the Koopman operator (see Section \ref{sec:koopman}) were given in Rowley \etal~[\citen{rowley2009spectral}], which explains the meaning of DMD for a nonlinear system (see also the review articles by Mezic [\citen{mezic2013koopman}] and Tu \etal~[\citen{tu2014dynamic}]).  There have been a number of different formulations and interpretations of DMD since then [\citen{Kutz2016book}], which are mentioned in this section.

In many ways, DMD may be viewed as combining favorable aspects of both the POD and the discrete Fourier transform (DFT)~[\citen{rowley2009spectral,chen2011variants}], resulting in spatio-temporal coherent structures identified purely from data.  
Because DMD is rooted firmly in linear algebra, the method is highly extensible, spurring considerable algorithmic developments.  
Moreover, as DMD is purely a data-driven algorithm without the requirement for governing equations, it has been widely applied beyond fluid dynamics: in finance~
[\citen{Mann2016qf}], video processing~[\citen{Grosek2014arxiv,kutzRPCA1,erichson2015}], epidemiology~[\citen{Proctor2015ih}], robotics~[\citen{Berger2014ieee}], and neuroscience~[\citen{brunton2016extracting}]. 
As with many modal decomposition techniques, DMD is most often applied as a diagnostic to provide physical insight into a system.  The use of DMD for future-state prediction, estimation, and control is generally more challenging and less common in the literature~[\citen{Kutz2016book}].

\subsection{Description}

\subsubsection*{Algorithm}

\begin{inputs}
A set of snapshot pairs from fluids
experiments or simulations, where the two snapshots in each pair are separated
by a constant interval of time. Often, this will come from a time-series of data.
\end{inputs}

\begin{outputs}
DMD eigenvalues and modes. The modes are spatial structures that oscillate and/or grow/decay at rates given by the corresponding eigenvalues. These come from the eigendecomposition of a {\it best-fit} linear operator that approximates the dynamics present in the data. 
\end{outputs}

We begin by collecting snapshots of data and arranging
them as columns of matrices ${\boldsymbol{X}}$ and $\dmdnext{\boldsymbol{X}}$, such that 
\begin{equation}
  {\boldsymbol{X}} = \left[ {\boldsymbol{x}}(t_1)~ {\boldsymbol{x}}(t_2)~ \cdots ~{\boldsymbol{x}}(t_m) \right]
  \in \mathbb{R}^{n\times m}
  \quad \text{and} \quad
  {\dmdnext{\boldsymbol{X}}} = \left[ {\boldsymbol{x}}(t_2)~ {\boldsymbol{x}}(t_3)~ \cdots ~{\boldsymbol{x}}(t_{m+1}) \right]
  \in \mathbb{R}^{n\times m}.
\end{equation}  
In DMD, we approximate the relationship between the snapshots in a linear manner
such that
\begin{equation}
  \label{eq:dmd-def}
  \dmdnext{\boldsymbol{X}} = \boldsymbol{A}\boldsymbol{X}.
\end{equation}
The matrix~${\boldsymbol{A}}$ may be defined by ${\boldsymbol{A}}=\dmdnext{\boldsymbol{X}}\boldsymbol{X}^+$, where $\boldsymbol{X}^+$
denotes the pseudoinverse of~$\boldsymbol{X}$. 
The DMD eigenvalues and modes are then defined as the
eigenvalues and eigenvectors of~${\boldsymbol{A}}$ [\citen{tu2014dynamic}].  It is
common that the number of snapshots is smaller than the number of components of
each snapshot ($m \ll n$).  In this case, it is
not efficient to compute ${\boldsymbol{A}}$ explicitly, so we normally use an algorithm
such as the following, similar to Algorithm~2 in
Tu \etal~[\citen{tu2014dynamic}]:

\begin{itemize}
   \item Perform the reduced SVD\footnote{Since $\boldsymbol{X} \in \mathbb{R}^{n\times m}$, we have $\boldsymbol{U}\in \mathbb{R}^{m\times m}$, $\boldsymbol{V}\in \mathbb{R}^{n \times n}$, and $\boldsymbol{\Sigma}\in \mathbb{R}^{m\times n}$.  Note that in this case, $\boldsymbol{U}^T = \boldsymbol{U}^{-1}$ and $\boldsymbol{V}^T = \boldsymbol{V}^{-1}$.} (Section \ref{sec:SVD}) of $\boldsymbol{X}$, letting $\boldsymbol{X} = \boldsymbol{U}\boldsymbol{\Sigma}\boldsymbol{V}^T$.
   \item (Optional) Truncate the SVD by considering only the first $r$ columns of $\boldsymbol{U}$ and $\boldsymbol{V}$, and the first $r$ rows and columns of $\boldsymbol{\Sigma}$ (with the singular values ordered by size), to obtain $\boldsymbol{U}_r$, $\boldsymbol{\Sigma}_r$, and $\boldsymbol{V}_r$.
   \item Let $\widetilde{\boldsymbol{A}} = \boldsymbol{U}_r^T {\boldsymbol{A}} \boldsymbol{U}_r 
   = \boldsymbol{U}_r^T\dmdnext{\boldsymbol{X}}\boldsymbol{V}_r\boldsymbol{\Sigma}_r^{-1} \in \mathbb{R}^{r\times r}$ and find the eigenvalues $\mu_j$ and eigenvectors $\widetilde{\boldsymbol{v}}_j$ of $\widetilde{\boldsymbol{A}}$, with $\widetilde{\boldsymbol{A}}\widetilde{\boldsymbol{v}}_j = \mu_j \widetilde{\boldsymbol{v}}_j$,
   \item Every nonzero $\mu_i$ is a DMD eigenvalue, with corresponding DMD
     mode given by $\boldsymbol{v}_i = \mu_i^{-1} \dmdnext{\boldsymbol{X}}\boldsymbol{V}_r\mSigma_r^{-1}\widetilde{\boldsymbol{v}}_i$.
     It is common to compute the {\it projected DMD modes}, which are simply
     $\boldsymbol{P}\boldsymbol{v}_i= \boldsymbol{U}_r\widetilde{\boldsymbol{v}}_i$, where $\boldsymbol{P}=\boldsymbol{U}_r\boldsymbol{U}_r^T$ is the orthogonal projection
     onto the first $r$ POD modes of the data in~$\boldsymbol{X}$.
\end{itemize}

Note that matrix $\boldsymbol{A}$ in \eq (\ref{eq:dmd-def}) is related to operator $\exp(\boldsymbol{A}\Delta t)$ in \eq (\ref{eq:expAt}), with $\Delta t = t_{i+1}-t_i$.  Hence, the eigenvalues are related by
\begin{equation}
   \lambda_j = \frac{1}{\Delta t} \log(\mu_j).
\end{equation}
Using this relationship, the growth/decay rates and frequencies of the DMD modes can be inferred by examining the real and imaginary components of $\lambda_j$.

In addition to the DMD algorithms presented above, there are a number of variants.  Some of them are discussed below in Notes.  For further details, refer to Rowley and Dawson [\citen{Rowley:ARFM17}], Tu \etal~[\citen{tu2014dynamic}], and Kutz \etal~[\citen{Kutz2016book}].

\subsubsection*{Notes}

\paragraph{Computationally efficient algorithms}
A fast method to perform DMD in real time on large datasets was recently proposed in Hemati \etal~[\citen{Hemati2014streaming}]. 
A library of tools for computing variants of DMD is available at \url{https://github.com/cwrowley/dmdtools}.  A parallelized implementation of DMD (as well as other system identification/modal decomposition techniques) is described in Belson \etal~[\citen{belson2013modred}] at \url{https://github.com/belson17/modred}.

\paragraph{Sparsity}
It is often desirable to represent a dataset {\it sparsely}, i.e., in terms of a
small number of DMD modes. 
 Because DMD modes are not orthogonal and have
no objective {\it ranking} (as POD modes have), this is not an
easy task.  A number of variants of DMD have been proposed to provide such a
sparse representation: for instance, the {\it optimized DMD} [\citen{chen2011variants}] and the {\it optimal mode decomposition} [\citen{wynn2013omd}] are two such
variants.  An algorithm called {\it sparsity-promoting DMD} has also been proposed [\citen{jovanovic2014dmdsp}], with Matlab code available at
\url{http://www.ece.umn.edu/users/mihailo/software/dmdsp/}.  
Note that finding a sparse representation of a dataset is different from applying DMD to sparse data. On this latter problem, it has been shown that compressed sensing can be used to apply DMD to temporally [\citen{tu2014compressed}] or spatially [\citen{Brunton2015jcd,Gueniat2015pof}] sparse data, and that randomized methods can be used for enhanced computational efficiency [\citen{Erichson2015arxiv}].  

\paragraph{Systems with inputs/control}
Some systems have external {\it inputs} that are known (unlike process noise): for
instance, the input would be the signal to the
actuator in a flow control situation.  DMD has recently been extended to handle such systems [\citen{Proctor2016siads}].

\paragraph{Nonlinear systems}
As mentioned above, the connections with the Koopman operator give meaning to
DMD when applied to a nonlinear system; however, when nonlinearity is important,
DMD often gives unreliable results, unless a sufficiently large set of
measurements is used in the snapshots.  For instance, it can help to include the
square of the velocities in each snapshot, in addition to the velocities
themselves.  Recent work [\citen{williams2014edmd}] clarifies these issues and
extends DMD to allow for a better global approximation of Koopman modes and
eigenvectors, as well as Koopman eigenfunctions. 
Data-driven approximation of Koopman eigenfunctions can be used as a set of universal coordinates, through which disparate datasets may be related [\citen{Williams2015epl}].

\paragraph{Connections with other methods}
In many situations, DMD is equivalent to a Discrete Fourier Transform [\citen{rowley2009spectral,chen2011variants}]. 
In addition, DMD shares algorithmic similarities with a number of other techniques, such as the eigensystem realization algorithm [\citen{tu2014dynamic}], and linear inverse modeling [\citen{schmid2010DMD,tu2014dynamic}] (a technique used in climate science).  If DMD is applied to linearized flow about its steady state, the extracted DMD modes capture the global modes (also see Section \ref{sec:global}).

\subsubsection*{Strengths and Weaknesses}

\begin{strengths}
\end{strengths}
\begin{itemize}
\item DMD does not require any {\it a priori} assumptions or knowledge of the underlying dynamics.  DMD is an entirely data-driven amalysis.
\item DMD can be applied to many types of data, or even concatenations of disparate data sources.
\item Under certain conditions, DMD gives a finite-dimensional approximation to the Koopman operator, which is an infinite-dimensional linear operator that can be used to describe nonlinear dynamics (see Section~\ref{sec:koopman}). 
\item DMD modes can isolate specific dynamic structures (associated with a particular frequency).
\item DMD has proven to be quite {\it customizable}, in the sense that a number of proposed modifications address the weaknesses outlined below.
\end{itemize}

\begin{weaknesses}
\end{weaknesses}
\begin{itemize}
\item It can be difficult (or at least subjective) to determine which modes are
  the most physically relevant (i.e., there is no single {\it correct} way to rank
  eigenvalue importance, unlike other methods such as POD).
\item DMD typically requires time-resolved data to identify dynamics, although extensions exist [\citen{tu2014compressed,tu2014dynamic}].
\item If DMD is used for system identification (without any modifications), the resulting model will be linear.
\item DMD can be unreliable for nonlinear systems.  In particular, for a
  nonlinear system, we must be
  careful to choose a sufficiently rich set of measurements (in each snapshot).  Without care,
  the connection with the Koopman operator and the underlying dynamical system may be lost.
   Furthermore, for nonlinear systems with complex (e.g., chaotic) dynamics, 
   there are further complications that could limit the applicability of DMD and related
   algorithms.
\item The outputs of DMD can be sensitive to noisy data, which was shown empirically [\citen{duke2012error}] and analytically [\citen{dawson2016dmdnoise}].  The effect of process noise (that is, a disturbance that affects the dynamics of the system) has also been investigated [\citen{bagheri2014noise}].  However, there are algorithms that are robust to sensor noise [\citen{dawson2016dmdnoise,hemati2015tls}].  
\item DMD should generally be used only for autonomous systems (i.e., the
  governing equations should have no time dependence or external inputs, unless these are explicitly accounted for [\citen{Proctor2016siads}]).
\item DMD modes are not orthogonal. This has a number of drawbacks: for
  instance, if the modes are used as a basis/coordinate system for a reduced-order model, the model will have additional terms due to the spatial inner product between different modes being non-zero. Note that a recent variant, recursive DMD [\citen{noack2015recursive}], considers orthogonalized DMD modes. 
  \item DMD relies fundamentally on the separation of variables, as does POD, and hence does not readily extend to traveling wave problems.  
  \item DMD does not typically work well for systems with highly intermittent dynamics. However, multi-resolution [\citen{Kutz2016siads}] and time-delay [\citen{Brunton2016havok}] variants show promise for overcoming this weakness.
\end{itemize}

\subsection{Illustrative Example}

\subsubsection*{Jet in Crossflow}

\begin{figure}[htb]
\center{\includegraphics[width= 0.65\textwidth]{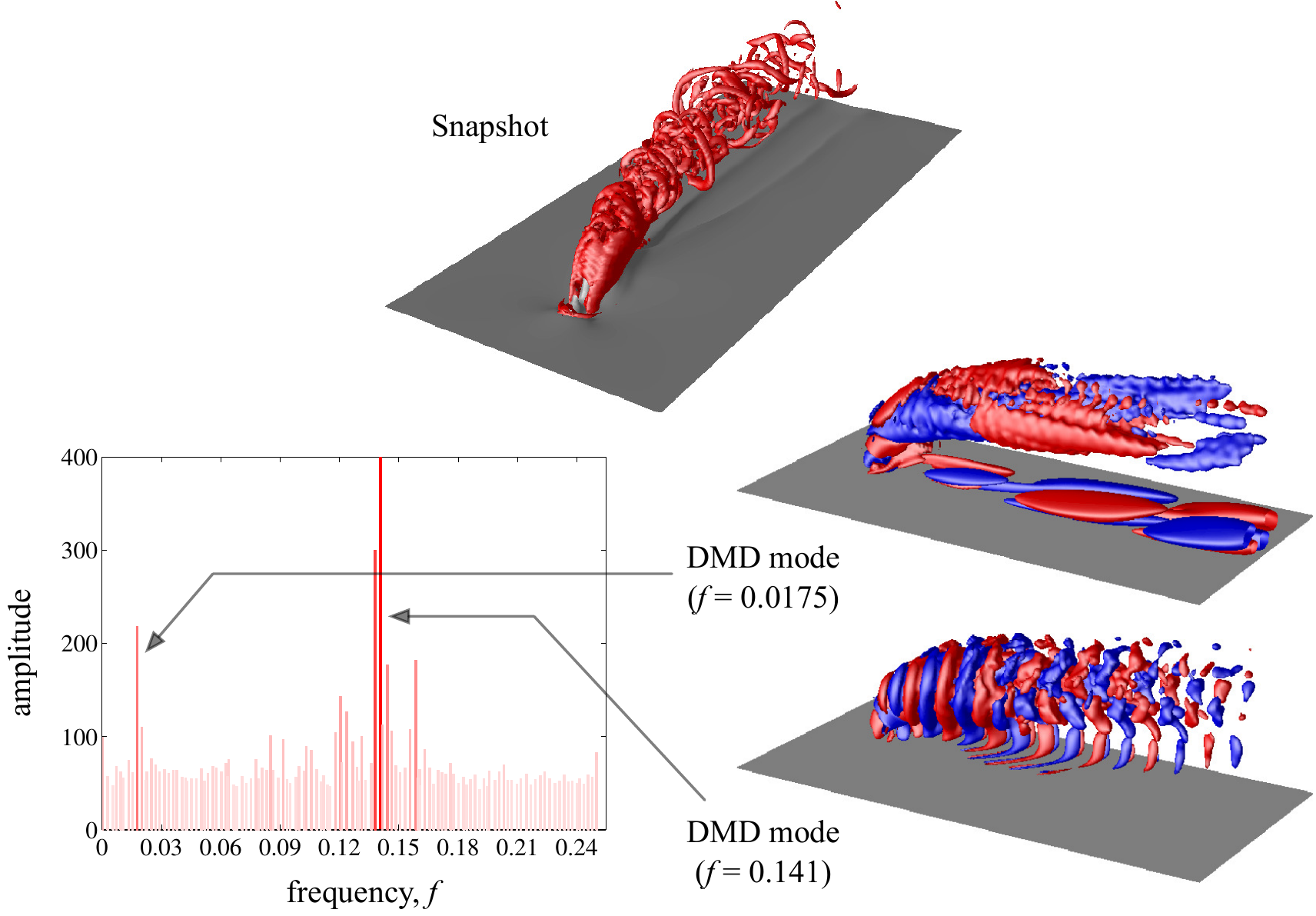}}
\caption{Snapshot of jet-in-crossflow data used for DMD [\citen{rowley2009spectral}], shown with $\lambda_2$ contour in the top plot.   The frequency and amplitude of the DMD modes are shown in the bottom left plot, while the bottom right figures show the modes corresponding to (dimensionless) frequencies of 0.0175 and 0.141, visualized with contours of streamwise velocity.  Reprinted with permission from Cambridge University Press.}
\label{fig:dmd_rowley}      
\end{figure}

We show an example from Rowley \etal~[\citen{rowley2009spectral}], where DMD is applied to three-dimensional jet-in-crossflow direct numerical simulation (DNS) data. A typical snapshot of the complex flow is visualized in \fig \ref{fig:dmd_rowley} (top) using the $\lambda_2$ criterion. The results of applying DMD to a sequence of 251 snapshots are shown in \fig \ref{fig:dmd_rowley} (bottom). The bottom-left plot shows the frequency and amplitude of each DMD mode, with two modes having large amplitudes visualized in the bottom plots. Note that these modes capture very different flow structures, each having a different characteristic frequency of oscillation identified by DMD.

\subsubsection*{Canonical Separated Flow with Control}

\begin{figure}[htb]
\center{\includegraphics[width= 0.95\textwidth]{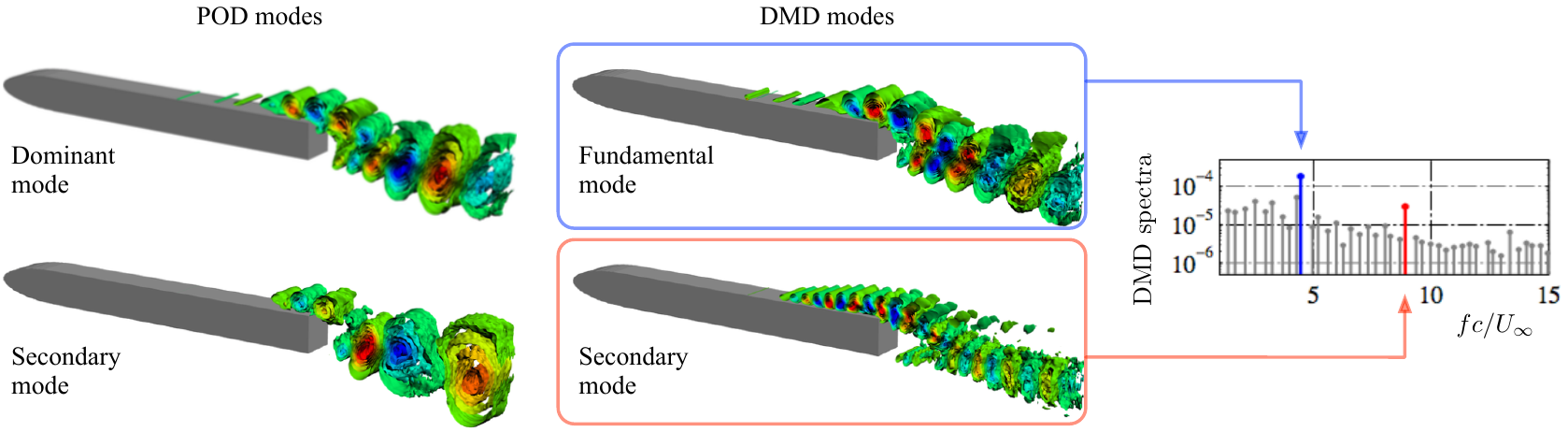}}
 \caption{Comparison of POD and DMD modes for separated three-dimensional turbulent flow over a flat-plate wing with synthetic-jet actuation on the top surface [\citen{Tu:thesis}].  The fundamental and secondary DMD modes correspond to the actuation frequencies of 4.40 (blue) and its superharmonic (red).  Reprinted with permission from J. Tu.}
\label{fig:dmd_tu}      
\end{figure}

Let us present a second example [\citen{Tu:thesis}] of DMD (and POD) analysis performed on three-dimensional separated turbulent flow over a finite-thickness plate with elliptical leading edge at $Re = 100,000$. 
The plate is aligned with the freestream, and separation is induced by imposing a steady blowing/suction boundary condition above the plate.  The flow field is obtained from LES and the wake dynamics in the example is modified by a synthetic jet actuator placed on the top surface of the wing.  First, we consider the dominant and secondary POD modes, shown in \fig \ref{fig:dmd_tu} (left).  We observe that some interactions between the actuation input (at 0.6 chord location) and the wake are captured by these energetically dominant modes.  Also shown on the right are the DMD modes corresponding to the actuation frequency and its superharmonic.  While the fundamental DMD mode is similar to the dominant POD mode, the secondary mode is able to clearly identify the synchronization of the actuator input with the downstream wake.  In contrast, POD modes are comprised of spatial structures having a distribution of temporal frequencies that makes pinpointing the POD mode to a specific frequency difficult.

\subsection{Outlook}

While DMD has quickly become a widely used method for analyzing fluid flow data, 
there remain challenges and applications that have yet to be fully addressed. 
Connections between DMD and the Koopman operator indicate its potential to model (and control) nonlinear systems.
However, choosing suitable observables
to give an accurate finite-dimensional approximation to the Koopman operator 
generally remains an open question. 
Algorithmic improvements that have been and should continue to be made will
allow DMD to remain a practical tool to analyze increasingly large fluid flow datasets.


\section*{Operator-Based Modal Analysis Methods}

In the latter part of this paper, we discuss modal analysis methods that are focused on the operator that describes the state dynamics, which is in contrast to the data-based methods.  In particular, we cover Koopman analysis, global linear stability analysis, and resolvent analysis.  Koopman analysis is a theoretical framework to study a finite-dimensional nonlinear dynamical system as an infinite-dimensional linear dynamical system.  Linear global and resolvent analyses reveal stability and input-output characteristics of the system by examining its linear state operator.


\section{Koopman Analysis}
\label{sec:koopman}

Koopman analysis provides an important alternative perspective to classical dynamical systems theory [\citen{guckenheimer_holmes}] for the description of complex systems.  The Koopman operator was introduced in the early 1930s [\citen{Koopman1931pnas}] to show how the dynamics of Hamiltonian systems could be described by an infinite-dimensional linear operator on the space of observable functions of the state of the system.  Recently, this theory has been at the center of efforts for the data-driven characterization of complex systems [\citen{Mezic2005nd}].  There is particular interest in obtaining finite-rank approximations to the linear Koopman operator.  Practically, dynamic mode decomposition (Section \ref{sec:dmd}) is the most implemented numerical framework for Koopman mode decomposition in fluids [\citen{schmid2008,rowley2009spectral,schmid2010DMD}].  There have been a number of excellent in-depth reviews on Koopman analysis recently [\citen{Budivsic2012chaos,mezic2013koopman}].

Koopman analysis not only provides a set of modes, of which DMD modes are a subset, but also a set of eigenvalues that determine the modal dynamics and a set of Koopman eigenfunctions that serve as intrinsic observable functions.  In many contexts, Koopman analysis provides an equation-free method [\citen{Kevrekidis2003cms}] to extract coherent structures and dynamics from data measurements of a complex system; these coherent structures are related to POD modes in fluids [\citen{Holmes96}].

\subsection{Description}

\subsubsection*{Algorithm}

\begin{inputs}
Nonlinear dynamical system:
\begin{eqnarray}
\boldsymbol{x}_{i+1} = \boldsymbol{f}(\boldsymbol{x}_i).
\end{eqnarray}
The state $\boldsymbol{x}$ typically lives in a vector space, such as $\mathbb{R}^n$ or $\mathbb{C}^n$, although the theory is defined more generally on curved manifolds.  
Although the dynamical system above is written in discrete time, the theory also holds for continuous dynamical systems. 
\end{inputs}

\begin{outputs}
Infinite-dimensional linear operator $\boldsymbol{U}_t$ that describes the evolution of scalar observables $g(\boldsymbol{x})$ on state-space:
\begin{eqnarray}
\boldsymbol{U}_t g = g\circ \boldsymbol{f}.
\end{eqnarray}
Here, $\circ$ denotes function composition\footnote{e.g., $\sin(x)\circ x^2 = \sin(x^2)$.}.  
This operator advances {\it{all}} functions in the Hilbert space of measurement functions, and holds for {\it{all}} states $\boldsymbol{x}$.  An implication is that measurements are advanced in time:
\begin{eqnarray}
\boldsymbol{U}_tg(\boldsymbol{x}_i) = g\circ\boldsymbol{f}(\boldsymbol{x}_i) = g(\boldsymbol{x}_{i+1}).
\end{eqnarray}
\end{outputs}

An alternative description, related to DMD, is:
\begin{inputs}
Data snapshots of observables on the state-space of a dynamical system.
\end{inputs}

\begin{outputs}
Modal decomposition and linear dynamical system describing modal evolution (see Section \ref{sec:dmd}; DMD).
\end{outputs}

\subsubsection*{Notes}

\paragraph{Approximation by DMD}
A major applied goal of Koopman analysis is the identification of eigenfunctions from data.  
These eigenfunctions provide an intrinsic coordinate system, along which the dynamics appear linear.  
The Koopman mode decomposition is usually approximated via the dynamic mode decomposition (DMD), described in Section \ref{sec:dmd}.    
DMD is essentially a linear regression of data onto dynamics, and augmenting the measurement data with nonlinear functions of the state may enable a more accurate approximation of the Koopman operator~[\citen{williams2014edmd}] and resulting eigenfunctions.  
Another promising algorithm computes Koopman eigenfunctions based on the diffusion operator~[\citen{Giannakis2015arxiv}].  

\paragraph{Hamiltonian Systems}
For Hamiltonian systems, the Koopman operator is unitary, meaning that the inner product of any two observable functions remains the same before and after the operator.  Unitarity is a familiar concept, as the discrete Fourier transform (DFT) and the POD basis both provide unitary coordinate transformations.  In the original paper of Koopman~[\citen{Koopman1931pnas}], connections were drawn between the Koopman eigenvalue spectrum and both conserved quantities and integrability.

\subsubsection*{Strengths and Weaknesses}

\begin{strengths}
\end{strengths}
\begin{itemize}
\item Koopman analysis provides an alternative operator-theoretic perspective to dynamical systems and allows nonlinear systems to be represented and analyzed using linear techniques.   
\item The Koopman operator is amenable to standard spectral decomposition in terms of eigenvalues and eigenvectors.  
\item Powerful techniques in control theory apply to linear systems and may improve nonlinear control performance via Koopman linear systems [\citen{Kaiser:arXiv17}].
\end{itemize}

\begin{weaknesses}
\end{weaknesses}
\begin{itemize}
\item By introducing the Koopman operator, we trade finite-dimensional nonlinear dynamics for infinite-dimensional linear dynamics.  Dealing with infinite-dimensional operators and obtaining low-order approximations is challenging, although there are a larger number of computational techniques to analyze linear operators than there are for differential geometry on manifolds.  
\item It may be difficult to find Koopman eigenfunctions, which define an intrinsic observable measurement coordinate system.  Without Koopman eigenfunctions, it may be difficult or impossible to obtain a finite-dimensional subspace of the Hilbert space of measurement functions that remains closed under the Koopman operator.   However, if discovered, a Koopman invariant subspace defines a measurement system where measurements are propagated by a finite-dimensional linear dynamical system. 
\item The integration of control theory and Koopman analysis remains to be fully developed.  
\end{itemize}

\subsection{Illustrative Example}

\subsubsection*{Simple Nonlinear Dynamical System}

\begin{figure}
\begin{center}
\includegraphics[width=0.66\textwidth]{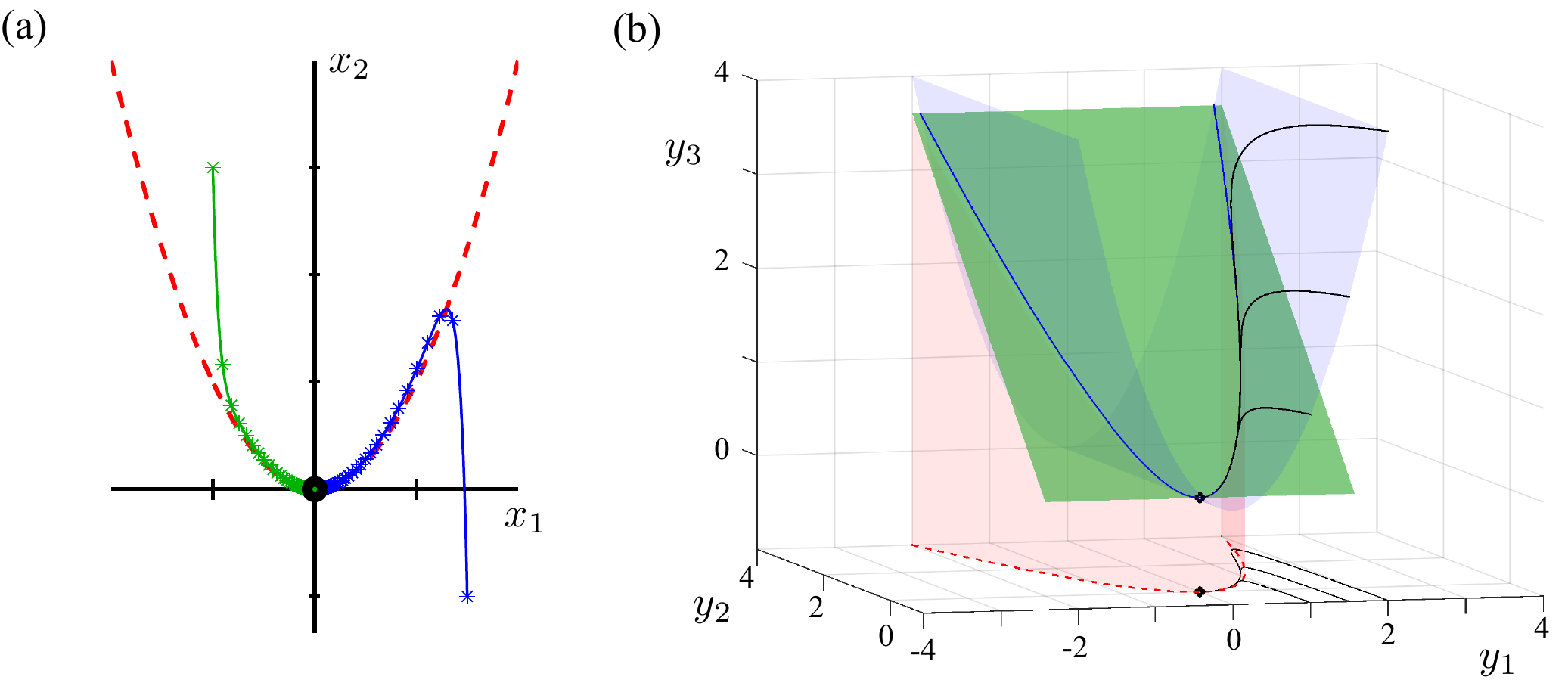}
\caption{Example illustrating how the nonlinear system in Eq.~\eqref{eq:KoopNL} shown in (a) may be transformed into the linear system in Eq.~\eqref{eq:KoopLIN} shown in (b) through an appropriate choice of measurements $\boldsymbol{y}$ of the state $\boldsymbol{x}$.  In this case, the measurement subspace given by $(y_1,y_2,y_3)=(x_1,x_2,x_1^2)$ provides a matrix representation $\boldsymbol{K}$ of the Koopman operator.  Modified from Ref.~[\citen{Brunton2016plosone}].}
\label{Fig:KoopExample}
\end{center}
\end{figure}

As an example, consider the following simple nonlinear dynamical system in two variables with a single fixed point at the origin [\citen{tu2014dynamic,Brunton2016plosone}]:
\begin{equation}
\begin{split}
\dot{x}_1 & =  \mu x_1,\\
\dot{x}_2 & = \lambda(x_2-x_1^2).
\label{eq:KoopNL}
\end{split}
\end{equation}
For $\lambda\ll\mu<0$ the fixed point at the origin is stable and there is a slow manifold given by $x_2=x_1^2$; trajectories quickly attract onto this manifold before converging to the origin (\fig \ref{Fig:KoopExample}a).  
Introducing a nonlinear change of coordinates, given by $(y_1,y_2,y_3) = (x_1,x_2,x_1^2)$, the nonlinear dynamics become linear:
\begin{eqnarray}
\begin{bmatrix}\dot{y}_1\\ \dot{y}_2 \\ \dot{y}_3\end{bmatrix} = \underbrace{\begin{bmatrix}\mu & 0 & 0 \\ 0 & \lambda & -\lambda\\ 0 & 0 & 2\mu\end{bmatrix}}_{\boldsymbol{K}} \begin{bmatrix}y_1\\ y_2 \\ y_3\end{bmatrix}.\label{eq:KoopLIN}
\end{eqnarray}

The nonlinear system in Eq.~\eqref{eq:KoopNL} and the linear Koopman system in Eq.~\eqref{eq:KoopLIN} are shown in \fig \ref{Fig:KoopExample}.  
The system in Eq.~\eqref{eq:KoopLIN} is defined on a vector of observables $\boldsymbol{y} = (y_1,y_2,y_3)^T$, which defines a Koopman-invariant subspace.  The infinite-dimensional Koopman operator restricted to this subspace defines a $3\times 3$ matrix $\boldsymbol{K}$.  With this linear representation, it is possible to analytically predict the state of the system at any future time (i.e., $\boldsymbol{y}(t) = \exp(\boldsymbol{K}t)\boldsymbol{y}(0)$), or extend textbook linear estimation and control techniques to nonlinear systems. 

The example above may appear exceedingly simple, but there are a few direct applications of Koopman analysis for more complex systems, as it is challenging to discover the coordinate transformations that linearize the problem.  
In fluid dynamics, Bagheri~[\citen{Bagheri:2013}] analyzed the Koopman mode decomposition of the flow past a cylinder, relating the decomposition to POD--Galerkin with a shift mode~[\citen{Noack:JFM03}].  
Related work investigates the Liouville equation~[\citen{Gaspard:PRE95}], which has resulted in cluster-based reduced-order models in fluid systems~[\citen{Kaiser:JFM14}].  
In power electronics, Koopman mode decomposition has been used to model and predict instabilities~[\citen{susuki2011nonlinear,susuki2012nonlinear}].   
However, DMD is used in most fluid applications to approximate the Koopman mode decomposition, although the quality of this approximation depends on the measurements used for DMD~[\citen{Kutz2016book}].  
True Koopman linearization of complex systems in fluid dynamics relies on the choice of good observable functions that provide linear embeddings for the nonlinear dynamics, as in the example above.  
The extended DMD~[\citen{williams2014edmd}] provides one approach to augment DMD with nonlinear measurements of the state, potentially improving the approximation of Koopman eigenfunctions.

\subsection{Outlook}

The promise of Koopman analysis hinges on the discovery of good observable functions that provide a coordinate transformation in which measurements behave linearly.  
Identifying these Koopman eigenfunctions from data is both a major goal and central challenge moving forward, making it a focus of research efforts.  
Dynamic mode decomposition, which is a workhorse of Koopman analysis, implicitly uses linear observable functions.  In fluid flows, these often take the form of direct velocity field measurements from numerical simulations or particle image velocimetry (PIV) in experiments.  In other words, the observable function is an identity map on the fluid flow state.  This set of linear observables is often too limited to describe the rich dynamics observed in fluid systems.  Recently, DMD has been extended to include a richer set of nonlinear observable functions, providing the ability to effectively analyze nonlinear systems [\citen{williams2014edmd}].  
To avoid overfitting, sparse regression is emerging as a principled technique to select active terms in dynamical system~[\citen{Brunton2016pnas}], and similar techniques may be used to improve the identification of observable functions.  Recent techniques also construct Koopman eigenfunctions using a regularized advection-diffusion operator~[\citen{Giannakis2015arxiv}] or approximate the Koopman operator in delay coordinates~[\citen{Brunton2016havok,Arbabi2016arxiv}]. 
Discovering Koopman eigenfunctions is a central computational issue to obtain closed systems that may be used for nonlinear control and estimation, and the ultimate success of this analysis will depend on our ability to accurately and efficiently approximate these eigenfunctions from data.  
It is likely that developments in machine learning, such as deep learning, will continue to advance these capabilities.



\section{Global Linear Stability Analysis}
\label{sec:global}

Global linear stability analysis solves the linearized Navier--Stokes equations pertinent to an underlying {\em base flow} with multiple inhomogeneous spatial directions. The introduction of a modal formulation for the linear perturbations converts the linearized equations of motion (initial value problem) to an eigenvalue problem with respect to the {\it base flow}, which needs to be an exact steady (or unsteady) solution of the equations of fluid motion.  The term {\it global} is used to distinguish the analysis from the classic {\it local} linear stability theory [\citen{Drazin81,Schmid01,JuniperHanifiTheofilis}], where the base flow is independent of two coordinate directions.  The eigenvalue problem delivers information on the {\it spectrum} of the linearized Navier--Stokes operator at asymptotically large times. Much like in classic local linear stability theory, the eigenvalue determines whether the corresponding eigenvector grows (exponentially) in time or space and also delivers information on the frequency of the linear perturbation (Section \ref{sec:eig} and \fig \ref{fig:eig_plane}). The eigenvector describes the spatial structure of the linear perturbation and may be used to reconstruct the full flow field at conditions consistent with the linear approximation. 

Global modes are solutions to the two or three-dimensional partial derivative eigenvalue problems to which the linearized Navier--Stokes equations may be recast. They describe disturbances developing upon base flows that vary inhomogeneously in two ({\it{biglobal}}) or three ({\it{triglobal}}) spatial directions [\citen{TheofilisPrAeS2003}]. Thus, global modes are the counterparts, in base flows that vary in multiple spatial dimensions, of well-known linear instabilities developing in one-dimensional base flows such as the inviscid Kelvin--Helmholtz instabilities, viscous Tollmien--Schlichting, and crossflow modes. When the two or three-dimensional base flow contains portions of flow that could be analyzed with local theory, the spatial distribution of the global eigenfunction is seen to contain structures related to the eigenfunction delivered by local analysis, e.g., Tollmien--Schlichting waves in the boundary layer on a flat plate [\citen{AkervikEtAl2008}] and on the downstream wall of an open cavity [\citen{TheofilisColoniusAIAA2003}].  The applications of global stability analysis to study linearly stability of two and three-dimensional base flow have been reviewed recently [\citen{Theofilis:ARFM11}] and are continuously expanding.

\subsection{Description}

\subsubsection*{Algorithm}

\begin{inputs}
Steady or time-periodic laminar base flow, inhomogeneous in two or all three spatial directions, at a given Reynolds number and Mach number.  The base flow can be a stable or an unstable steady state.
\end{inputs}

\begin{outputs}
Global linear modal stability theory delivers global modes (spatial distribution pattern) and the associated growth rates and frequencies of small-amplitude perturbations about the base state.  
\end{outputs}

Unlike the aforementioned POD, Balanced POD, or DMD techniques, global stability analysis is not based on the snapshots of the flow field.
Global modes are found numerically by discretizing the Navier--Stokes equations, which result from the decomposition of any flow quantity $\boldsymbol{q}$ into a base flow ${\boldsymbol {q}}_0$ and small-amplitude perturbations $\boldsymbol {q}'$ (i.e., $||\boldsymbol {q}'||/|| {\boldsymbol {q}}_0 || \ll 1$):
\begin{equation}
   \boldsymbol q(\boldsymbol{\xi},t) = {\boldsymbol {q}}_0(\boldsymbol{\xi},t) + \boldsymbol {q}'(\boldsymbol{\xi},t).
   \label{eq:linear_decomp}
\end{equation}
We can then examine the growth or decay of the perturbations with respect to the base state ${\boldsymbol {q}}_0$.  Substitution of the flow variable in the form of \eq (\ref{eq:linear_decomp}) into the Navier--Stokes equations and neglecting the quadratic terms of small perturbation $\boldsymbol{q}'$ yields the linearized Navier--Stokes equations.

Now, as an example, let us assume the base flow ${\boldsymbol {q}}_0 = {\boldsymbol {q}}_0(\boldsymbol{\xi})$ to be steady and homogeneous (e.g., periodic in space) in the $\zeta$ direction and perform a biglobal stability analysis.  In this case, the perturbation can assume the form of
 \begin{equation}
 \begin{split}
\boldsymbol q'(\xi,\eta,\zeta,t) =\hat{\boldsymbol {  q} }(\xi,\eta)e^{i(\beta \zeta - \omega t)}+\text{complex conjugate},
 \end{split}
 \label{modal}
\end{equation}
where $\beta$ is a real wavenumber in the $\zeta$ direction.  The real and imaginary components of the complex number $\omega = \omega_r + i \omega_i$ respectively correspond to the {\it frequency and the growth/decay rate} of the {\it amplitude function of the global mode} $\hat{\boldsymbol{q}} (\xi,\eta)$.  There are two main approaches to find the global stability modes and the associated eigenvalues.  Namely, they are the {\it matrix-based approach} and the {\it time-stepping approach}, described below.  Various techniques have also been summarized in the review by Theofilis [\citen{Theofilis:ARFM11}], where the relative merits of matrix-based and time-stepping techniques are discussed.

\subsubsection*{Matrix-Based Approach}

The matrix-based approach determines the global modes by solving a generalized eigenvalue problem.  The substitution of the assumed form of \eq (\ref{modal}) into the linearized Navier--Stokes equations results in
\begin{equation}
\boldsymbol{A} ( {\boldsymbol{q}}_0;\beta) {\hat {\boldsymbol{q}}}=\omega {\boldsymbol{B}} \hat{ \boldsymbol{q}},
\label{modal2}
\end{equation}
which is written in terms of a generalized eigenvalue problem where $\omega$ is the eigenvalue and $\hat{\boldsymbol q}$ is the global stability mode (eigenvector).  The matrix $\boldsymbol{A}$ is dependent on the base state ${\boldsymbol {q}}_0$ as well as the wavenumber $\beta$ and flow parameters such as the Reynolds number and Mach number.  In general, $\boldsymbol{B}$ is invertible for compressible flow but is not invertible for incompressible flow due to the incompressibility constraint.  For detailed discussions on the linearized Navier--Stokes equations, see Schmid and Henningson [\citen{Schmid01}] and Theofilis [\citen{Theofilis:ARFM11}].

The matrices $\boldsymbol{A}$ and $\boldsymbol{B}$ are approximately of size $n_\text{grid} n_\text{var} \times n_\text{grid} n_\text{var}$, where $n_\text{grid}$ and $n_\text{var}$ are the number of grid points and number of variables, respectively.  If the operators are of modest size, the matrices may be formed and stored on memory.  For larger size operators, the eigenvalues can be determined without storing the operators but by only relying on matrix-vector operations (matrix-free approach) [\citen{Theofilis:ARFM11, Zhang:PF16, Sun:TCFDXX, Sun:JFM17}].  We can also distribute the matrix entries over several processors on a cluster and use appropriate libraries (e.g., ScaLAPACK) for the linear algebra operations [\citen{Rodriguez:AIAAJ09}].

\subsubsection*{Time-Stepping Approach}

In the {\it time-stepping approach} [\citen{Theofilis:ARFM11, Tezuka:AIAAJ06, Bagheri:JFM09, Liu:JFM16}], a code is employed to compute the linear operators of the Navier--Stokes equations (Fr\'echet derivative [\citen{Keller:MC75}]), without the need to store the matrix. Since typically the numerical discretization of the operators does not expand the solution on bases which automatically satisfy the boundary conditions for the perturbations, spurious modes can be introduced.  We need to extract the physically meaningful modes from the numerically obtained part of the eigenspectrum.

\subsubsection*{Notes}

\paragraph{Computational effort}

With either the matrix-based approach or the time-stepping approach, the computational effort required to find eigenvalues and eigenfunctions can be very high.  Hence the use of sparse yet high-order spatial discretization methods can be beneficial [\citen{BresColonius, ParedesCMAME:13}]. If the flow of interest has symmetry, it can be exploited to reduce the computational cost of solving two and three-dimensional eigenvalue problems [\citen{TatsumiYoshimura,ParedesTheofilisJFS,Liu:JFM16}].

\paragraph{Choice of base flow}

For stable flows, the steady state solution to the Navier--Stokes equations can be found by integrating the solution in time until all unsteadiness is eliminated.  If the unstable steady state is needed, techniques such as the selective frequency damping method [\citen{Akervik:PF06}] or a Newton--Krylov type iterative solver [\citen{Tuckerman:IMA00,Kelley:arxiv}] can be utilized.  We note that time-periodic flow can also be utilized as the base flow [\citen{Barkley:JFM96}], as we will see in the first illustrative example below [\citen{HeGioriaPerezTheofilisJFM}].

Occasionally researchers compute global modes for base flows that are not themselves solutions of the governing equations (e.g., the mean of a turbulent flow).  In these cases, the resulting modes are not associated with a question of stability of the base flow, but may be useful in modeling and identifying the frequency content of large-scale coherent structures [\citen{Sun:TCFDXX, Edstrand:JFM16}].  Additional details with a cautionary note are provided by Sipp and Lebedev [\citen{Sipp:JFM07}].

\paragraph{Global non-modal/transient-growth analysis} 
Global modes may represent either stable or unstable disturbances with respect to the base flow. When unstable modes are identified, the base flow itself is said to be asymptotically unstable, meaning that at least one perturbation will grow exponentially in time and the equilibrium is not expected to be observed in nature.  As discussed in Section \ref{sec:pspec}, systems that are non-normal can exhibit significant growth of linear combinations of modal perturbations, known as transient growth, a phenomenon which may occur even when the global modes are all stable [\citen{Schmid:ARFM07,Trefethen05,Luchini:JFM00}].  Non-normality of the governing linear operator implies that global transient growth analysis may deliver qualitatively different short-time behavior of linear combinations of perturbations when modal analysis predicts only asymptotically stable perturbations.  Global transient growth analysis has unraveled initial optimal perturbations that are qualitatively different from their corresponding global eigenmodes on the cylinder [\citen{AbdessemedSharmaSherwinTheofilisPF}], low pressure turbine blades [\citen{AbdessemedSharmaTheofilisJFM,SharmaAbdessemedSherwinTheofilis}] and several stalled NACA airfoils [\citen{HeGioriaPerezTheofilisJFM}].

\subsubsection*{Strengths and Weaknesses}

\begin{strengths}
\end{strengths}
\begin{itemize} 
   \item A spectrum of (discrete and continuous) eigenmodes can be determined, especially with the matrix-based approach.
   \item Modal analysis, based on the solution of the eigenvalue problem (see Section \ref{sec:eig_svd}), determines whether unstable modes exist.  Exponential growth implies that flow dynamics will be asymptotically dominated by the characteristics of the most unstable eigenmode (or the least stable mode if stable).
   \item If only stable modes are found, global non-modal/transient growth analysis, using linear combination of the leading members of the eigenvalue spectrum, or solution of a related SVD problem (see Section \ref{sec:eig_svd}), can determine the level of energy growth of small-amplitude perturbations over a short time horizon.
\end{itemize}

\begin{weaknesses}
\end{weaknesses}
\begin{itemize} 
   \item Global linear stability analysis is inherently linear.
   \item The base flow provides the spatially variable coefficients of the underlying partial derivative eigenvalue/initial-value problem. Any algorithm for the solution of the latter problems can be influenced by the quality (accuracy) of the underlying base flow.
   \item The analysis requires a base flow which is an exact solution of the equations of motion. Although counter-examples exist in both classic [\citen{GasterKitWygnanski}] and global linear stability theory [\citen{ParedesTerhaarOberleithnerTheofilisPaschereit}], analysis of mean turbulent flow requires validation of the turbulence closures employed and typically aims at prediction of frequencies and spatial structure of coherent turbulent structures. 
\end{itemize}

\subsection{Illustrative Examples}

\subsubsection*{Large-Scale Separation Cells on Stalled Airfoils}

\begin{figure}[htb]  
\centering
\includegraphics[width=0.85\linewidth]{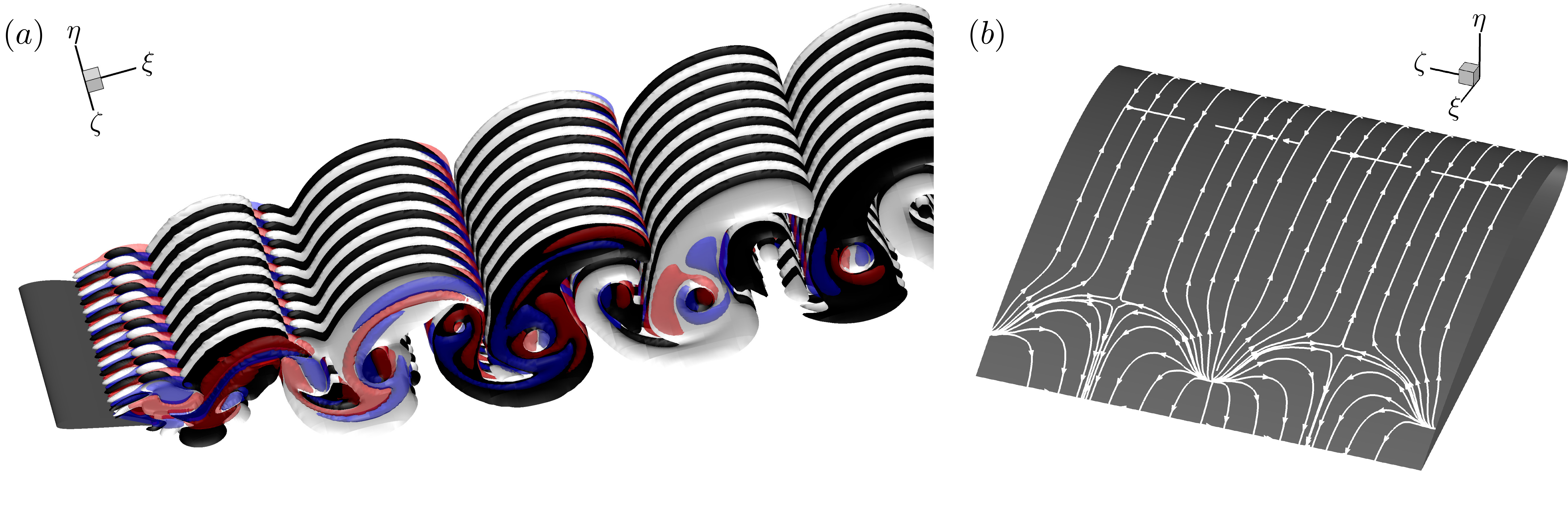}
\caption{Three-dimensional short-wavelength instability mode determined from global stability analysis for a NACA 4415 airfoil at $Re = 500$ and $\alpha = 20^\circ$.  (a) Streamwise vorticity ($\omega_\zeta$) isosurface of the short-wavelength eigenmode and (b) the corresponding wall-streamlines, superposed upon the time-periodic base state [\citen{HeGioriaPerezTheofilisJFM}].  Reprinted with permission from Cambridge University Press.}
\label{fig:sc}
\end{figure}

He {\it et al.}~[\citen{HeGioriaPerezTheofilisJFM}] have employed a suite of matrix-forming and time-stepping techniques to re-examine large-scale separation patterns on the suction side of wings in near-stall flight conditions. The origins of these structures was explained through systematic application of primary and secondary linear global eigenvalue and transient growth stability analysis to massively separated spanwise homogeneous laminar flows over spanwise-periodic wings of different thickness and camber. At low chord Reynolds numbers, the dominant flow structure arising from either primary modal or non-modal mechanisms is associated with two- or three-dimensional Kelvin--Helmholtz eigenmodes. As the wing aspect ratio is shortened, the stationary three-dimensional global eigenmode discovered by Theofilis {\it et al.}~[\citen{TheofilisHeinDallmann}] is amplified stronger than the Kelvin--Helmholtz mode and stall cells may arise through this primary modal linear amplification mechanism [\citen{Rodriguez:TCFD11}]. As the Reynolds number increases, linear amplification of the two-dimensional Kelvin--Helmholtz global eigenmode leads to a time-periodic wake which, in turn, is linearly unstable with respect to two distinct classes of three-dimensional secondary (Floquet) eigenmodes, which peak at short- and long-spanwise wavelengths, respectively. The short-wavelength Floquet eigenmode is the stronger of the two at moderate and high Reynolds numbers. The three-dimensional spatial structure of the short-wavelength mode can be seen in \fig \ref{fig:sc}(a), while the wall-streamline pattern resulting from its linear superposition upon the underlying time-periodic base flow gives rise to stall-cell-like patterns on the wing surface, as they can be seen in \fig \ref{fig:sc}(b).

\subsubsection*{Finite-Span Open Cavity Flow}

\begin{figure}[htb]  
\centering
\includegraphics[width=0.8\linewidth]{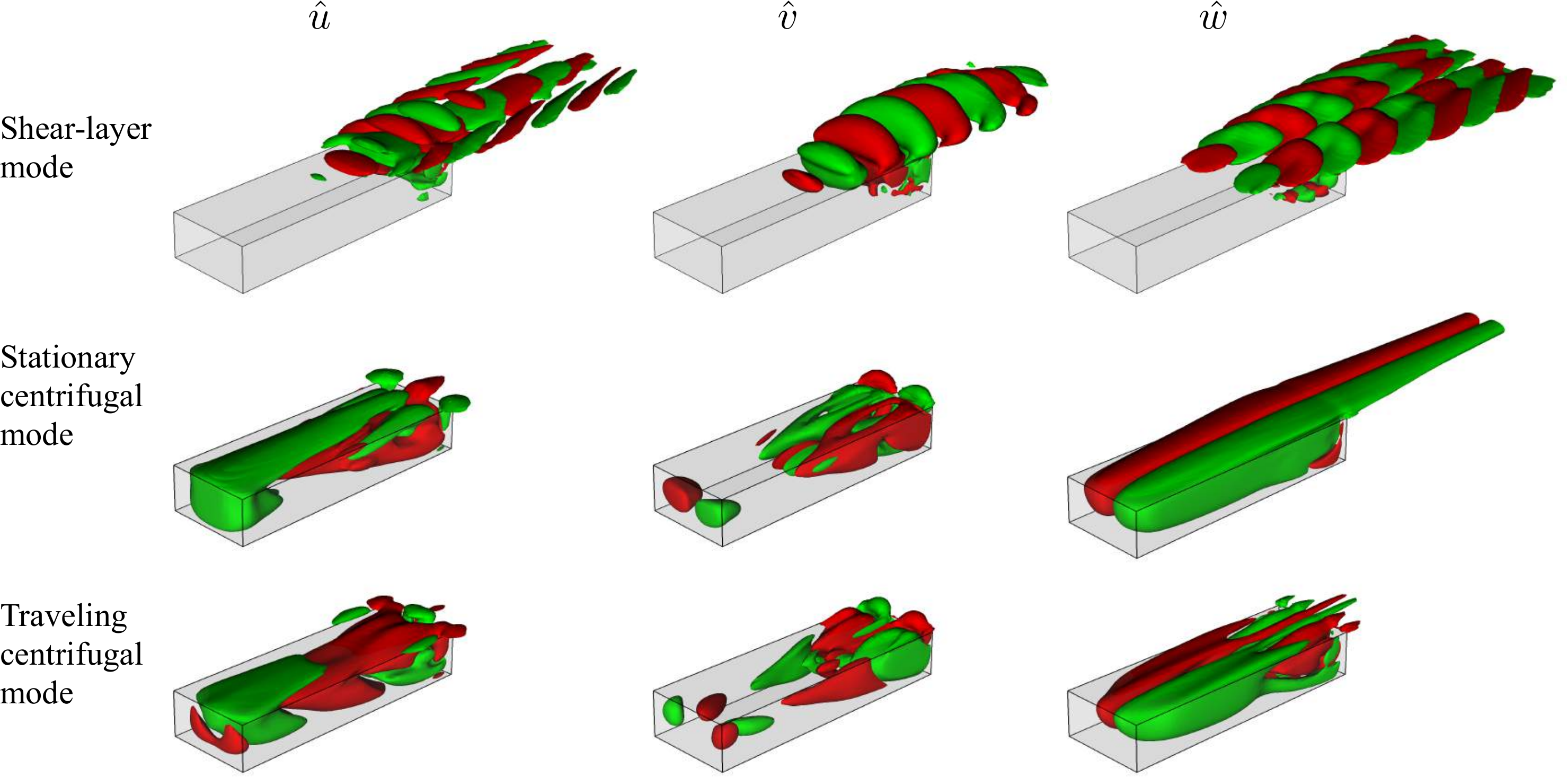}
\vspace{3mm}
\caption{Amplitude functions of the components of the three-dimensional perturbation velocity components of the leading eigenvectors in lateral-wall bounded open rectangular cavity flow at $Re=1050$ [\citen{Liu:JFM16}].  Reprinted with permission from Cambridge University Press.}
\label{fig:3d_modes}
\end{figure}

The second example leverages the time-stepper based approach to gain insights into the critical conditions and spatial characteristics of distinct classes of eigenmodes of incompressible flow over a rectangular three-dimensional lateral-wall-bounded open cavity.  The analysis has been performed in a triglobal setting [\citen{Liu:JFM16}]. The leading traveling shear-layer mode responsible for transition in this geometry as well as the next in significance stationary and traveling centrifugal modes, respectively, are shown in \fig \ref{fig:3d_modes} at the slightly subcritical Reynolds number $Re=1050$. Detailed temporal biglobal analysis of the related spanwise homogeneous open cavity has also been performed [\citen{BresColonius, Sun:JFM17}] and the effect of the presence of lateral walls on the linear instability mechanisms in this class of flows can now be quantified on the basis of global stability theory. 

\subsection{Outlook}

With the availability of enhanced computational resources, the use of temporal and spatial biglobal stability analysis is becoming prevalent to examine a range of fluid flows.  An area in which the theory is expected to become particularly useful is laminar-turbulent transition prediction in hypersonic flow, in support of increasing efforts to understand and control flow phenomena critical to the design of next-generation vehicles. Unlike typical incompressible or supersonic conditions at which turbulent flow prevails in flight, hypersonic flow is mostly laminar and the underlying base flows are exact solutions of the equations of motion which can be computed either using a continuum assumption [\citen{GaitondeEtAl2002}] or by employing Direct Simulation Monte Carlo (DSMC) methods [\citen{TumukluLiLevin:AIAAJ16,TumukluLevinTheofilis,TumukluPerezLevinTheofilis,TumukluPerezTheofilisLevin}].  As an example, linear global instability analysis has recently addressed hypersonic flow over the HIFiRE-5 elliptic cone [\citen{ParedesTheofilisJFS,ParedesGosseTheofilisKimmel}]. The spatial biglobal eigenvalue problem has been solved and four classes of (leading) centerline, attachment-line, crossflow and second Mack modes have been recovered as two-dimensional global mode amplitude functions, underlying the potential of the theory to provide predictions of laminar-turbulent transition at hypersonic speeds. 

In addition to taking advantage of the symmetry present in flows of interest, we can also incorporate the parabolized assumption [\citen{Herbert1997}] into global stability analysis.  For instance, three-dimensional inhomogeneous base flows that depend strongly on two and weakly on the third spatial direction is an extension of the {\it non-local} stability analysis, based on the parabolized stability equations.  Examples of successful application include the system of trailing vortices behind an aircraft wing [\citen{BroadhurstTheofilisSherwin,BroadhurstSherwin,ParedesTheofilisRodriguezTendero}], the wake of an isolated roughness element in supersonic flow [\citen{DeTullioParedesSandhamTheofilisJFM}], streaks in a boundary layer [\citen{MartinParedesTCFD2016}], corner flows [\citen{GalionisHallTCFD}] and models of duct intakes [\citen{GalionisHallJFM}]. Computationally expensive triglobal stability analysis is emerging to provide insights into the influence of three-dimensionality on the stability of flows [\citen{Tezuka:AIAAJ06,Bagheri:JFM09,Liu:JFM16}].  It is anticipated that there will be growing use of these approaches to analyze increasingly complex flows.  

As a companion to the global stability analysis, adjoint global stability analysis [\citen{LuchiniBottaroARFM}] can provide efficient responses to questions on flow receptivity, sensitivity, and control. Global modes are the right eigenvectors of the eigenvalue problem, while the left eigenvectors are the {\it adjoint global modes}. The adjoint linearized Navier--Stokes equations, which are at the heart of linear instability analysis and linear control methodologies [\citen{LuchiniBottaroARFM}], can be solved with some added effort to the global eigenvalue problem. At its simplest implementation, a single solution of the global eigenvalue problem delivers the eigenvalues, as well as linear perturbations and their adjoint variables as the right and left eigenvectors, respectively.


\section{Resolvent Analysis}
\label{sec:resolvent}

\begin{figure}[htb]
\centering
\includegraphics[width=0.3\textwidth]{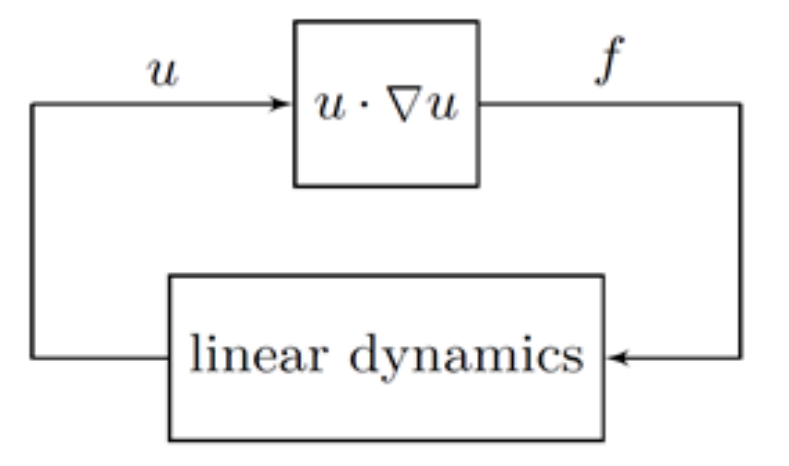}
\caption{Schematic of a divergence-free projection of the resolvent formulation of the incompressible Navier--Stokes equations, showing the nonlinear term, $\boldsymbol{f}$, providing input forcing to the linear dynamics (described by the resolvent operator), with velocity output, $\boldsymbol{u}$ (i.e., ${\boldsymbol q}'$ in Eq.~(\ref{eq:reseq1})) [\citen{McKeonPoF13}].  Reprinted with permission from AIP Publishing.}
\label{fig:resolventsystem}
\end{figure}

The {\it resolvent} is defined in the context of fluid mechanics as the linear operator relating an input forcing to a linear system to the output.  For the complete incompressible Navier--Stokes equations, the output is the (divergence-free) velocity field and the input to the resolvent (linear dynamics) is provided by the nonlinear term, as shown in \fig \ref{fig:resolventsystem}. Resolvent analysis is therefore a complement to eigenvalue and global stability analyses, which are linear analyses with perturbations assumed to be sufficiently small such that the nonlinear advective forcing term (second order in the perturbation) can be neglected.  More generally, inputs to resolvent analysis can include actuation and other kinds of forcing terms added to the governing equations, and outputs can be any observable.

Resolvent analysis relies on the pseudospectrum of the linearized Navier--Stokes operator rather than the spectrum itself (see Section~\ref{sec:pspec}).  For stable base flows, the damped modes of the linear system can be superposed to construct the response to stationary (real-frequency) inputs.  Even for these stable flows, when the operator is non-normal (most fluid systems linearized about spatially non-uniform flows give rise to non-normal operators), stationary inputs can be amplified.  For these systems, there are typically a limited number of highly amplified inputs, implying that the resolvent can be effectively approximated at low rank. An SVD of the resolvent (in the discrete setting) can be used to identify these inputs and their corresponding outputs, and to provide a low-rank approximation of the input-output dynamics of the full system. High-gain inputs are useful for flow control efforts by showing where actuation will produce the largest effect.

\subsection{Description}

\subsubsection*{Algorithm}

\begin{inputs}
A real-valued frequency $\omega$, and a linear time-invariant input-output system (operator) including its boundary conditions, for example the linearized Navier--Stokes equations with an appropriate assumed base or mean velocity profile ${\overline {\boldsymbol q}}$.  These must typically have been discretized in space using a finite-difference, finite-element, spectral, or other numerical method.
\end{inputs}

\begin{outputs}
A ranked set of modes (total number equal to the number of degrees of freedom of the discretized system) that represent the spatial pattern of inputs, outputs, and the positive gain (amplification) from input to output.  High-gain modes represent stationary (statistically steady) inputs that are most amplified, at the specified frequency.
\end{outputs}

Consider $\boldsymbol{q} = \overline{\boldsymbol{q}} + \boldsymbol{q}'$ and the spatially discretized governing equations for the perturbation variable expressed in the compact form
\begin{equation}
{\boldsymbol M} {\partial {\boldsymbol q}^{\prime} \over \partial t} + {\boldsymbol A} {\boldsymbol q}^{\prime} = {\boldsymbol f}\left( {\boldsymbol q}^{\prime} \right),
\label{eq:reseq1}
\end{equation}
where the vector ${\boldsymbol q}^{\prime}$ represents the fluctuating discrete flow variables, the (typically sparse) matrices ${\boldsymbol M}$ and ${\boldsymbol A} = {\boldsymbol A}(\overline{\boldsymbol{q}})$ represent the linearization about a time-invariant flow, ${\overline {\boldsymbol q}}$, and ${\boldsymbol f}$ denotes the remaining nonlinear terms, as well as any additional forcing terms added to the equations. Such terms can include, for example, stochastic forcing (synthetic turbulence) added to inflow regions of spatially developing flows.  For statistically steady (stationary) flows, we may work in the frequency domain.  By expressing
\begin{equation}
   \boldsymbol{q}' = \hat{\boldsymbol{q}} e^{i \omega t}
   \quad \text{and} \quad
   \boldsymbol{f} = \hat{\boldsymbol{f}} e^{i \omega t} ,
\end{equation}
for temporal frequency $\omega \in \mathbb{R}$, \eq (\ref{eq:reseq1}) can be written as
\begin{equation}
\left( i \omega {\boldsymbol M}  + {\boldsymbol A} \right) {\hat {\boldsymbol q} } = {\hat {\boldsymbol f}}.
\end{equation}
Proceeding formally, we write ${\hat{\boldsymbol q} } = {\boldsymbol R} {\hat {\boldsymbol f}}$, where ${\boldsymbol R} 
= \left( i \omega {\boldsymbol M}  + {\boldsymbol A} \right)^{-1}$ is the {\it resolvent operator}.  For a particular frequency, $\omega$, the maximum amplification $G$ (defined in a chosen norm, ${\hat {\boldsymbol q}}^* {\boldsymbol Q} {\hat {\boldsymbol q}}$, where ${\boldsymbol Q}$ is a positive definite matrix) of {\it any} forcing may be found
\begin{equation}
G = \max_{{\hat {\boldsymbol f}}} {{\hat {\boldsymbol q}}^* {\boldsymbol Q} {\hat {\boldsymbol q}} \over {\hat{\boldsymbol f}}^* {\boldsymbol Q} {\hat {\boldsymbol f}}} 
= \max_{{\hat {\boldsymbol f}}} { {\hat {\boldsymbol f}}^* {\boldsymbol R}^* {\boldsymbol Q} {\boldsymbol R} {\hat{\boldsymbol f}} \over {\hat{\boldsymbol f}}^* {\boldsymbol Q} {\hat {\boldsymbol f}}}.
\end{equation}
The solution to this maximization problem can be found through the SVD of ${\boldsymbol R}$.  The largest singular value represents the (squared) gain and the left and right singular vectors represent the optimal response and forcing, respectively.

In the context of the incompressible Navier-Stokes equations, the resolvent, $\boldsymbol{R}$, is the linear operator that relates velocity perturbation $\boldsymbol{u}'$ and pressure perturbation $p'$ to the nonlinear term, $\boldsymbol{f}$,
\begin{equation}
  \label{eq:res-def}
  \boldsymbol{q}' =
  \left[ \begin{array}{c}\boldsymbol{u}'\\p'\end{array}\right] = \boldsymbol{R}\boldsymbol{f}.
\end{equation}
Note that a base velocity profile is required to formulate the resolvent and to define the perturbations.

Spatial homogeneity can be exploited to reduce the computational burden by first decomposing the linear system into the Fourier basis in the homogeneous spatial directions.  This is also useful for identifying dynamics associated with individual wavenumber components of interest in the data. For simplicity, we consider here an example flow that is homogeneous in the streamwise and spanwise directions and statistically stationary in time, such that the wavenumber/frequency triplet $\boldsymbol{k} = (k_x, k_z, \omega)$ associated with oblique downstream traveling waves can be defined. Accordingly, we have
\begin{equation}
  \label{eq:res-H}
\boldsymbol{R}_{\boldsymbol{k}} =
\left(-i\omega \left[ \begin{array}{cc}\boldsymbol{I} &\\& 0\end{array}\right]
- \left[\begin{array}{cc}\boldsymbol{L}_{\boldsymbol{k}} & -\nabla\\ \nabla^T & 0\end{array} \right] \right)^{-1}
\left[ \begin{array}{c}\boldsymbol{I}\\0\end{array}\right],
\end{equation}
where $\boldsymbol{L}_{\boldsymbol{k}}$ is the linear Navier--Stokes (advection and diffusion) operator evaluated at $\boldsymbol{k}$, $\nabla$ and $\nabla^T$ are the gradient and divergence operators, and $\boldsymbol{I}$ is the identity operator. As such, the resolvent operator can be obtained by relatively minor extension of code for the more common Orr--Sommerfeld--Squire analysis. Of course, the system may be formulated in terms of either velocity-vorticity or primitive variables.  For an analogous derivation of the resolvent for compressible flow, see Jeun \etal~[\citen{NicholsJovanovic16}]. 

Two closely related formulations of the resolvent analysis are common in the literature. In the {\it{input-output}} formulation, an output quantity is defined as the state vector premultiplied by a matrix that is chosen to extract an observable of interest [\citen{NicholsJovanovic16}]. Similarly, the forcing is premultiplied by a (not necessarily) different matrix to restrict the input-output relation to forcings of interest. In the {\it{linear frequency response}} approach [\citen{Garnaut13}], the gain is defined as the quotient of the input and output in terms of their energy defined through (not necessarily different) norms.

For a given $\boldsymbol{k}$, the SVD procedure described above in Section~\ref{sec:eig_svd} returns right singular vectors corresponding to the {\it most dangerous} (most amplified) inputs or forcing modes, left singular vectors corresponding to the associated response modes, with gains given by the corresponding singular values. Example code for resolvent analysis of turbulent pipe flow in primitive variables is available online for general use [\citen{Github13}].  The analysis proceeds depending on the specific objective.  For example, the globally most amplified input can be sought, the forcing and response modes compared with observations, or data projected onto the resolvent basis [\citen{McKeon2010,McKeonPoF13}].

\subsubsection*{Notes}

\paragraph{Origin} The resolvent is a familiar construction in the study of forced, ordinary differential equations, and more generally, linear operators and their spectra.  It arises in control theory and in eigenvalue/eigenvector perturbation analysis.  Identified by Schmid and Henningson [\citen{Schmid01}], its use as a tool to study structures in transitional and turbulent flows seems to have been stimulated by Farrell and Ioannou [\citen{Farrell93}], who considered a stochastically forced Navier--Stokes system, and follows contemporaneous work on understanding transient growth in stable base flows with pseudospectra [\citen{Trefethen:Science93}].

\paragraph{Choice of base flow} Similar to the linear global stability analysis discussed in the previous section, the choice of the base flow ${\boldsymbol{q}_0}$ is important in resolvent analysis, because operator $\boldsymbol{A}$ and forcing input $\boldsymbol{f}$ are established with respect to the base state $\boldsymbol{q}_0$.  If the (stable or unstable) steady-state solution to the Navier--Stokes equations is available, it can be used as the base flow [\citen{Schmid01,Trefethen:Science93}] and the nonlinear term can be assumed to be negligible within the context of linear analysis.  In this case, the input forcing must be provided from an external source. We also note that resolvent analysis has been performed about the time-average flow $\overline{\boldsymbol{q}}$.  In such case, the nonlinear term is not negligible but is considered as an internal source of the forcing mechanism [\citen{McKeonPoF13}], as illustrated in \fig \ref{fig:resolventsystem}.  For this reason, resolvent analysis has been used to examine how fluctuations including those from nonlinear effects are amplified or attenuated with respect to the time-average flow.  When applying resolvent analysis to the temporal mean of inherently unsteady or turbulent flows, it is advisable to examine whether the operator $\boldsymbol{A} = \boldsymbol{A}(\boldsymbol{q}_0)$ has stable eigenvalues to separate the forced response from the unforced response. 

\paragraph{Implementation in Different Geometries} The use of resolvent analysis is increasing in fluid dynamics to study both transitional and fully turbulent flows.  Flows that have been analyzed in terms of the most amplified (forced) modes include, amongst others, canonical flows such as boundary layers, Couette and Poiseuille flows [\citen{Jovanovic05,McKeon2010}], jets [\citen{Garnaut13,Towne15,NicholsJovanovic16,Beneddine:JFM17}], backward-facing step [\citen{Beneddinestep16}], and cavity flows [\citen{Gomezcavity16}].

\paragraph{Computational Costs}
Resolvent modes can be routinely and efficiently computed for base flows that are homogeneous in two spatial dimensions (one-dimensional base flow). Resolvent analysis for flows that are homogeneous in one spatial dimension (two-dimensional base flow) incurs significantly more computing cost [\citen{Papadakis14}]. With current computational resources, fully inhomogeneous flows represent an unmet challenge, although flows with narrowband perturbations can be tackled [\citen{Gomezcavity16}].

\paragraph{Wall Boundary Conditions and Control}
Resolvent analysis admits any linear control formulation through the boundary conditions. For example, passive (a compliant surface [\citen{Luharcompliant15}]), open-loop (dynamic roughness [\citen{Jacobidynamic11}]), and active (opposition control [\citen{Luharcontrol14}]) control inputs at the wall have been investigated in wall turbulence.

\paragraph{Wavepackets and Coherent Structures}
While formally a tool for linear systems, resolvent analysis can provide insights into coherent structures in turbulent flow by analyzing the response of the system associated with the linear resolvent operator formed using the turbulent mean flow. This is not necessarily equivalent to linearization, since the nonlinear forcing is required to sustain an otherwise stable system.  Note that the mean velocity itself can be recovered via an extension of the analysis shown in \fig \ref{fig:resolventsystem}. Resolvent modes can be identified and the corresponding inputs can be thought to represent how such coherent structures are forced through nonlinear (triadic) interactions amongst modes at other frequencies, or through other stochastic inputs to the system (for example noise added at an inflow in a DNS to excite turbulence).  A particular success has been the recovery of packets of hairpin vortices in wall turbulence from resolvent modes [\citen{Sharma13}].

The resolvent-based approach is also particularly useful to analyze convectively unstable configurations that exhibit no intrinsic dynamics. Despite the absence of an unstable linear global mode (see Section \ref{sec:global}) in such cases, coherent structures are often observed in such flows. A typical example is a turbulent jet under most operating conditions. Here, continuous forcing through the background turbulence sustains large-scale coherent structures in the jet shear-layer over a wide range of frequencies. The resolvent analysis allows identification of these wavepackets as optimal responses, and associates each response mode to its corresponding optimal input via the gain of the input-output relation (see second example below).

\subsubsection*{Strengths and Weaknesses}

\begin{strengths}
\end{strengths}
\begin{itemize}
\item Resolvent analysis can use the mean flow as the base state (instead of the exact solution to the Navier--Stokes equations), even in the case of turbulent flows.
\item When resolvent analysis is applied to turbulent mean flows, large-scale coherent structures can be interpreted as modal solutions that are sustained through forcing by the turbulent background.
\item The resolvent operator can be formulated to describe the linear dynamics associated with the governing Navier--Stokes equations.
\item Resolvent analysis identifies the form of the most amplified inputs, the corresponding output, and the gain.
\item The resolvent operator is often low-rank, with underlying mathematical structure that can be used to understand coherent flow structure and permitting exploitation of state-of-the-art matrix approximation techniques.
\end{itemize}

\begin{weaknesses}
\end{weaknesses}
\begin{itemize}
\item Resolvent analysis identifies the most amplified inputs and outputs, whereas the product of the amplification (singular value) and the weight of the nonlinear forcing determines the observations.  Information on the nonlinear forcing term is therefore required to predict the most energetic observed modes and synthesize nonlinear models.
\end{itemize}

\subsection{Illustrative Examples}

\subsubsection*{Turbulent Channel Flow}

\begin{figure}[htb]
\centering
  \includegraphics[width=0.685\textwidth]{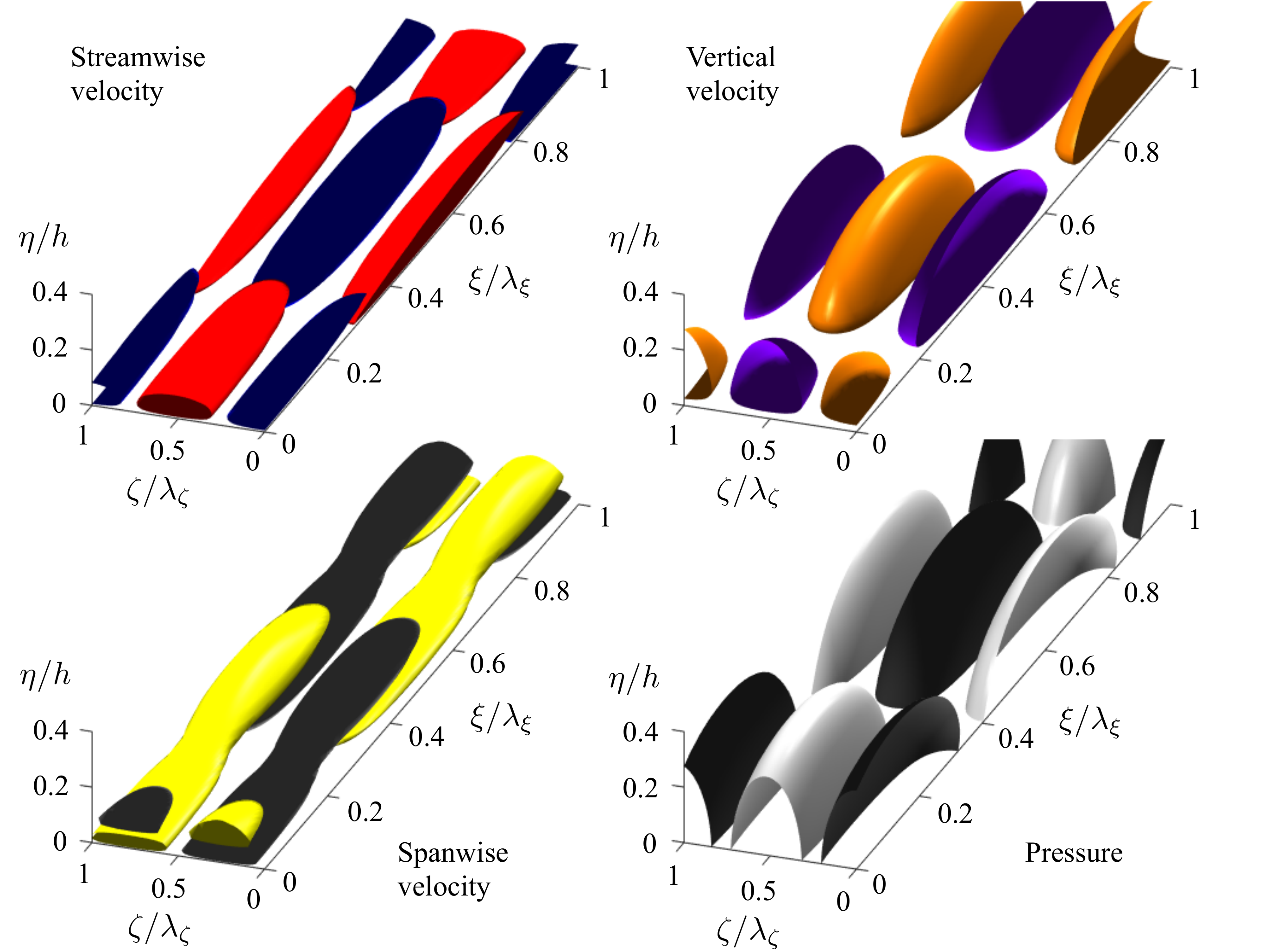}
  \caption{Isocontours at half the maximum value for each component (warm/light colors positive value, cold/dark colors negative value) of the real part of the first resolvent response mode pair $(k_\xi, \pm k_\zeta,c=\omega/k_\xi)=(1h, 6h, 0.5 h U_{CL})$ in channel flow at friction Reynolds number $h^+ = 3000$. Here $U_{CL}$ is the centerline velocity and $h^+$ is the channel half-height normalized by the viscous length scale.}
\label{fig:resmode}
\end{figure}

We show in \fig \ref{fig:resmode} an example of the typical form of first, i.e. most amplified, resolvent response modes in turbulent channel flow. Here, a left-and right-going pair of modes (i.e., $\pm k_\zeta$) are considered with wavenumbers and wavespeed representative of the very-large scale motions in wall turbulence. For this mode, the amplitudes of the three velocity components have relative magnitudes $(|u|,|v|,|w|) = (1,0.05,0.17)$, i.e., the turbulent kinetic energy of the mode is mostly associated with the streamwise velocity fluctuations. The isocontours of velocity correspond to a quasi-streamwise vortex (inclined roll) structure in the cross-stream plane familiar from observations of fully-developed turbulence at all scales, but especially reminiscent of the very large scale motions, suggesting that the resolvent mode is a useful simplified model for that structure at least.

\subsubsection*{Turbulent Jet}

\begin{figure}[htb]
\centering
   \begin{overpic}[width=0.55\textwidth]{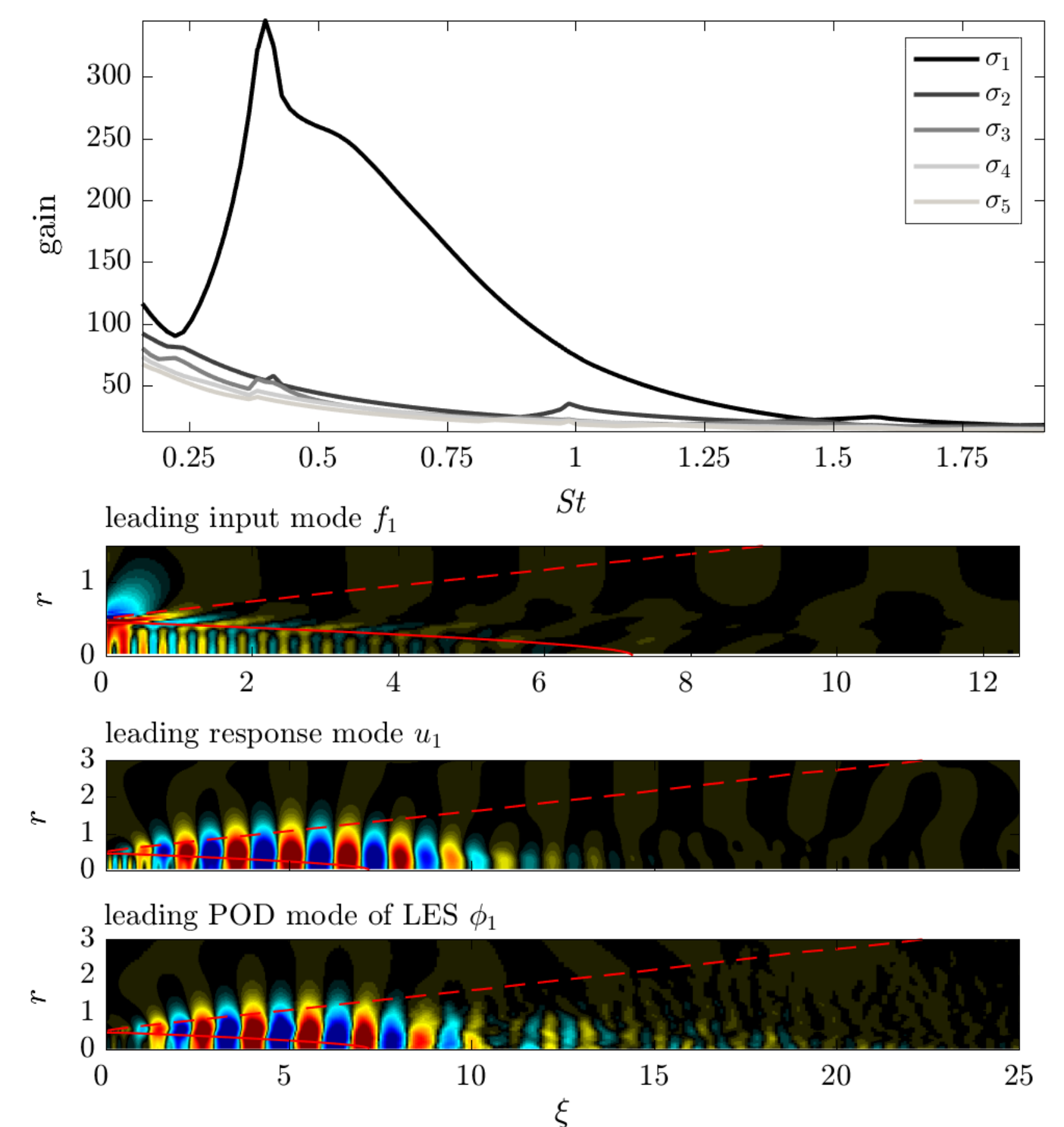}
   \put(0,98){\scriptsize (a)}
   \put(0,53.5){\scriptsize (b)}
   \put(0,34.5){\scriptsize (c)}
   \put(0,18){\scriptsize (d)}
   \end{overpic}
	\caption{Resolvent analysis of a turbulent jet mean flow. The gains $\sigma$ of the five leading axisymmetric resolvent modes over Strouhal number are reported in (a). As an example, the pressure component of the leading input mode $f_1$ for $St=0.5$ is shown in (b). The corresponding response $u_1$ shown in (c) closely resembles the leading POD mode $\boldsymbol{\phi}_1$ deduced from LES data in (d).}
	\label{fig:jet_resolvent}
\end{figure}
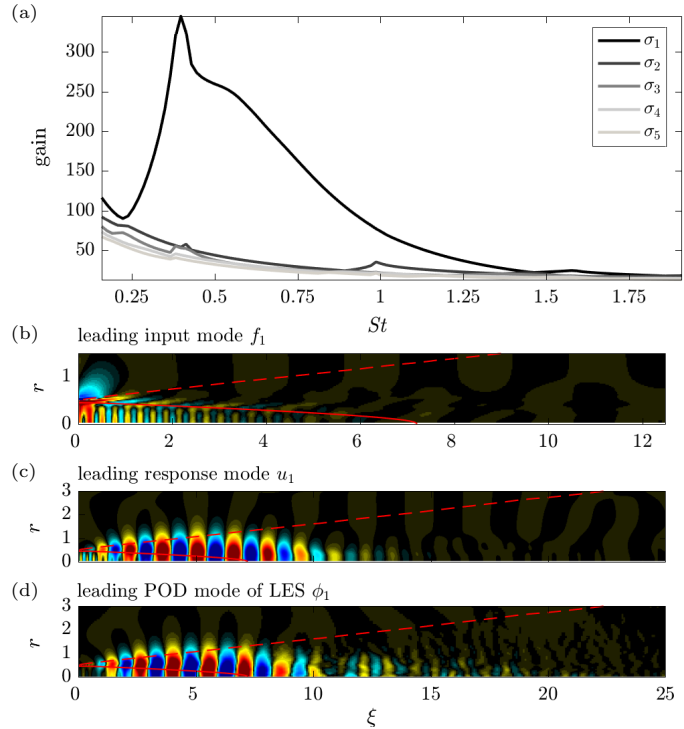

In \fig \ref{fig:jet_resolvent}(a), the gain of first five resolvent modes associated with a turbulent jet is given as a function of the Strouhal number $St$ (dimensionless frequency in acoustic units). The gain of the leading resolvent mode clearly dominates over a large frequency interval. Two features of the gain curve are of special physical interest. The sharp peak at $St\approx0.4$ is associated with an acoustic resonance in the potential core [\citen{SchmidtEtAl16}], and the underlying broad-band response peak with the preferred mode of forced jets [\citen{Garnaut13}]. The forcing structure at $St=0.5$ in \fig \ref{fig:jet_resolvent}(b) reveals that the response depicted in \fig \ref{fig:jet_resolvent}(c) is most efficiently sustained by an upstream forcing distribution that utilizes the Orr-mechanism in the shear-layer and direct acoustic forcing in the potential core to achieve optimality. The optimal response closely resembles the coherent structure educed from a LES database using Spectral POD (see Section \ref{sec:pod}) based on the spectral cross-correlation matrix.

\subsection{Outlook}

The use of resolvent analysis to develop parametric sensitivity studies is an especially promising direction for future research.  The utility of the resolvent approach in developing dynamical models rests on identifying important nonlinear interactions; determining the correct weighting of the nonlinear forcing.  These weights can be obtained or approximated in various ways: for example by direct calculation of the forcing from full-field snapshots [\citen{Towne15,TowneERMD15}], projection of the snapshots onto resolvent response modes [\citen{Moarref13,Moarref14,SharmaECS16}], by scaling arguments [\citen{SharmaPTAXX}], or by direct solution of the quadratic programming problem defining the coupling between response modes and wavenumbers [\citen{McKeonPoF13}], the subject of ongoing work.

Resolvent modes have recently been shown to provide an efficient basis to represent exact coherent solutions of the Navier--Stokes equations [\citen{SharmaECS16}], suggesting that the analysis may both provide an inexpensive approximate continuation method within families of solutions and aid the search for new exact solutions.  Formal connections between resolvent analysis, Koopman analysis, DMD and exact nonlinear traveling wave solutions have recently been outlined [\citen{SharmaMezicM16}].  This is out of scope of this manuscript, but may be of interest to expert practitioners.


\section{Concluding Remarks}
\label{sec:remarks}

We have provided an overview of the modal decomposition/analysis techniques that are widely used to examine a variety of fluid flows.  Specifically, we have presented POD, Balanced POD, DMD, Koopman analysis, global linear stability analysis, and resolvent analysis.  The modal structures extracted by these techniques can shed light on different aspects of the flow field with some shared similarities.  For the data-based methods, POD analysis captures the most energetic modes and the balanced POD analysis captures the most controllable and observable modes, providing balancing and adjoint modes.  DMD (and Koopman analysis) extracts the dynamic modes along with their growth rates and frequencies from the flow field data.  While POD and DMD methods can use flow field data from numerical simulation and experimental measurements, Balanced POD requires adjoint flow data, which makes it applicable to numerical studies only.  We have also presented operator-based analysis techniques of Koopman analysis, global linear stability analysis, and resolvent analysis.  The first rigorously connects DMD to nonlinear dynamical systems.  The latter two techniques can examine growth or decay characteristics of perturbations with respect to a given base or mean flow.  These two approaches require discretized operators from the Navier--Stokes equations to perform the modal stability analysis.  

The examples shown in this paper focused mostly on fluid flow analysis.  However, modes obtained from these techniques can also be used to develop reduced-order models, which can capture the dynamics of the flow with significantly lower computational cost.  Such models can be useful in closed-loop flow control [\citen{Samimy:JFM07, Ilak:PF08, Ahuja:JFM10, Noack11, Brunton:AMR15, Kaiser:JFM14, Nair:arXiv17}] as well as aerodynamic design [\citen{Bui-Thanh:AIAAJ04,LeGresleyThesis}].   A large number of references were provided throughout the paper so that readers can seek additional insights as needed.  We hope that this document can serve as a stepping stone for the readers to become familiar with various modal analysis techniques, analyze a variety of complex flow physics problems, and further advance the developments of these modal analysis techniques.


\section*{Acknowledgments}

This paper was one of the major outcomes from the AIAA Discussion Group (Fluid Dynamics Technical Committee) entitled ``Modal Decomposition of Aerodynamics Flows'' organized by KT and Dr.~Douglas R. Smith, AFOSR (Program Officer, Unsteady Aerodynamics and Turbulent Flows). 

DRS has been instrumental in creating, organizing, and supporting this Discussion Group at every stage. He in fact has been one of the contributors to the present overview paper from conception to final editing. We gratefully acknowledge the contributions made by DRS to this document and for his continued support of much of this work.

The authors also thank the fruitful discussions with the members of the discussion group and greatly acknowledge the generous support from the following agencies: Air Force Office of Scientific Research (AFOSR), Army Research Office (ARO), Defense Advanced Research Projects Agency (DARPA), Department of Energy (DoE), Deutsche Forschungsgemeinschaft (DFG), European Office of Aerospace Research and Development (AFOSR/EOARD), National Science Foundation (NSF), and Office of Naval Research (ONR). 

The authors acknowledge the anonymous referees for providing insightful comments on the manuscript.  KT is grateful to his research group members for providing extensive feedback throughout the preparation of this paper.


\appendix
\section{Appendix}
\label{sec:App}

We present an example of formatting the flow field data into a matrix form $\boldsymbol{X}$ in preparation to perform the modal decomposition analysis.  For simplicity, we assume that the grid is structured with uniform spacing and the data is sampled with constant time step.  Here, we take data from a two-dimensional velocity field $\boldsymbol{q} = (u,v)$ and construct $\boldsymbol{X}$.  In general, the data to be examined can be from an unstructured or non-uniform grid, if scaled appropriately, or processed to reside on a uniform grid.  The formulation discussed here can be easily extended to scalar variables or three-dimensional vector fields.

\begin{figure}[tb]
\centering
\includegraphics[width=0.4\textwidth]{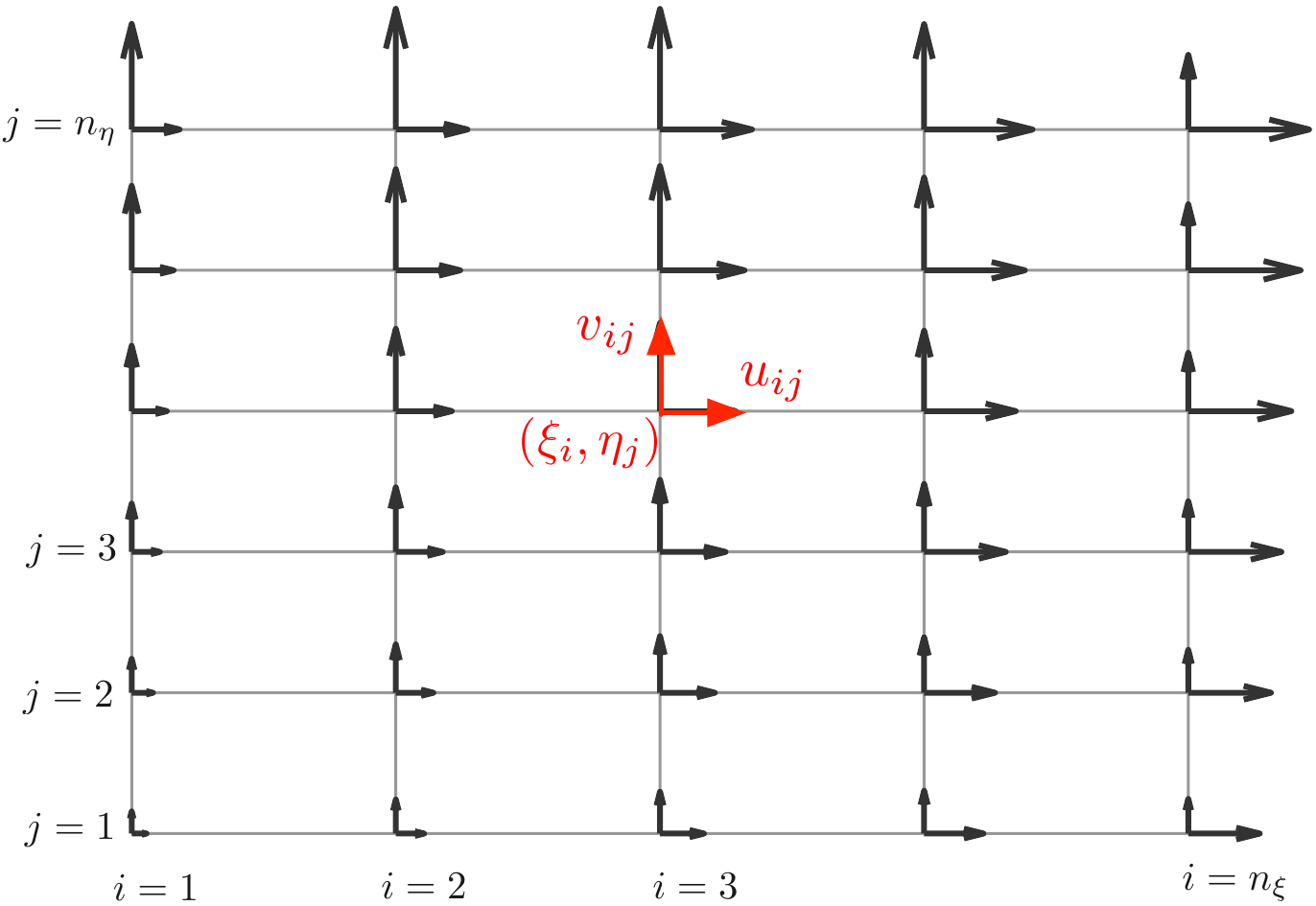}
  \caption{An example of discrete velocity data on a uniform collocated grid. }
\label{fig:coord}
\end{figure}

Consider the velocity field $\boldsymbol{q}$ made available from simulation or experiment
\begin{equation}
   \boldsymbol{q}_{ij} = (u_{ij},v_{ij}) = (u(\xi_i,\eta_j),v(\xi_i,\eta_j)),
\end{equation}
where the spatial coordinates $(\xi_i,\eta_j)$ with $i=1,\dots,n_\xi$ and $j=1,\dots,n_\eta$ are for a collocated setup, as illustrated in \fig \ref{fig:coord}.  To construct the data matrix $\boldsymbol{X}$, we need to organize the data in the form of a column vector at each time instance.  An example of stacking the snapshot data to form a column vector $\boldsymbol{x}(t)$ at time $t = t_k$ can be written as
\begin{equation}
   {\scriptsize{
   \left[
   \begin{array}{ccc}
   u_{11} & \dots & u_{n_\xi 1} \\
   \vdots & \ddots& \vdots \\
   u_{1 n_\eta} & \dots & u_{n_\xi n_\eta}  \\ \hline
   v_{11} & \dots & v_{n_\xi 1} \\
   \vdots & \ddots& \vdots \\
   v_{1 n_\eta} & \dots & v_{n_\xi n_\eta}  
   \end{array}
   \right]_{t=t_k} \! \! \! \! \! \! \! \! \!
   \normalsize
   \begin{array}{c}\stackrel{\tt{stack}}{\longrightarrow}\\ 
   \stackrel{\tt{unstack}}{\longleftarrow} \end{array}
   }}
   ~~
   {\scriptsize{
   \boldsymbol{x}(t_k) \equiv 
   \left[ \! \! \! 
   \begin{array}{c}
   \left( \! \! \begin{array}{c} u_{11}\\ \vdots \\ u_{1n_\eta} \end{array} \! \! \right)\\
   \vdots \\
   \left( \! \! \! \! \begin{array}{c} u_{n_\xi 1}\\ \vdots \\ ~u_{n_\xi n_\eta}~ \end{array} \! \! \! \! \right)\\ \hline
   \left( \! \! \begin{array}{c} v_{11}\\ \vdots \\ v_{1n_\eta} \end{array} \! \right)\\
   \vdots \\
   \left( \! \! \! \! \begin{array}{c} v_{n_\xi 1}\\ \vdots \\ ~v_{n_\xi n_\eta}~ \end{array} \! \! \! \! \right)
   \end{array}
   \! \! \! 
   \right]_{t=t_k} 
   \! \! \! \! \! \! \! \! \! \! \! \!
   \in \mathbb{R}^{n}
   }}
   \label{eq:stack}
\end{equation}
Note that the above stacking and unstacking operations are for a two-dimensional vector, which yield a column vector size $n=2 n_x n_y$.  The unstacking operation is useful for visualizing the decomposition modes once decomposition is performed on the flow field data.  For scalar data, we only need to consider a particular component.  For three-dimensional vector data, we can append the third component to the above formulation.  Once the column data vectors are available at the desired time levels, we can collect them into a data matrix
\begin{equation}
   \boldsymbol{X} = \left[ \boldsymbol{x}(t_1)~ \boldsymbol{x}(t_2) ~\dots~ \boldsymbol{x}(t_m)\right] 
   \in \mathbb{R}^{n\times m}. 
\end{equation}
In general, this data matrix is a tall and skinny matrix for fluid flows ($n \gg m$).
With this data matrix at hand, we are ready to perform modal decompositions through POD, Balanced POD, and DMD.


\bibliographystyle{aiaa}
\bibliography{refs_all}

\begin{thebibliography}{100}
\newcommand{\enquote}[1]{``#1''}

\bibitem{Holmes96}
Holmes, P., Lumley, J.~L., Berkooz, G., and Rowley, C.~W., {\em Turbulence,
  coherent structures, dynamical systems and symmetry\/}, Cambridge Univ.
  Press, 2nd ed., 2012.

\bibitem{Rowley:ARFM17}
Rowley, C.~W. and Dawson, S. T.~M., \enquote{Model Reduction for Flow Analysis
  and Control,} {\em Annu. Rev. Fluid Mech.\/}, Vol.~49, 2017, pp.~387--417.

\bibitem{Kutz2013book}
Kutz, J.~N., {\em Data-Driven Modeling \& Scientific Computation: Methods for
  Complex Systems \& Big Data\/}, Oxford University Press, 2013.

\bibitem{Kutz2016book}
Kutz, J.~N., Brunton, S.~L., Brunton, B.~W., and Proctor, J.~L., {\em Dynamic
  Mode Decomposition: Data-Driven Modeling of Complex Systems\/}, SIAM, 2016.

\bibitem{Theofilis:ARFM11}
Theofilis, V., \enquote{Global linear instability,} {\em Annu. Rev. Fluid
  Mech.\/}, Vol.~43, 2011, pp.~319--352.

\bibitem{Schmid01}
Schmid, P.~J. and Henningson, D.~S., {\em Stability and transition in shear
  flows\/}, Springer, 2001.

\bibitem{Taira:Nagare11a}
Taira, K., \enquote{Proper orthogonal decomposition in fluid flow analysis: 1.
  Introduction,} {\em J. Japan Soc. Fluid Mech. (Nagare)\/}, Vol.~30, 2011 (in
  Japanese), pp.~115--123.

\bibitem{Taira:Nagare11b}
Taira, K., \enquote{Proper orthogonal decomposition in fluid flow analysis: 2.
  Applications,} {\em J. Japan Soc. Fluid Mech. (Nagare)\/}, Vol.~30, 2011 (in
  Japanese), pp.~263--271.

\bibitem{Samimy03}
Samimy, M., Breuer, K.~S., Leal, L.~G., and Steen, P.~H., editors, {\em A
  gallery of fluid motion\/}, Cambridge Univ. Press, 2003.

\bibitem{Brown:JFM74}
Brown, G.~L. and Roshko, A., \enquote{On density effects and large structures
  in turbulent mixing layers,} {\em J. Fluid Mech.\/}, Vol.~64, No.~4, 1974,
  pp.~775--816.

\bibitem{Strouhal:APC78}
Strouhal, V., \enquote{On one particular way of tone generation,} {\em Annalen
  der Physik und Chemie (Leipzig), ser. 3\/}, Vol.~5, 1878, pp.~216--251.

\bibitem{Rayleigh:PM79}
Rayleigh, L., \enquote{Acoustical observations,} {\em Phil. Mag. and J. of
  Sci.\/}, Vol.~7, No.~42, 1879, pp.~149--162.

\bibitem{Benard:CAS08}
B\'enard, H., \enquote{Formation de centres de giration \`a l'arri\`ere d'un
  obstacle en mouvem\`ent,} {\em Comp. rend. Acad. Sci.\/}, Vol.~147, 1908.

\bibitem{Karman:GN11}
{von K\'arm\'an}, T., \enquote{\"Uber den Mechanismus des Widerstandes, den ein
  bewegter K\"orper in einer Fl\"usseigkeit erf\"ahrt,} {\em Gott. Nachr.\/},
  1911, pp.~509--517.

\bibitem{Taneda:JPSJ56}
Taneda, S., \enquote{Experimental investigation of the wakes behind cylinders
  and plates at low {R}eynolds numbers,} {\em J. Phys. Soc. Japan\/}, Vol.~11,
  1956, pp.~302--307.

\bibitem{Coutanceau:JFM77a}
Coutanceau, M. and Bouard, R., \enquote{Experimental determination of the main
  features of the viscous flow in the wake of a circular cylinder in uniform
  translation. {P}art 1. {S}teady flow,} {\em J. Fluid Mech.\/}, Vol.~79,
  No.~2, 1977, pp.~231--256.

\bibitem{Canuto:JFM15}
Canuto, D. and Taira, K., \enquote{Two-dimensional compressible viscous flow
  around a circular cylinder,} {\em J. Fluid Mech.\/}, Vol.~785, 2015,
  pp.~349--371.

\bibitem{Helmholtz1868}
von Helmholtz, H., \enquote{On discontinuous movements of fluids,} {\em Phil.
  Mag.\/}, Vol.~36, 1868, pp.~337--346.

\bibitem{Kelvin1871}
Kelvin, L., \enquote{Hydrokinetic solutions and observations,} {\em Phil.
  Mag.\/}, Vol.~42, 1871, pp.~362--377.

\bibitem{Rosenhead1931}
Rosenhead, L., \enquote{The formation of vortices from a surface of
  discontinuity,} {\em Proc. Roy. Soc. London A\/}, Vol.~134, No. 823, 1931,
  pp.~170--192.

\bibitem{Winant:JFM74}
Winant, C.~D. and Browand, F.~K., \enquote{Vortex pairing : the mechanism of
  turbulent mixing-layer growth at moderate {R}eynolds number,} {\em J. Fluid
  Mech.\/}, Vol.~63, No.~2, 1974, pp.~237--255.

\bibitem{Zaman:JFM80}
Zaman, K. B. M.~Q. and Hussain, A. K. M.~F., \enquote{Vortex pairing in a
  circular jet under controlled excitation. Part 1. General jet response,} {\em
  J. Fluid Mech.\/}, Vol.~101, No.~3, 1980, pp.~449--491.

\bibitem{Melander:JFM88}
Melander, M.~V., Zabusky, N.~J., and Mcwilliams, J.~C., \enquote{Symmetric
  vortex merger in two dimensions: causes and conditions,} {\em J. Fluid
  Mech.\/}, Vol.~195, 1988, pp.~303--340.

\bibitem{Bonnet:EF98}
Bonnet, J.~P., Delville, J., Glauser, M.~N., Antonia, R.~A., Bisset, D.~K.,
  Cole, D.~R., Fiedler, H.~E., Garem, J.~H., Hilberg, D., Jeong, J., Kevlahan,
  N. K.~R., Ukeiley, L.~S., and Vincendeaux, E., \enquote{Collaborative testing
  of eddy structure identification methods in free turbulent shear flows,} {\em
  Exp. Fluids\/}, Vol.~25, 1998, pp.~197--228.

\bibitem{Taira:JFM09}
Taira, K. and Colonius, T., \enquote{Three-dimensional flows around
  low-aspect-ratio flat-plate wings at low {R}eynolds numbers,} {\em J. Fluid
  Mech.\/}, Vol.~623, 2009, pp.~187--207.

\bibitem{Colonius:CMAME08}
Colonius, T. and Taira, K., \enquote{A fast immersed boundary method using a
  nullspace approach and multi-domain far-field boundary conditions,} {\em
  Comput. Methods Appl. Mech. Engrg.\/}, Vol.~197, 2008, pp.~2131--2146.

\bibitem{Lumley1967}
Lumley, J.~L., \enquote{The structure of inhomogeneous turbulent flows,} {\em
  Proceedings of the International Colloquium on the Fine Scale Structure of
  the Atmosphere and its Influence on Radio Wave Propagation\/}, edited by
  A.~M. Yaglam and V.~I. Tatarsky, Doklady Akademii Nauk SSSR, Moscow, Nauka,
  1967.

\bibitem{Sirovich:QAM87}
Sirovich, L., \enquote{Turbulence and the dynamics of coherent structures,
  {P}arts {I}-{III},} {\em Q. Appl. Math.\/}, Vol.~XLV, 1987, pp.~561--590.

\bibitem{Trefethen97}
Trefethen, L.~N. and Bau, D., {\em Numerical linear algebra\/}, SIAM, 1997.

\bibitem{Horn85}
Horn, R.~A. and Johnson, C.~R., {\em Matrix analysis\/}, Cambridge Univ. Press,
  1985.

\bibitem{Golub96}
Golub, G.~H. and Loan, C. F.~V., {\em Matrix computations\/}, Johns Hopkins
  Univ. Press, 3rd ed., 1996.

\bibitem{Saad11}
Saad, Y., {\em Numerical methods for large eigenvalue problems\/}, SIAM, 2nd
  ed., 2011.

\bibitem{Trefethen05}
Trefethen, L.~N. and Embree, M., {\em Spectra and pseudospectra\/}, Princeton
  Univ. Press, 2005.

\bibitem{Drazin81}
Drazin, P.~G. and Reid, W.~H., {\em Hydrodynamic stability\/}, Cambridge Univ.
  Press, 1981.

\bibitem{Trefethen:Science93}
Trefethen, L.~N., Trefethen, A.~E., Reddy, S.~C., and Driscoll, T.~A.,
  \enquote{Hydrodynamic stability without eigenvalues,} {\em Science\/},
  Vol.~261, No. 5121, 1993, pp.~578--584.

\bibitem{Schmid:ARFM07}
Schmid, P.~J., \enquote{Nonmodal stability theory,} {\em Annu. Rev. Fluid
  Mech.\/}, Vol.~39, 2007, pp.~129--162.

\bibitem{Eckart:Psycho36}
Eckart, C. and Young, G., \enquote{The approximation of one matrix by another
  of lower rank,} {\em Psychometrika\/}, Vol.~1, No.~3, 1936, pp.~211--218.

\bibitem{LapackUsersGuide}
Anderson, E., Bai, Z., Bischof, C., Blackford, L.~S., Demmel, J., Dongarra, J.,
  Croz, J.~D., Greenbaum, A., Hammarling, S., McKenney, A., and Sorensen, D.,
  {\em {LAPACK} Users' Guide\/}, {SIAM}, 1999.

\bibitem{ScaLapackUsersGuide}
Blackford, L.~S., Choi, J., Cleary, A., D'Azevedo, E., Demmel, J., Dhillon, I.,
  Dongarra, J., Hammarling, S., Henry, G., Petitet, A., Stanley, K., Walker,
  D., and Whaley, R.~C., {\em {ScaLAPACK} Users' Guide\/}, {SIAM}, 1997.

\bibitem{Lehoucq98}
Lehoucq, R.~B., Sorensen, D.~C., and Yang, C., {\em {ARPACK} Users' Guide:
  Solution of Large-Scale Eigenvalue Problems with Implicitly Restarted
  {Arnoldi} Methods\/}, SIAM, 1998.

\bibitem{Trefethen:AN99}
Trefethen, L.~N., \enquote{Computation of pseudospectra,} {\em Acta
  Numerica\/}, Vol.~8, 1999, pp.~247--295.

\bibitem{Higham:BIT93}
Higham, D.~J. and Trefethen, L.~N., \enquote{Stiffness of {ODE}s,} {\em BIT\/},
  Vol.~33, 1993, pp.~285--303.

\bibitem{Karhunen1946}
Karhunen, K., \enquote{Zur Spektraltheorie Stochasticher Prozesse,} {\em
  Annales Academiae Scientiarum Fennicae\/}, Vol.~37, 1946.

\bibitem{Loeve1955}
Lo{\`e}ve, M., {\em Probability Theory\/}, Princeton, N.J.: D Van Nostrand,
  1955.

\bibitem{Berkooz1993}
Berkooz, G., Holmes, P., and Lumley, J.~L., \enquote{The proper orthogonal
  decomposition in the analysis of turbulent flows,} {\em Annu. Rev. Fluid
  Mech.\/}, Vol.~25, 1993, pp.~539--575.

\bibitem{cordier:hal-00417819}
Cordier, L. and Bergmann, M., \enquote{{Proper Orthogonal Decomposition: an
  overview},} {\em {Lecture series 2002-04, 2003-03 and 2008-01 on
  post-processing of experimental and numerical data, Von Karman Institute for
  Fluid Dynamics, 2008.}\/}, {VKI}, 2008, p. 46 pages.

\bibitem{Herzog86}
Herzog, S., {\em The Large Scale Structure in the Near-Wall Region of Turbulent
  Pipe Flow\/}, Ph.D. thesis, Cornell University, 1986.

\bibitem{Aubry:JFM88}
Aubry, N., Holmes, P., Lumley, J.~L., and Stone, E., \enquote{The dynamics of
  coherent structures in the wall region of a turbulent boundary layer,} {\em
  J. Fluid Mech.\/}, Vol.~192, 1988, pp.~115--173.

\bibitem{Moin:JFM82}
Moin, P. and Kim, J., \enquote{Numerical investigation of turbulent channel
  flow,} {\em J. Fluid Mech.\/}, Vol.~118, 1982, pp.~341--377.

\bibitem{Glauser87}
Glauser, M.~N. and George, W.~K., \enquote{An orthogonal decomposition of the
  axisymmetric jet mixing layer utilizing cross-wire velocity measurements,}
  {\em Symposium on Turbulent Shear Flows\/}, Toulouse, 1987.

\bibitem{Ukeiley:AIAAJ93}
Ukeiley, L., Glauser, M., and Wick, D., \enquote{Downstream evolution of proper
  orthogonal decomposition eigenfunctions in a lobed mixer,} {\em AIAA J.\/},
  Vol.~31, 1993, pp.~1392--1397.

\bibitem{HLBR-11}
Holmes, P., Lumley, J.~L., Berkooz, G., and Rowley, C., \enquote{Turbulence,
  Coherent Structures, Dynamical Systems and Symmetry,} 2012.

\bibitem{Algazi1969}
Algazi, V. and Sakrison, D., \enquote{On the Optimality of the Karhunen-Loeve
  Expansion,} {\em IEEE Trans. Inform. Theory\/}, Vol.~25, 1968, pp.~319--321.

\bibitem{Noack11}
Noack, B.~R., Morzynski, M., and Tadmor, G., editors, {\em Reduced-order
  modelling for flow control\/}, Springer, 2011.

\bibitem{Ukeiley:JFM01}
Ukeiley, L., Cordier, L., Manceau, R., Delville, J., Glauser, M., and Bonnet,
  J.~P., \enquote{Examination of large-scale structures in a turbulent plane
  mixing layer. Part 2. Dynamical systems model,} {\em J. Fluid Mech.\/},
  Vol.~441, 2001, pp.~67--108.

\bibitem{Noack:JFM03}
Noack, B.~R., Afanasiev, K., Morzynski, M., Tadmor, G., and Thiele, F.,
  \enquote{A hierarchy of low-dimensional models for the transient and
  post-transient cylinder wake,} {\em J. Fluid Mech.\/}, Vol.~497, 2003,
  pp.~335--363.

\bibitem{Noack:JFM05}
Noack, B.~R., Papas, P., and Monkewitz, P.~A., \enquote{The need for a
  pressure-term representation in empirical {Galerkin} models of incompressible
  shear flows,} {\em J. Fluid Mech.\/}, Vol.~523, 2005, pp.~339--365.

\bibitem{Rowley:PhysicaD04}
Rowley, C.~W., Colonius, T., and Murray, R.~M., \enquote{Model reduction for
  compressible flows using {POD} and {Galerkin} projection,} {\em Physica D\/},
  Vol.~189, 2004, pp.~115--129.

\bibitem{Nair:arXiv17}
Nair, A.~G., Brunton, S.~L., and Taira, K., \enquote{Networked oscillator based
  modeling and control of unsteady wakes,} {\em in review\/}, 2017.

\bibitem{Glavaski:IEEE98}
Glava\v{s}ki, S., Marsden, J.~E., and Murray, R.~M., \enquote{Model reduction,
  centering, and the {Karhunen--Loeve} expansion,} {IEEE} Conf.~Decision and
  Control, 1998.

\bibitem{Noack:JFM16}
Noack, B.~R., \enquote{From snapshots to model expansions - bridging low
  residuals and pure frequencies,} {\em J. Fluid Mech.\/}, Vol.~802, 2016,
  pp.~1--4.

\bibitem{Lumley1970}
Lumley, J.~L., {\em Stochastic Tools in Turbulence\/}, Academic Press, New
  York, 1970.

\bibitem{George88}
George, W.~K., \enquote{Insight into the dynamics of coherent structures from a
  proper orthogonal decomposition,} {\em Near Wall Turbulence\/}, edited by
  S.~Kline and N.~Afghan, Hemisphere, 1988.

\bibitem{CitrinitiGeorge2000}
Citriniti, J.~H. and George, W.~K., \enquote{Reconstruction of the global
  velocity field in the axisymmetric mixing layer utilizing the proper
  orthogonal decomposition,} {\em J. Fluid Mech.\/}, Vol.~418, 2000,
  pp.~137--166.

\bibitem{PicardDelville2000}
Picard, C. and Delville, J., \enquote{Pressure velocity coupling in a subsonic
  round jet,} {\em International Journal of Heat and Fluid Flow\/}, Vol.~21,
  No.~3, 2000, pp.~359--364.

\bibitem{Suzuki2006}
Suzuki, T. and Colonius, T., \enquote{Instability waves in a subsonic round jet
  detected using a near-field phased microphone array,} {\em J. Fluid Mech.\/},
  Vol.~565, No.~1, 2006, pp.~197--226.

\bibitem{Gudmundsson2011}
Gudmundsson, K. and Colonius, T., \enquote{Instability wave models for the
  near-field fluctuations of turbulent jets,} {\em J. Fluid Mech.\/}, Vol.~689,
  2011, pp.~97--128.

\bibitem{SchmidtEtAl16}
Schmidt, O., Towne, A., Colonius, T., Cavalieri, A., Jordan, P., and Br\`{e}s,
  G., \enquote{Wavepackets and trapped acoustic modes in a Mach 0.9 turbulent
  jet: a global stability analysis,} {\em (submitted)\/}, 2016.

\bibitem{TowneEtAl2017inPreparation}
Towne, A., Schmidt, O.~T., and Colonius, T., \enquote{Instability wave models
  for the near-field fluctuations of turbulent jets,} {\em in preparation for
  J. Fluid Mech.\/}, 2017.

\bibitem{Welch1967}
Welch, P., \enquote{The use of fast {Fourier} transform for the estimation of
  power spectra: a method based on time averaging over short, modified
  periodograms,} {\em IEEE Transactions on audio and electroacoustics\/},
  Vol.~15, No.~2, 1967, pp.~70--73.

\bibitem{Sieber:JFM16}
Sieber, M., Paschereit, C.~O., and Oberleithner, K., \enquote{Spectral proper
  orthogonal decomposition,} {\em J. Fluid Mech.\/}, Vol.~792, 2016,
  pp.~798--828.

\bibitem{Munday:AIAA14}
Munday, P.~M. and Taira, K., \enquote{Wall-normal vorticity injection in
  separation control of {NACA} 0012 airfoil,} {AIAA Paper 2014-2685}, 2014.

\bibitem{Kajishima17}
Kajishima, T. and Taira, K., {\em Computational fluid dynamics: incompressible
  turbulent flows\/}, Springer, 2017.

\bibitem{Munday:AIAAJXX}
Munday, P.~M. and Taira, K., \enquote{Quantifying wall-normal and angular
  momentum injections in airfoil separation control,} {\em AIAA J.\/}, 2017 (in
  review).

\bibitem{murray2009}
Murray, N., S\"allstr\"om, E., and Ukeiley, L., \enquote{Properties of Subsonic
  Open Cavity Flow Fields,} {\em Phys. Fluids\/}, Vol.~21, No.~9, 2009,
  pp.~1661.

\bibitem{Rowley-ijbc05}
Rowley, C.~W., \enquote{Model reduction for fluids using balanced proper
  orthogonal decomposition,} {\em Int.\ J. Bifurcation Chaos\/}, Vol.~15,
  No.~3, 2005, pp.~997--1013.

\bibitem{Camphouse2008}
Camphouse, R.~C., Myatt, J.~H., Schmit, R.~F., Glauser, M.~N., Ausseur, J.~M.,
  Andino, M.~Y., and Wallace, R.~D., {\em A snapshot decomposition method for
  reduced order modeling and boundary feedback control\/}, AIAA Paper
  2008-4195, 2008.

\bibitem{Jorgensen:TCFD03}
J{\o}rgensen, B.~H., S{\o}rensen, J.~N., and Br{\o}ns, M.,
  \enquote{Low-Dimensional Modeling of a Driven Cavity Flow with Two Free
  Parameters,} {\em Theo. Comput. Fluid Dyn.\/}, Vol.~16, 2003, pp.~299--317.

\bibitem{delSastre:06}
del Sastre, P.~G. and Bermejo, R., \enquote{The {POD} Technique for Computing
  Bifurcation Diagrams: A Comparison among Different Models in Fluids,} {\em
  Numerical Mathematics and Advanced Applications\/}, edited by A.~B.
  de~Castro, D.~G\'omez, P.~Quintela, and P.~Salgado, 6th {E}uropean Conference
  on Numerical Mathematics and Advanced Applications, 2005, pp. 880--888.

\bibitem{Gordeyev2013}
Gordeyev, S. and Thomas, F.~O., \enquote{A Temporal Proper Decomposition (TPOD)
  for Closed-Loop Flow Control,} {\em Exp. Fluids\/}, Vol.~54, 2013, pp.~1477.

\bibitem{Gordeyev2014}
Gordeyev, S., De~Lucca, N., Jumper, E., Hird, K., Juliano, T.~J., Gregory,
  J.~W., Thordahl, J., and Wittich, D.~J., \enquote{Comparison of Unsteady
  Pressure Fields on Turrets with Different Surface Features using Pressure
  Sensitive Paint,} {\em Exp. Fluids\/}, Vol.~55, 2014, pp.~1661.

\bibitem{TowneEtAl2015}
Towne, A., Colonius, T., Jordan, P., C. A.~V.~G., and Br{\`e}s, G.~A.,
  \enquote{Stochastic and nonlinear forcing of wavepackets in a Mach 0.9 jet,}
  {\em 21st {AIAA/CEAS} Aeroacoustics Conference\/}, 2015, p. 2217.

\bibitem{SemeraroEtAl2016}
Semeraro, O., Jaunet, V., Jordan, P., Cavalieri, A.~V.~G., and Lesshafft, L.,
  \enquote{Stochastic and harmonic optimal forcing in subsonic jets,} {\em 22nd
  {AIAA/CEAS} Aeroacoustics Conference, AIAA Paper\/}, Vol. 2935, 2016.

\bibitem{Samimy:JFM07}
Samimy, M., Debiasi, M., Caraballo, E., Serrani, A., Yuan, X., Little, J., and
  Myatt, J.~H., \enquote{Feedback control of subsonic cavity flows using
  reduced-order models,} {\em J. Fluid Mech.\/}, Vol.~579, 2007, pp.~315--346.

\bibitem{Kaiser:JFM14}
Kaiser, E., Noack, B.~R., Cordier, L., Spohn, A., Segond, M., Abel, M.,
  Daviller, G., \:Osth, J., Krajnovi\'c, S., and Niven, R.~K.,
  \enquote{Cluster-based reduced-order modelling of a mixing layer,} {\em J.
  Fluid Mech.\/}, Vol.~754, 2014, pp.~365--414.

\bibitem{Moore-81}
Moore, B.~C., \enquote{Principal component analysis in linear systems:
  Controllability, observability, and model reduction,} {\em IEEE Trans.\
  Automat.\ Contr.\/}, Vol.~26, No.~1, 1981, pp.~17--32.

\bibitem{JuPa-85}
Juang, J.-N. and Pappa, R.~S., \enquote{An eigensystem realization algorithm
  for modal parameter identification and model reduction,} {\em J. Guid.\
  Contr.\ Dyn.\/}, Vol.~8, No.~5, 1985, pp.~620--627.

\bibitem{Ma:TCFD09}
Ma, Z., Ahuja, S., and Rowley, C.~W., \enquote{Reduced order models for control
  of fluids using the Eigensystem Realization Algorithm,} {\em Theo. Comp.
  Fluid Dyn.\/}, Vol.~25, No.~1, 2009, pp.~233--247.

\bibitem{willcox:2002}
Willcox, K. and Peraire, J., \enquote{Balanced Model Reduction via the Proper
  Orthogonal Decomposition,} {\em AIAA J.\/}, Vol.~40, No.~11, 2002,
  pp.~2323--2330.

\bibitem{Ilak:PF08}
Ilak, M. and Rowley, C.~W., \enquote{Modeling of transitional channel flow
  using balanced propoer orthogonal decomposition,} {\em Phys. Fluids\/},
  Vol.~20, 2008, pp.~034103.

\bibitem{LallMarsden-02}
Lall, S., Marsden, J.~E., and Glava\v{s}ki, S., \enquote{A subspace approach to
  balanced truncation for model reduction of nonlinear control systems,} {\em
  Int. J. Robust Nonlinear Control\/}, Vol.~12, 2002, pp.~519--535.

\bibitem{ilak2010gl}
Ilak, M., Bagheri, S., Brandt, L., Rowley, C.~W., and Henningson, D.~S.,
  \enquote{Model reduction of the nonlinear complex Ginzburg-Landau equation,}
  {\em SIAM J. App. Dyn. Sys.\/}, Vol.~9, No.~4, 2010, pp.~1284--1302.

\bibitem{Ahuja:JFM10}
Ahuja, S. and Rowley, C.~W., \enquote{Feedback control of unstable steady
  states of flow past a flat plate using reduced-order estimators,} {\em J.
  Fluid Mech.\/}, Vol.~645, 2010, pp.~447--478.

\bibitem{Barbagallo:2009}
Barbagallo, A., Sipp, D., and Schmid, P.~J., \enquote{Closed-loop control of an
  open cavity flow using reduced-order models,} {\em J. Fluid Mech.\/},
  Vol.~641, 2009, pp.~1--50.

\bibitem{bagheri:2009}
Bagheri, S., Brandt, L., and Henningson, D.~S., \enquote{Input-output analysis,
  model reduction and control of the flat-plate boundary layer,} {\em J. Fluid
  Mech.\/}, Vol.~620, 2009, pp.~263--298.

\bibitem{semeraro2011feedback}
Semeraro, O., Bagheri, S., Brandt, L., and Henningson, D.~S., \enquote{Feedback
  control of three-dimensional optimal disturbances using reduced-order
  models,} {\em J. Fluid Mech.\/}, Vol.~677, 2011, pp.~63--102.

\bibitem{semeraro2013TSwaves}
Semeraro, O., Bagheri, S., Brandt, L., and Henningson, D.~S.,
  \enquote{Transition delay in a boundary layer flow using active control,}
  {\em J. Fluid Mech.\/}, Vol.~731, 2013, pp.~288--311.

\bibitem{Flinois2015unstableBPOD}
Flinois, T. L.~B., Morgans, A.~S., and Schmid, P.~J., \enquote{Projection-free
  approximate balanced truncation of large unstable systems,} {\em Phys. Rev.
  E\/}, Vol.~92, Aug 2015, pp.~023012.

\bibitem{derghan2011frequential}
Dergham, G., Sipp, D., Robinet, J.-C., and Barbagallo, A., \enquote{Model
  reduction for fluids using frequential snapshots,} {\em Phys. Fluids\/},
  Vol.~23, No.~6, 2011.

\bibitem{tu2012improved}
Tu, J.~H. and Rowley, C.~W., \enquote{An improved algorithm for balanced POD
  through an analytic treatment of impulse response tails,} {\em J. Comput.
  Phys.\/}, Vol.~231, No.~16, 2012, pp.~5317--5333.

\bibitem{yu2015randomized}
Yu, D. and Chakravorty, S., \enquote{A randomized proper orthogonal
  decomposition technique,} {\em 2015 American Control Conference (ACC)\/},
  IEEE, 2015, pp. 1137--1142.

\bibitem{halko2011random}
Halko, N., Martinsson, P.-G., and Tropp, J.~A., \enquote{Finding structure with
  randomness: Probabilistic algorithms for constructing approximate matrix
  decompositions,} {\em SIAM Review\/}, Vol.~53, No.~2, 2011, pp.~217--288.

\bibitem{schmid2008}
Schmid, P.~J. and Sesterhenn, J., \enquote{Dynamic mode decomposition of
  numerical and experimental data,} {\em 61st Annual Meeting of the APS
  Division of Fluid Dynamics\/}, American Physical Society, 2008.

\bibitem{schmid2010DMD}
Schmid, P.~J., \enquote{Dynamic mode decomposition of numerical and
  experimental data,} {\em J. Fluid Mech.\/}, Vol.~656, 2010, pp.~5--28.

\bibitem{rowley2009spectral}
Rowley, C.~W., Mezi{\'c}, I., Bagheri, S., Schlatter, P., and Henningson,
  D.~S., \enquote{Spectral analysis of nonlinear flows,} {\em J. Fluid
  Mech.\/}, Vol.~641, No.~1, 2009, pp.~115--127.

\bibitem{mezic2013koopman}
Mezi{\'c}, I., \enquote{Analysis of fluid flows via spectral properties of the
  {Koopman} operator,} {\em Annu. Rev. Fluid Mech.\/}, Vol.~45, 2013,
  pp.~357--378.

\bibitem{tu2014dynamic}
Tu, J.~H., Rowley, C.~W., Luchtenburg, D.~M., Brunton, S.~L., and Kutz, J.~N.,
  \enquote{On dynamic mode decomposition: theory and applications,} {\em J.
  Comput. Dyn.\/}, Vol.~1, No.~2, 2014, pp.~391--421.

\bibitem{chen2011variants}
Chen, K.~K., Tu, J.~H., and Rowley, C.~W., \enquote{Variants of dynamic mode
  decomposition: boundary condition, {Koopman}, and {Fourier} Analyses,} {\em
  J. Nonlinear Sci.\/}, Vol.~22, No.~6, 2012, pp.~887--915.

\bibitem{Mann2016qf}
Mann, J. and Kutz, J.~N., \enquote{Dynamic mode decomposition for financial
  trading strategies,} {\em Quantitative Finance\/}, 2016, pp.~1--13.

\bibitem{Grosek2014arxiv}
Grosek, J. and Kutz, J.~N., \enquote{Dynamic Mode Decomposition for Real-Time
  Background/Foreground Separation in Video,} {\em arXiv preprint
  arXiv:1404.7592\/}, 2014.

\bibitem{kutzRPCA1}
Kutz, J.~N., Grosek, J., and Brunton, S.~L., \enquote{Dynamic Mode
  Decomposition for Robust PCA with Applications to Foreground/Background
  Subtraction in Video Streams and Multi-Resolution Analysis,} {\em in CRC
  Handbook on Robust Low-Rank and Sparse Matrix Decomposition: Applications in
  Image and Video Processing, T. Bouwmans Ed.\/}, 2015.

\bibitem{erichson2015}
Erichson, N.~B. and Donovan, C., \enquote{Randomized low-rank Dynamic Mode
  Decomposition for motion detection,} {\em Computer Vision and Image
  Understanding\/}, Vol.~146, 2016, pp.~40--50.

\bibitem{Proctor2015ih}
Proctor, J.~L. and Eckhoff, P.~A., \enquote{Discovering dynamic patterns from
  infectious disease data using dynamic mode decomposition,} {\em International
  health\/}, Vol.~7, No.~2, 2015, pp.~139--145.

\bibitem{Berger2014ieee}
Berger, E., Sastuba, M., Vogt, D., Jung, B., and Amor, H.~B.,
  \enquote{Estimation of perturbations in robotic behavior using dynamic mode
  decomposition,} {\em J. Adv. Robotics\/}, Vol.~29, No.~5, 2015, pp.~331--343.

\bibitem{brunton2016extracting}
Brunton, B.~W., Johnson, L.~A., Ojemann, J.~G., and Kutz, J.~N.,
  \enquote{Extracting spatial--temporal coherent patterns in large-scale neural
  recordings using dynamic mode decomposition,} {\em J. Neurosc. Meth.\/},
  Vol.~258, 2016, pp.~1--15.

\bibitem{Hemati2014streaming}
Hemati, M.~S., Williams, M.~O., and Rowley, C.~W., \enquote{Dynamic mode
  decomposition for large and streaming datasets,} {\em Phys. Fluids\/},
  Vol.~26, No.~11, 2014, pp.~111701.

\bibitem{belson2013modred}
Belson, B.~A., Tu, J.~H., and Rowley, C.~W., \enquote{A parallelized model
  reduction library,} {\em ACM T. Math. Software\/}, 2013.

\bibitem{wynn2013omd}
Wynn, A., Pearson, D.~S., Ganapathisubramani, B., and Goulart, P.~J.,
  \enquote{Optimal mode decomposition for unsteady flows,} {\em J. Fluid
  Mech.\/}, Vol.~733, 2013, pp.~473--503.

\bibitem{jovanovic2014dmdsp}
Jovanovi{\'c}, M.~R., Schmid, P.~J., and Nichols, J.~W.,
  \enquote{Sparsity-promoting dynamic mode decomposition,} {\em Phys.
  Fluids\/}, Vol.~26, No.~2, 2014.

\bibitem{tu2014compressed}
Tu, J.~H., Rowley, C.~W., Kutz, J.~N., and Shang, J.~K., \enquote{Spectral
  analysis of fluid flows using sub-{Nyquist}-rate {PIV} data,} {\em Exp.
  Fluids\/}, Vol.~55, No.~9, 2014, pp.~1--13.

\bibitem{Brunton2015jcd}
Brunton, S.~L., Proctor, J.~L., Tu, J.~H., and Kutz, J.~N., \enquote{Compressed
  sensing and dynamic mode decomposition,} {\em J. Comput. Dyn.\/}, Vol.~2,
  No.~2, 2015, pp.~165--191.

\bibitem{Gueniat2015pof}
Gueniat, F., Mathelin, L., and Pastur, L., \enquote{A dynamic mode
  decomposition approach for large and arbitrarily sampled systems,} {\em Phys.
  Fluids\/}, Vol.~27, No.~2, 2015, pp.~025113.

\bibitem{Erichson2015arxiv}
Erichson, N.~B., Brunton, S.~L., and Kutz, J.~N., \enquote{Compressed Dynamic
  Mode Decomposition for Real-Time Object Detection,} {\em submitted to Journal
  of Real-Time Image Processing\/}, 2015.

\bibitem{Proctor2016siads}
Proctor, J.~L., Brunton, S.~L., and Kutz, J.~N., \enquote{Dynamic mode
  decomposition with control,} {\em SIAM Journal on Applied Dynamical
  Systems\/}, Vol.~15, No.~1, 2016, pp.~142--161.

\bibitem{williams2014edmd}
Williams, M.~O., Kevrekidis, I.~G., and Rowley, C.~W., \enquote{A Data-Driven
  Approximation of the {Koopman} Operator: Extending Dynamic Mode
  Decomposition,} {\em J. Nonlinear Sci.\/}, 2015, pp.~1--40.

\bibitem{Williams2015epl}
Williams, M.~O., Rowley, C.~W., Mezi{\'c}, I., and Kevrekidis, I.~G.,
  \enquote{Data fusion via intrinsic dynamic variables: An application of
  data-driven {K}oopman spectral analysis,} {\em EPL (Europhysics Letters)\/},
  Vol.~109, No.~4, 2015, pp.~40007.

\bibitem{duke2012error}
Duke, D., Soria, J., and Honnery, D., \enquote{An error analysis of the dynamic
  mode decomposition,} {\em Exp. Fluids\/}, Vol.~52, No.~2, 2012, pp.~529--542.

\bibitem{dawson2016dmdnoise}
Dawson, S. T.~M., Hemati, M.~S., Williams, M.~O., and Rowley, C.~W.,
  \enquote{Characterizing and correcting for the effect of sensor noise in the
  dynamic mode decomposition,} {\em Exp. Fluids\/}, Vol.~57, No.~3, 2016,
  pp.~1--19.

\bibitem{bagheri2014noise}
Bagheri, S., \enquote{Effects of weak noise on oscillating flows: linking
  quality factor, {Floquet} modes, and {Koopman} spectrum,} {\em Phys.
  Fluids\/}, Vol.~26, No.~9, 2014, pp.~094104.

\bibitem{hemati2015tls}
Hemati, M.~S., Rowley, C.~W., Deem, E.~A., and Cattafesta, L.~N.,
  \enquote{De-biasing the dynamic mode decomposition for applied {Koopman}
  spectral analysis of noisy datasets,} {\em Theo. Comp. Fluid Dyn.\/}, 2017
  (accepted).

\bibitem{noack2015recursive}
Noack, B.~R., Stankiewicz, W., Morzy\'nski, M., and Schmid, P.~J.,
  \enquote{Recursive dynamic mode decomposition of a transient and
  post-transient cylinder wake flows,} {\em J. Fluid Mech.\/}, Vol.~809, 2016,
  pp.~843--872.

\bibitem{Kutz2016siads}
Kutz, J.~N., Fu, X., and Brunton, S.~L., \enquote{Multi-Resolution Dynamic Mode
  Decomposition,} {\em SIAM Journal on Applied Dynamical Systems\/}, Vol.~15,
  No.~2, 2016, pp.~713--735.

\bibitem{Brunton2016havok}
Brunton, S.~L., Brunton, B.~W., Proctor, J.~L., Kaiser, E., and Kutz, J.~N.,
  \enquote{Chaos as an intermittently forced linear system,} {\em Nature
  Comm.\/}, Vol.~8, No.~19, 2017.

\bibitem{Tu:thesis}
Tu, J.~H., {\em Dynamic mode decomposition: theory and applications\/}, Ph.D.
  thesis, Princeton University, 2013.

\bibitem{guckenheimer_holmes}
Holmes, P. and Guckenheimer, J., {\em Nonlinear oscillations, dynamical
  systems, and bifurcations of vector fields\/}, Vol.~42 of {\em Applied
  Mathematical Sciences\/}, Springer-Verlag, Berlin, Heidelberg, 1983.

\bibitem{Koopman1931pnas}
Koopman, B.~O., \enquote{Hamiltonian Systems and Transformation in {H}ilbert
  Space,} {\em Proc. Nat. Academy Sci.\/}, Vol.~17, No.~5, 1931, pp.~315--318.

\bibitem{Mezic2005nd}
Mezi{\'c}, I., \enquote{Spectral properties of dynamical systems, model
  reduction and decompositions,} {\em Nonlinear Dynamics\/}, Vol.~41, No. 1-3,
  2005, pp.~309--325.

\bibitem{Budivsic2012chaos}
Budi{\v{s}}i{\'c}, M., Mohr, R., and Mezi{\'c}, I., \enquote{Applied Koopmanism
  a),} {\em Chaos: An Interdisciplinary Journal of Nonlinear Science\/},
  Vol.~22, No.~4, 2012, pp.~047510.

\bibitem{Kevrekidis2003cms}
Kevrekidis, I.~G., Gear, C.~W., Hyman, J.~M., Kevrekidis, P.~G., Runborg, O.,
  and Theodoropoulos, C., \enquote{Equation-Free, Coarse-Grained Multiscale
  Computation: Enabling Microscopic Simulators to Perform System-Level
  Analysis,} {\em Communications in Mathematical Science\/}, Vol.~1, No.~4,
  2003, pp.~715--762.

\bibitem{Giannakis2015arxiv}
Giannakis, D., \enquote{Data-driven spectral decomposition and forecasting of
  ergodic dynamical systems,} {\em arXiv preprint arXiv:1507.02338\/}, 2015.

\bibitem{Kaiser:arXiv17}
Kaiser, E., Kutz, J.~N., and Brunton, S.~L., \enquote{Data-driven discovery of
  Koopman eigenfunctions for control,} {\em in review\/}, 2017.

\bibitem{Brunton2016plosone}
Brunton, S.~L., Brunton, B.~W., Proctor, J.~L., and Kutz, J.~N.,
  \enquote{Koopman observable subspaces and finite linear representations of
  nonlinear dynamical systems for control,} {\em PLoS ONE\/}, Vol.~11, No.~2,
  2016, pp.~e0150171.

\bibitem{Bagheri:2013}
Bagheri, S., \enquote{Koopman-mode decomposition of the cylinder wake,} {\em J.
  Fluid Mech.\/}, Vol.~726, 2013, pp.~596--623.

\bibitem{Gaspard:PRE95}
Gaspard, P., Nicolis, G., Provata, A., and Tasaki, S., \enquote{Spectral
  signature of the pitchfork bifurcation: {L}iouville eqaution approach,} {\em
  Phys. Rev. E\/}, Vol.~51, No.~1, 1995, pp.~74--94.

\bibitem{susuki2011nonlinear}
Susuki, Y. and Mezic, I., \enquote{Nonlinear Koopman modes and coherency
  identification of coupled swing dynamics,} {\em IEEE Transactions on Power
  Systems\/}, Vol.~26, No.~4, 2011, pp.~1894--1904.

\bibitem{susuki2012nonlinear}
Susuki, Y. and Mezic, I., \enquote{Nonlinear Koopman modes and a precursor to
  power system swing instabilities,} {\em IEEE Transactions on Power
  Systems\/}, Vol.~27, No.~3, 2012, pp.~1182--1191.

\bibitem{Brunton2016pnas}
Brunton, S.~L., Proctor, J.~L., and Kutz, J.~N., \enquote{Discovering governing
  equations from data by sparse identification of nonlinear dynamical systems,}
  {\em Proc. Nat. Academy Sci.\/}, Vol.~113, No.~15, 2016, pp.~3932--3937.

\bibitem{Arbabi2016arxiv}
Arbabi, H. and Mezi{\'c}, I., \enquote{Ergodic theory, Dynamic Mode
  Decomposition and Computation of Spectral Properties of the Koopman
  operator,} {\em arXiv preprint arXiv:1611.06664\/}, 2016.

\bibitem{JuniperHanifiTheofilis}
Juniper, M.~P., Hanifi, A., and Theofilis, V., \enquote{Modal stability
  theory,} {\em Appl. Mech. Rev.\/}, Vol.~66, No.~2, 2014, pp.~024804.

\bibitem{TheofilisPrAeS2003}
Theofilis, V., \enquote{Advances in global linear instability of nonparallel
  and three-dimensional flows,} {\em Prog. Aero. Sciences\/}, Vol.~39, No.~4,
  2003, pp.~249--315.

\bibitem{AkervikEtAl2008}
{\AA}kervik, E., Ehrenstein, U., Gallaire, F., and Henningson, D.~S.,
  \enquote{Global two-dimensional stability measures of the flat plate
  boundary-layer flow,} {\em Eur. J. Mech. B / Fluids\/}, Vol.~27, 2008,
  pp.~501--513.

\bibitem{TheofilisColoniusAIAA2003}
Theofilis, V. and Colonius, T., \enquote{An algorithm for the recovery of 2-
  and {3-D BiGlobal} Instabilities of Compressible Flow Over 2-D Open
  Cavities,} AIAA Paper 2003-4143, 2003.

\bibitem{Zhang:PF16}
Zhang, W. and Samtaney, R., \enquote{{BiGlobal} linear stability analysis on
  low-{Re} flow past an airfoil at high angle of attack,} {\em Phys. Fluids\/},
  Vol.~28, 2016, pp.~044105.

\bibitem{Sun:TCFDXX}
Sun, Y., Taira, K., Cattafesta, L.~N., and Ukeiley, L.~S., \enquote{Spanwise
  effects on instabilities of compressible flow over a long rectangular
  cavity,} {\em Theor. Comput. Fluid Dyn.\/}, 2016 (in press).

\bibitem{Sun:JFM17}
Sun, Y., Taira, K., Cattafesta, L.~N., and Ukeiley, L.~S., \enquote{Biglobal
  instabilities of spanwise-periodic compressible open-cavity flows,} {\em J.
  Fluid Mech.\/}, Vol.~826, 2017, pp.~270--301.

\bibitem{Rodriguez:AIAAJ09}
Rodr\'iguez, D. and Theofilis, V., \enquote{Massively Parallel Solution of the
  {BiGlobal} Eigenvalue Problem Using Dense Linear Algebra,} {\em AIAA J.\/},
  Vol.~47, No.~10, 2009, pp.~2449--2459.

\bibitem{Tezuka:AIAAJ06}
Tezuka, A. and Suzuki, K., \enquote{Three-dimensional global linear stability
  analysis of flow around a spheroid,} {\em AIAA J.\/}, Vol.~44, No.~8, 2006,
  pp.~1697--1708.

\bibitem{Bagheri:JFM09}
Bagheri, S., Schlatter, P., Schmid, P.~J., and Henningson, D.~S.,
  \enquote{Global stability of a jet in crossflow,} {\em J. Fluid Mech.\/},
  Vol.~624, 2009, pp.~33--44.

\bibitem{Liu:JFM16}
Liu, Q., G\'omez, F., and Theofilis, V., \enquote{Linear instability analysis
  of low-$Re$ incompressible flow over a long rectangular finite-span open
  cavity,} {\em J. Fluid Mech.\/}, Vol.~799, 2016, pp.~R2.

\bibitem{Keller:MC75}
Keller, H.~B., \enquote{Approximation methods for nonlinear problems with
  application to two-point boundary value problems,} {\em Math. Comput.\/},
  Vol.~29, No. 130, 1975, pp.~464--474.

\bibitem{BresColonius}
Bres, G.~A. and Colonius, T., \enquote{Three-dimensional instabilities in
  compressible flow over open cavities,} {\em J. Fluid Mech.\/}, Vol.~599,
  2008, pp.~309--339.

\bibitem{ParedesCMAME:13}
Paredes, P., Hermanns, M., Chainche, S.~L., and Theofilis, V., \enquote{Order
  $10^4$ speedup in global linear instability analysis using matrix formation,}
  {\em Comput. Methods Appl. Mech. Engrg.\/}, Vol.~253, 2013, pp.~287--304.

\bibitem{TatsumiYoshimura}
Tatsumi, T. and Yoshimura, T., \enquote{Stability of the laminar flow in a
  rectangular duct,} {\em J. Fluid Mech.\/}, Vol.~212, 1990, pp.~437--449.

\bibitem{ParedesTheofilisJFS}
Paredes, P. and Theofilis, V., \enquote{Centerline instabilities on the
  hypersonic international flight research experimentation {HIFiRE}-5 elliptic
  cone model,} {\em J. Fluids Struct.\/}, Vol.~53, 2015, pp.~36--49.

\bibitem{Akervik:PF06}
{\AA}kervik, E., Brandt, L., Henningson, D.~S., H{\oe}pffner, J., Marxen, O.,
  and Schlatter, P., \enquote{Steady solutions of the {N}avier-{S}tokes
  equations by selective frequency damping,} {\em Phys. Fluids\/}, Vol.~18,
  2006, pp.~068102.

\bibitem{Tuckerman:IMA00}
Tuckerman, L. and Barkley, D., \enquote{Bifurcation analysis for timesteppers,}
  {\em Numerical Methods for Bifurcation Problems and Large-Scale Dynamical
  Systems\/}, edited by E.~Doedel and L.~S. Tuckerman, Vol.~19, Springer, 2000,
  pp. 453--466.

\bibitem{Kelley:arxiv}
Kelley, C.~T., Kevrekidis, G.~I., and Qiao, L., \enquote{Newton--{Krylov}
  solvers for timesteppers.} {\em arXiv\/}, 2004.

\bibitem{Barkley:JFM96}
Barkley, D. and Henderson, R.~D., \enquote{Three-dimensional {F}loquet
  stability analysis of the wake of a circular cylinder,} {\em J. Fluid
  Mech.\/}, Vol.~322, 1996, pp.~215--241.

\bibitem{HeGioriaPerezTheofilisJFM}
He, W., Gioria, R.~S., P\'erez, J.~M., and Theofilis, V., \enquote{Linear
  instability of low {R}eynolds number massively separated flow around three
  {NACA} airfoils,} {\em J. Fluid Mech.\/}, Vol.~811, 2016, pp.~701--741.

\bibitem{Edstrand:JFM16}
Edstrand, A.~M., Davis, T.~B., Schmid, P.~J., Taira, K., and Cattafesta, L.~N.,
  \enquote{On the mechanism of trailing vortex wandering,} {\em J. Fluid
  Mech.\/}, Vol.~801, 2016, pp.~R1:1--11.

\bibitem{Sipp:JFM07}
Sipp, D. and Lebedev, A., \enquote{Global stability of base and mean flows: a
  general approach and its applications to cylinder and open cavity flows,}
  {\em J. Fluid Mech.\/}, Vol.~593, 2007, pp.~333--358.

\bibitem{Luchini:JFM00}
Luchini, P., \enquote{Reynolds-number-independent instability of the boundary
  layer over a flat surface: optimal perturbations,} {\em J. Fluid Mech.\/},
  Vol.~404, 2000, pp.~289--309.

\bibitem{AbdessemedSharmaSherwinTheofilisPF}
Abdessemed, N., Sharma, A.~S., Sherwin, S.~J., and Theofilis, V.,
  \enquote{Transient growth analysis of the flow past a circular,} {\em Phys.
  Fluids\/}, Vol.~21, 2009, pp.~044103.

\bibitem{AbdessemedSharmaTheofilisJFM}
Abdessemed, N., Sherwin, S.~J., and Theofilis, V., \enquote{Linear instability
  analysis of low pressure turbine flows,} {\em J. Fluid Mech.\/}, Vol.~628,
  2009, pp.~57--83.

\bibitem{SharmaAbdessemedSherwinTheofilis}
Sharma, A., Abdessemed, N., Sherwin, S.~J., and Theofilis, V.,
  \enquote{Transient growth mechanisms of low {R}eynolds number flow over a
  low-pressure turbine blade,} {\em Theo. Comput. Fluid Dyn.\/}, Vol.~25, 2011,
  pp.~19--30.

\bibitem{GasterKitWygnanski}
Gaster, M., Kit, E., and Wygnanski, I., \enquote{Large-scale structures in a
  forced turbulent mixing layer,} {\em J. Fluid Mech.\/}, Vol.~150, 1985,
  pp.~23--39.

\bibitem{ParedesTerhaarOberleithnerTheofilisPaschereit}
Paredes, P., Terhaar, S., Oberleithner, K., Theofilis, V., and Paschereit,
  C.~O., \enquote{Global and Local Hydrodynamic Stability Analysis as a Tool
  for Combustor Dynamics Modeling,} {\em {ASME} J. Eng. Gas Turb. Power\/},
  Vol.~138, 2016, pp.~021504.

\bibitem{TheofilisHeinDallmann}
Theofilis, V., Hein, S., and Dallmann, U., \enquote{On the origins of
  unsteadiness and three-dimensionality in a laminar separation bubble.} {\em
  Phil. Trans. Roy. Soc. London A\/}, Vol.~358, 2000, pp.~3229--3246.

\bibitem{Rodriguez:TCFD11}
Rodr\'iguez, D. and Theofilis, V., \enquote{On the birth of stall cells on
  airfoils,} {\em Theor. Comput. Fluid Dyn.\/}, Vol.~25, 2011, pp.~105--117.

\bibitem{GaitondeEtAl2002}
Gaitonde, D.~V., Canupp, P.~W., and Holden, M.~S., \enquote{Heat transfer
  predictions in a laminar hypersonic viscous/inviscid interaction,} {\em J.
  Thermophys. Heat Trans.\/}, Vol.~16, No.~4, 2002, pp.~481--489.

\bibitem{TumukluLiLevin:AIAAJ16}
Tumuklu, O., Li, Z., and Levin, D.~A., \enquote{Particle Ellipsoidal
  Statistical Bhatnagar--Gross--Krook Approach for Simulation of Hypersonic
  Shocks,} {\em AIAA J.\/}, Vol.~54, No.~12, 2016, pp.~3701--3716.

\bibitem{TumukluLevinTheofilis}
Tumuklu, O., Levin, D.~A., and Theofilis, V., \enquote{Assessment of Degree of
  Steadiness in Boundary Layer Shock-Interaction Flows,} AIAA Paper 2017-1614,
  2017.

\bibitem{TumukluPerezLevinTheofilis}
Tumuklu, O., P\'erez, J., Levin, D., and Theofilis, V., \enquote{On Linear
  Stability Analyses of Hypersonic Laminar Separated Flows in a DSMC Framework.
  Part {I}: Base Flow Computations in a Double Cone and a `Tick' Model,}
  57$^{\rm th}$ {IACAS} conference, {T}el {A}viv/{H}aifa, March 2017.

\bibitem{TumukluPerezTheofilisLevin}
Tumuklu, O., P\'erez, J., Theofilis, V., and Levin, D., \enquote{On Linear
  Stability Analyses of Hypersonic Laminar Separated Flows in a DSMC Framework.
  Part {II}: Residuals Algorithm and the Least Damped Global Modes,} 57$^{\rm
  th}$ {IACAS} conference, {T}el {A}viv/{H}aifa, March 2017.

\bibitem{ParedesGosseTheofilisKimmel}
Paredes, P., Gosse, R., Theofilis, V., and Kimmel, R., \enquote{Linear modal
  instabilities of hypersonic flow over an elliptic cone,} {\em J. Fluid
  Mech.\/}, Vol.~804, 2016, pp.~442--466.

\bibitem{Herbert1997}
Herbert, T., \enquote{Parabolized Stability Equations,} {\em Annu. Rev. Fluid
  Mech.\/}, Vol.~29, 1997, pp.~245--283.

\bibitem{BroadhurstTheofilisSherwin}
Broadhurst, M., Theofilis, V., and Sherwin, S.~J., \enquote{Spectral Element
  Stability Analysis of Vortical Flows,} 6th IUTAM Laminar-Turbulent Transition
  Symposium, Bangalore, India, Dec. 2004, 2006, pp. 153--158.

\bibitem{BroadhurstSherwin}
Broadhurst, M.~S. and Sherwin, S.~J., \enquote{The {P}arabolised {S}tability
  {E}quations for 3{D}-Flows: Implementation and {N}umerical {S}tability,} {\em
  Appl. Num. Math.\/}, Vol.~58, No.~7, 2008, pp.~1017--1029.

\bibitem{ParedesTheofilisRodriguezTendero}
Paredes, P., Theofilis, V., Rodr{\'\i}guez, D., and Tendero, J.~A.,
  \enquote{The {PSE-3D} instability analysis methodology for flows depending
  strongly on two and weakly on the third spatial dimension,} AIAA Paper
  2011-3752, 2011.

\bibitem{DeTullioParedesSandhamTheofilisJFM}
{De Tullio}, N., Paredes, P., Sandham, N., and Theofilis, V.,
  \enquote{Roughness-induced instability and breakdown to turbulence in a
  supersonic boundary-layer,} {\em J. Fluid Mech.\/}, Vol.~735, 2013,
  pp.~613--646.

\bibitem{MartinParedesTCFD2016}
Mart\'in, J. and Paredes, P., \enquote{Three-dimensional instability analysis
  of boundary layers perturbed by streamwise vortices,} {\em Theor. Comp. Fluid
  Dyn.\/}, 2016 (in press).

\bibitem{GalionisHallTCFD}
Galionis, I. and Hall, P., \enquote{Spatial stability of the incompressible
  corner flow,} {\em Theor. Comp. Fluid Dyn.\/}, Vol.~19, No.~2, 2005,
  pp.~77--113.

\bibitem{GalionisHallJFM}
Galionis, I. and Hall, P., \enquote{Stability of the flow in a slowly diverging
  rectangular duct,} {\em J. Fluid Mech.\/}, Vol.~555, 2006, pp.~43--58.

\bibitem{LuchiniBottaroARFM}
Luchini, P. and Bottaro, A., \enquote{Adjoint Equations in Stability Analysis,}
  {\em Annu.\ Rev.\ Fluid\ Mech.\/}, Vol.~46, 2014, pp.~493--517.

\bibitem{McKeonPoF13}
McKeon, B.~J., Sharma, A.~S., and Jacobi, I., \enquote{Experimental
  manipulation of wall turbulence: a systems approach,} {\em Phys. Fluids\/},
  Vol.~25, No. 031301, 2013.

\bibitem{NicholsJovanovic16}
Jeun, J., Nichols, J.~W., and Jovanovi{\'c}, M.~R., \enquote{Input-output
  analysis of high-speed axisymmetric isothermal jet noise,} {\em Phys.
  Fluids\/}, Vol.~28, No.~4, 2016, pp.~047101.

\bibitem{Garnaut13}
Garnaut, X., Lesshafft, L., Schmid, P., and Huerre, P., \enquote{The preferred
  mode of incompressible jets: linear frequency response analysis,} {\em J.
  Fluid Mech.\/}, Vol.~716, 2013, pp.~189--202.

\bibitem{Github13}
Luhar, M., Sharma, A.~S., and McKeon, B.~J., \enquote{Repository for all
  Matlab/Python code for the Navier-Stokes resolvent analysis,}
  https://github.com/mluhar/resolvent, 2013.

\bibitem{McKeon2010}
McKeon, B.~J. and Sharma, A.~S., \enquote{A critical layer model for turbulent
  pipe flow,} {\em J. Fluid Mech.\/}, Vol.~658, 2010, pp.~336--382.

\bibitem{Farrell93}
Farrell, B. and Ioannou, J., \enquote{Stochastic forcing of the linearized
  {N}avier-{S}tokes equations,} {\em Phys. Fluids\/}, Vol.~5, No.~11, 1993,
  pp.~2600--2609.

\bibitem{Jovanovic05}
Jovanovi\'c, M.~R. and Bamieh, B., \enquote{Componentwise energy amplification
  in channel flows,} {\em J. Fluid Mech.\/}, Vol.~534, 2005, pp.~145--183.

\bibitem{Towne15}
Towne, A., Colonius, T., and Schmidt, O., \enquote{Empirical resolvent mode
  decomposition,} {\em Bulletin of the APS\/}, No. G17.00006, 2015.

\bibitem{Beneddine:JFM17}
Beneddine, S., Yegavian, R., Sipp, D., and Leclaire, B., \enquote{Unsteady flow
  dynamics reconstruction from mean flow and point sensors: an experimental
  study,} {\em J. Fluid Mech.\/}, Vol.~824, 2017, pp.~174--201.

\bibitem{Beneddinestep16}
Beneddine, S., Sipp, D., Arnault, A., Dandois, J., and Lesshafft, L.,
  \enquote{Conditions for validity of mean flow stability analysis and
  application to the determination of coherent structures in a turbulent
  backward facing step flow,} {\em J. Fluid Mech.\/}, Vol.~798, 2016,
  pp.~485--504.

\bibitem{Gomezcavity16}
G\'omez, F., Blackburn, H.~M., Rudman, M., Sharma, A.~S., and Mc{K}eon, B.~J.,
  \enquote{A reduced-order model of three-dimensional unsteady flow in a cavity
  based on the resolvent operator,} {\em J. Fluid Mech.\/}, Vol.~798, 2016,
  pp.~R2.

\bibitem{Papadakis14}
Lu, L. and Papadakis, G., \enquote{An iterative method for the computation of
  the response of linearised Navier-Stokes equations to harmonic forcing and
  application to forced cylinder wakes,} {\em Int. J. Num. Meth. Fluids\/},
  Vol.~74, No.~11, 2014, pp.~794--817.

\bibitem{Luharcompliant15}
Luhar, M., Sharma, A.~S., and McKeon, B.~J., \enquote{A framework for studying
  the effect of compliant surfaces on wall turbulence,} {\em J. Fluid Mech.\/},
  Vol.~768, 2015, pp.~415--441.

\bibitem{Jacobidynamic11}
Jacobi, I. and McKeon, B.~J., \enquote{Dynamic roughness-perturbation of a
  turbulent boundary layer,} {\em J. Fluid Mech.\/}, Vol.~688, 2011,
  pp.~258--296.

\bibitem{Luharcontrol14}
Luhar, M., Sharma, A.~S., and McKeon, B.~J., \enquote{Opposition control within
  the resolvent analysis framework,} {\em J. Fluid Mech.\/}, Vol.~749, 2014,
  pp.~597--626.

\bibitem{Sharma13}
Sharma, A.~S. and McKeon, B.~J., \enquote{On coherent structure in wall
  turbulence,} {\em J. Fluid Mech.\/}, Vol.~728, 2013, pp.~196--238.

\bibitem{TowneERMD15}
Towne, A., Colonius, T., Jordan, P., Cavalieri, A.~V., and Bres, G.~A.,
  \enquote{Stochastic and nonlinear forcing of wavepackets in a {M}ach 0.9
  jet,} AIAA Paper 2015-2217, 2015.

\bibitem{Moarref13}
Moarref, R., Sharma, A.~S., Tropp, J.~A., and McKeon, B.~J.,
  \enquote{Model-based scaling and prediction of the streamwise energy
  intensity in high-{R}eynolds number turbulent channels,} {\em J. Fluid
  Mech.\/}, Vol.~734, 2013, pp.~275--316.

\bibitem{Moarref14}
Moarref, R., Jovanovic, M.~R., Sharma, A.~S., Tropp, J.~A., and McKeon, B.~J.,
  \enquote{A low-order decomposition of turbulent channel flow via resolvent
  analysis and convex optimization,} {\em Phys. Fluids\/}, Vol.~26, No. 051701,
  2014.

\bibitem{SharmaECS16}
Sharma, A.~S., Moarref, R., McKeon, B.~J., Park, J.~S., Graham, M., and Willis,
  A.~P., \enquote{Low-dimensional representations of exact coherent states of
  the {N}avier-{S}tokes equations from the resolvent model of wall turbulence,}
  {\em Phys. Rev. E\/}, Vol.~93, No. 021102(R), 2016.

\bibitem{SharmaPTAXX}
Sharma, A.~S., Moarref, R., and Mc{K}eon, B.~J., \enquote{Scaling and
  interaction of self-similar modes in models of high-{R}eynolds number wall
  turbulence,} {\em Phil. Trans. Roy. Soc. London A\/}, 2016 (in press).

\bibitem{SharmaMezicM16}
Sharma, A.~S., Mezi{\'{c}}, I., and McKeon, B.~J., \enquote{On the
  correspondence between {K}oopman mode decomposition, resolvent mode
  decomposition, and invariant solutions of the {N}avier-{S}tokes equations,}
  {\em Phys. Rev. Fluids\/}, Vol.~1, No.~3, 2016, pp.~032403(R).

\bibitem{Brunton:AMR15}
Brunton, S.~L. and Noack, B.~R., \enquote{Closed-loop turbulence control:
  progress and challenges,} {\em App. Mech. Rev.\/}, Vol.~67, No.~5, 2015,
  pp.~050801.

\bibitem{Bui-Thanh:AIAAJ04}
Bui-Thanh, T., Damodaran, M., and Wilcox, K., \enquote{Aerodynamic Data
  Reconstruction and Inverse Design Using Proper Orthogonal Decomposition,}
  {\em AIAA J.\/}, Vol.~42, No.~8, 2004, pp.~1505--1516.

\bibitem{LeGresleyThesis}
LeGresley, P.~A., {\em Application of proper orthogonal decomposition ({POD})
  to design decomposition methods\/}, Ph.D. thesis, Stanford Univ., 2005.

\end{thebibliography}

\end{document}